\documentclass[12pt]{article}

\usepackage{amsmath}
\usepackage{graphicx}
\usepackage{enumerate}
\usepackage{natbib}
\usepackage{epstopdf}
\usepackage{caption}
\usepackage{subcaption}
\usepackage[hyphens]{url}
\usepackage{hyperref}
\usepackage[hyphenbreaks]{breakurl}

\usepackage{mathtools}

\def\bs{\boldsymbol{s}}
\usepackage{txfonts}
\usepackage{cite}
\usepackage{listings}
\usepackage{booktabs}
\usepackage{amssymb}
\usepackage{comment}
\usepackage{xcolor}

\usepackage{cases}
\usepackage{accents}
\usepackage{enumitem}
\usepackage{multirow}
\usepackage{array}

\usepackage[utf8]{inputenc}
\usepackage{algorithm}
\usepackage{algpseudocode}
\usepackage[algo2e]{algorithm2e}
\usepackage{setspace}

\usepackage{tikz}
\DeclareRobustCommand\full  {\tikz[baseline=-0.6ex]\draw[thick] (0,0)--(0.5,0);}

\DeclareRobustCommand\dashed{\tikz[baseline=-0.6ex]\draw[thick,dashed] (0,0)--(0.54,0);}
\DeclareRobustCommand\chain {\tikz[baseline=-0.6ex]\draw[thick,dashdotted] (0,0)--(0.5,0);}
\usetikzlibrary{fit}
\tikzset{%
  highlight/.style={rectangle,rounded corners,draw,
    fill opacity=0.5,thick,inner sep=0pt}
}

\newcommand{\ubar}[1]{\underaccent{\bar}{#1}}

\newcommand{\dbtilde}[1]{\tilde{\tilde{#1}}}

\def\code#1{\texttt{#1}}
\newcommand\numberthis{\addtocounter{equation}{1}\tag{\theequation}}

\DeclarePairedDelimiter{\sceil}{\big\lceil}{\big\rceil}
\DeclarePairedDelimiter{\ceil}{\Big\lceil}{\Big\rceil}
\DeclarePairedDelimiter{\floor}{\Big\lfloor}{\Big\rfloor}

\DeclareMathOperator*{\argmin}{arg\,min}

\newtheorem{Thm}{\bf Theorem}

\newtheorem{Lem}{Lemma}

\def\b{\boldsymbol}

\def\ve{\varepsilon}
\def\bve{\boldsymbol{\varepsilon}}

\def\CS#1{\texttt{\textbackslash#1}} 

\newcommand*{\email}[1]{\href{mailto:#1}{\nolinkurl{#1}} }

\begin{document}
\sloppy

\def\spacingset#1{\renewcommand{\baselinestretch}%
{#1}\small\normalsize} \spacingset{1}

\title{\bf Detecting linear trend changes in data sequences}

\author{Hyeyoung Maeng\footnote{Department of Mathematical Sciences, Durham University. Email: \email{hyeyoung.maeng@durham.ac.uk}}  \ and Piotr Fryzlewicz\footnote{Department of Statistics, London School of Economics. Email: \email{p.fryzlewicz@lse.ac.uk}}}
\date{}
  \maketitle

\bigskip
\begin{abstract} 
We propose TrendSegment, a methodology for detecting multiple change-points corresponding to linear trend changes in one dimensional data. A core ingredient of TrendSegment is a new Tail-Greedy Unbalanced Wavelet transform: a conditionally orthonormal, bottom-up transformation of the data through an adaptively constructed unbalanced wavelet basis, which results in a sparse representation of the data. 
Due to its bottom-up nature, this multiscale decomposition focuses on local features in its early stages and on global features next which enables the detection of both long and short linear trend segments at once. To reduce the computational complexity, the proposed method merges multiple regions in a single pass over the data. We show the consistency of the estimated number and locations of change-points. The practicality of our approach is demonstrated through simulations and two real data examples, involving Iceland temperature data and sea ice extent of the Arctic and the Antarctic. Our methodology is implemented in the R package \code{trendsegmentR}, available from CRAN.
\end{abstract}

\noindent%
\vfill

\newpage
\spacingset{1.35} 

\section{Introduction} \label{sec1}

Multiple change-point detection is a problem of importance in many applications; recent examples include automatic detection of change-points in cloud data to maintain the performance and availability of an app or a website \citep{james2016leveraging}, climate change detection in tropical cyclone records \citep{robbins2011changepoints}, detecting exoplanets from light curve data \citep{fisch2018linear}, detecting changes in the DNA copy number \citep{olshen2004circular, jeng2012simultaneous, bardwell2017bayesian}, estimation of stationary intervals in potentially cointegrated stock prices \citep{matteson2013locally}, estimation of change-points in multi-subject fMRI data \citep{robinson2010change} and detecting changes in vegetation trends \citep{jamali2015detecting}.

This paper considers the change-point model
\begin{equation} \label{e1}
X_t = f_t + \ve_t,\quad  t=1, \ldots, T,
\end{equation}
where $f_t$ is a deterministic and piecewise-linear signal containing $N$ change-points, i.e. time indices at which the slope and/or the intercept in $f_t$ undergoes changes. 
These changes occur at unknown locations $\eta_1, \eta_2, \ldots, \eta_N$. 
In this article, we assume that the $\ve_t$'s are iid N(0, $\sigma^2$) and in the supplementary material, we show how our method can be extended to dependent and/or non-Gaussian noise such as $\varepsilon_t$ following a stationary Gaussian AR process or t-distribution. 
The true change-points $\{\eta_i\}_{i=1}^N$ are such that, 
\begin{align} \label{cp}
\begin{split}
& f_t= \theta_{\ell,1} + \theta_{\ell,2} \; t \; \text{ for } \; t\in[\eta_{\ell-1}+1, \eta_{\ell}],  \; \ell=1, \ldots, N+1 \\
& \text{ where } \; f_{\eta_\ell} + \theta_{\ell,2} \neq f_{\eta_\ell+1}  \; \text{ for } \; \ell=1, \ldots, N.
\end{split}
\end{align}
This definition permits both continuous and discontinuous changes in the linear trend. 

Our main interest is in the estimation of $N$ and $\eta_1, \eta_2, \ldots, \eta_N$ under some assumptions that quantify the difficulty of detecting each $\eta_i$; therefore, our aim is to segment the data into sections of linearity in $f_t$. 
In detail, a change-point located close to its neighbouring ones can only be detected when it has a large enough size of linear trend change, while a change-point capturing a small size of linear trend change requires a longer distance from its adjacent change-points to be detected.
Detecting linear trend changes is an important applied problem in a variety of fields, including climate change, as illustrated in Section \ref{sec5}. 

The change-point detection procedure proposed in this paper is referred to as TrendSegment; it is designed to work well in the presence of either long or short spacings between neighbouring change-points, or a mixture of both.
The engine underlying TrendSegment is a new Tail-Greedy Unbalanced Wavelet (TGUW) transform: a conditionally orthonormal, bottom-up transformation for univariate data sequences through an adaptively constructed unbalanced wavelet basis, which results in a sparse representation of the data.
In this article, we show that TrendSegment offers good performance in estimating the number and locations of change-points across a wide range of signals containing constant and/or linear segments. TrendSegment is also shown to be statistically consistent and computationally efficient.

In earlier related work regarding linear trend changes, \citet{bai1998estimating} consider the estimation of linear models with multiple structural changes by least-squares and present Wald-type tests for the null hypothesis of no change.
\citet{kim2009ell_1} and \citet{tibshirani2014adaptive} consider `trend filtering' with the $L_1$ penalty and \citet{fearnhead2019detecting} detect changes in the slope with an $L_0$ regularisation via a dynamic programming algorithm. 
\citet{spiriti2013knot} study two algorithms for optimising the knot locations in least-squares and penalised splines.
\citet{baranowski2019narrowest} propose a multiple change-point detection device termed Narrowest-Over-Threshold (NOT), which focuses on the narrowest segment among those whose contrast exceeds a pre-specified threshold. 
\citet{anastasiou2021detecting} propose the Isolate-Detect (ID) approach which continuously searches expanding data segments for changes.
\citet{yu2020localising} propose a two-step algorithm for detecting multiple change-points in piecewise polynomials with general degrees.

\citet{keogh2004segmenting} mention that sliding windows, top-down and bottom-up approaches are three principal categories which most time series segmentation algorithms can be grouped into. 
\citet{keogh2004segmenting} apply those three approaches to the detection of changes in linear trends in 10 different signals and discover that the performance of bottom-up methods is better than that of top-down methods and sliding windows, notably when the underlying signal has jumps, sharp cusps or large fluctuations.  
Bottom-up procedures have rarely been used in change-point detection. 
\citet{matteson2014nonparametric} use an agglomerative algorithm for hierarchical clustering in the context of change-point analysis.
\citet{keogh2004segmenting} merge adjacent segments of the data according to a criterion involving the minimum residual sum of squares (RSS) from a linear fit, until the RSS falls under
a certain threshold; but the lack of precise recipes for the choice of this threshold parameter causes the performance of this method to be somewhat unstable, as we report in Section
\ref{sec4}.

As illustrated later in this paper, our TGUW transform, which underlies TrendSegment, is designed to work well in detecting frequent change-points or abrupt local features in which many existing change-point detection methods for the piecewise-linear model fail.
The TGUW transform constructs, in a bottom-up way, an adaptive wavelet basis by consecutively merging neighbouring segments of the data starting from the finest level (throughout the paper, we refer to a wavelet basis as adaptive if it is constructed in a data-driven way). 
This enables it to identify local features at an early stage, before it proceeds to focus on more global features corresponding to longer data segments.

\citet{fryzlewicz2017tail} introduces the Tail-Greedy Unbalanced Haar (TGUH) transform, a bottom-up, agglomerative, data-adaptive transformation of univariate sequences that facilitates 
change-point detection in the piecewise-constant sequence model. 
The current paper extends this idea to adaptive wavelets other than adaptive Haar, which enables change-point detection in the piecewise-linear model (and, in principle, to higher-order piecewise polynomials, where the details can be found in Section G of the supplementary material). 
We emphasise that this extension from TGUH to TGUW is both conceptually and technically non-trivial, due to the fact that it is not a priori clear how to construct a suitable wavelet basis in TGUW for wavelets other than adaptive Haar; this is due to the non-uniqueness of the local orthonormal matrix transformation for performing each merge in TGUW, which does not occur in TGUH.
We solve this issue by imposing certain guiding principles in the way the merges are performed, which enables detecting not only long trend segments, but also frequent change-points including abrupt local features. 
The computational cost of TGUW is the same as TGUH. 
Important properties of the TGUW transform include orthonormality conditional on the merging order, nonlinearity and ``tail-greediness", and will be investigated in Section \ref{sec2}.
The TGUW transform is the first step of the TrendSegment procedure, which involves four steps.

The remainder of the article is organised as follows. 
Section \ref{sec2} gives a full description of the TrendSegment procedure and the relevant theoretical results are presented in Section \ref{sec3}. 
The supporting simulation studies are described in Section \ref{sec4} and our methodology is illustrated in Section \ref{sec5} through climate datasets. 
The proofs of our main theoretical results are in Appendix \ref{apx}. The supplementary material includes theoretical results for dependent and/or non-Gaussian noise, extension to piecewise-quadratic signal, details of robust threshold selection and extra simulation and data application results. 
The TrendSegment procedure is implemented in the R package \code{trendsegmentR}, available from CRAN.

\section{Methodology} \label{sec2}

\subsection{Summary of TrendSegment} \label{summ}

\begin{figure}[ht!] 
     \centering
    \begin{subfigure}[t]{0.49\textwidth} 
        \raisebox{-\height}{\includegraphics[width=\textwidth]{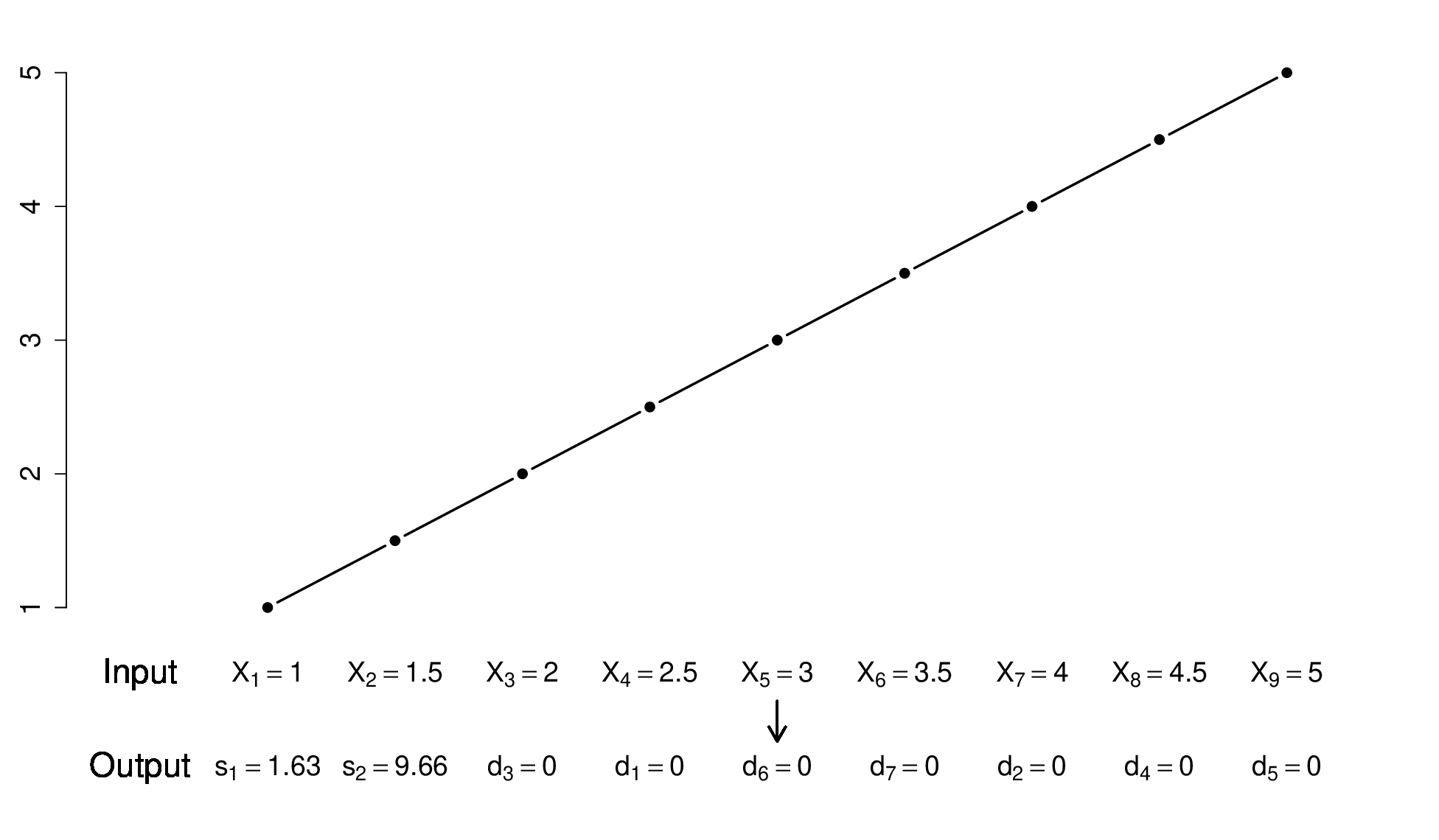}}
        \caption{No change-point without noise}
        \label{fig:tguw_ex1}
    \end{subfigure}
    \hfill
    \begin{subfigure}[t]{0.49\textwidth} 
        \raisebox{-\height}{\includegraphics[width=\textwidth]{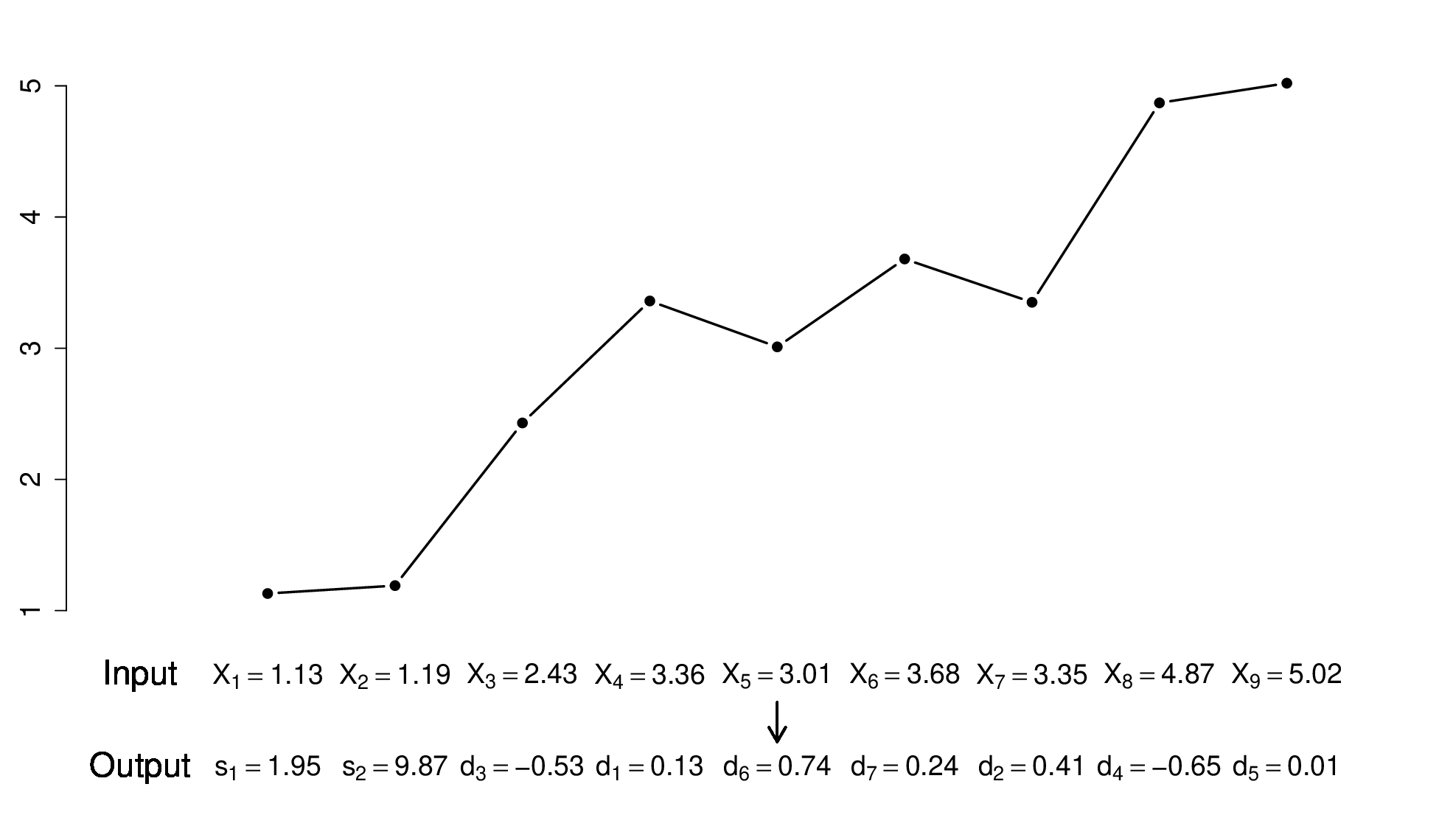}}
        \caption{No change-point with noise}
        \label{fig:tguw_ex2}
    \end{subfigure}

    \begin{subfigure}[t]{0.49\textwidth} 
        \raisebox{-\height}{\includegraphics[width=\textwidth]{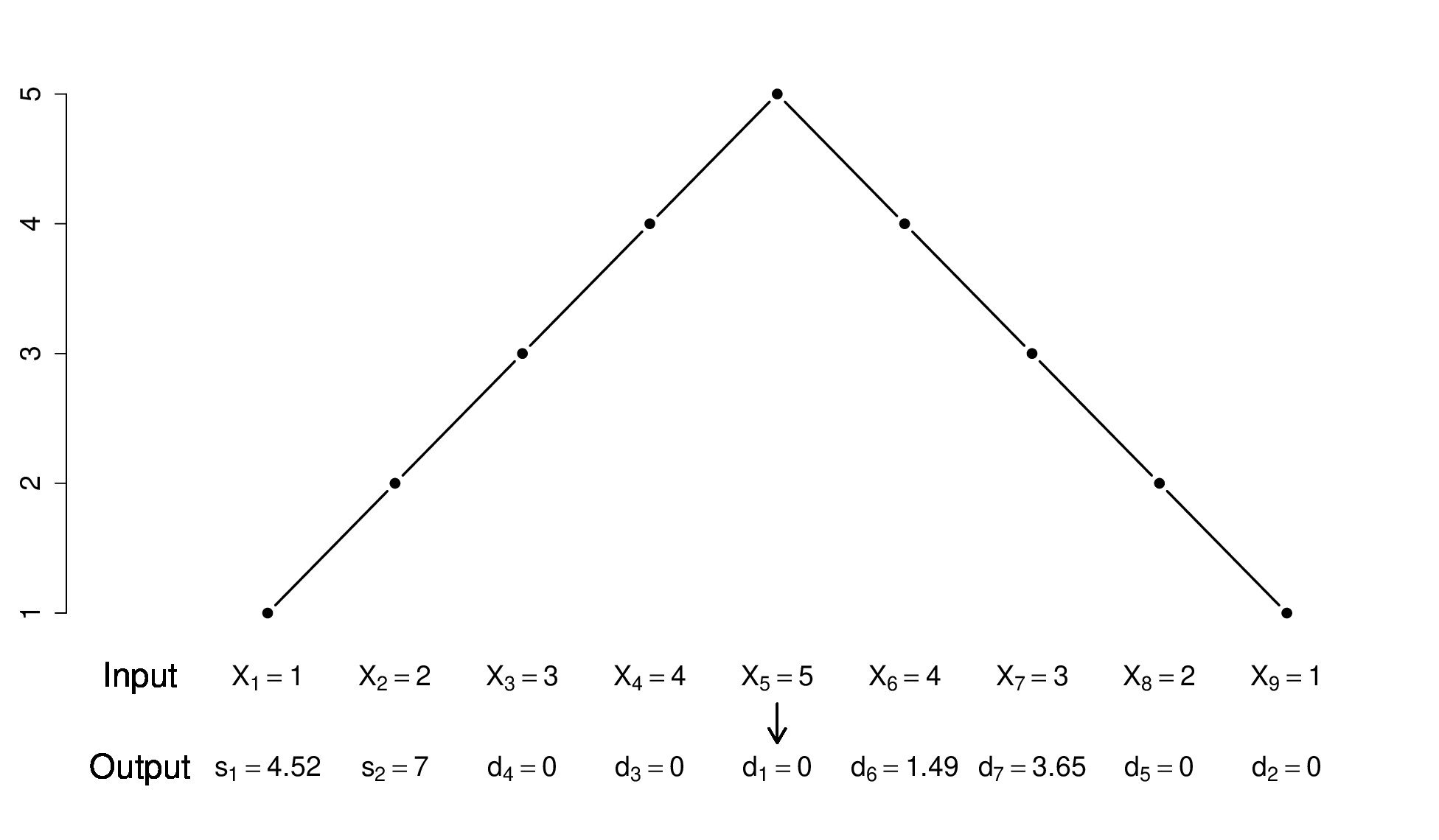}}
        \caption{One change-point without noise}
        \label{fig:tguw_ex3}
    \end{subfigure}
    \hfill
    \begin{subfigure}[t]{0.49\textwidth} 
        \raisebox{-\height}{\includegraphics[width=\textwidth]{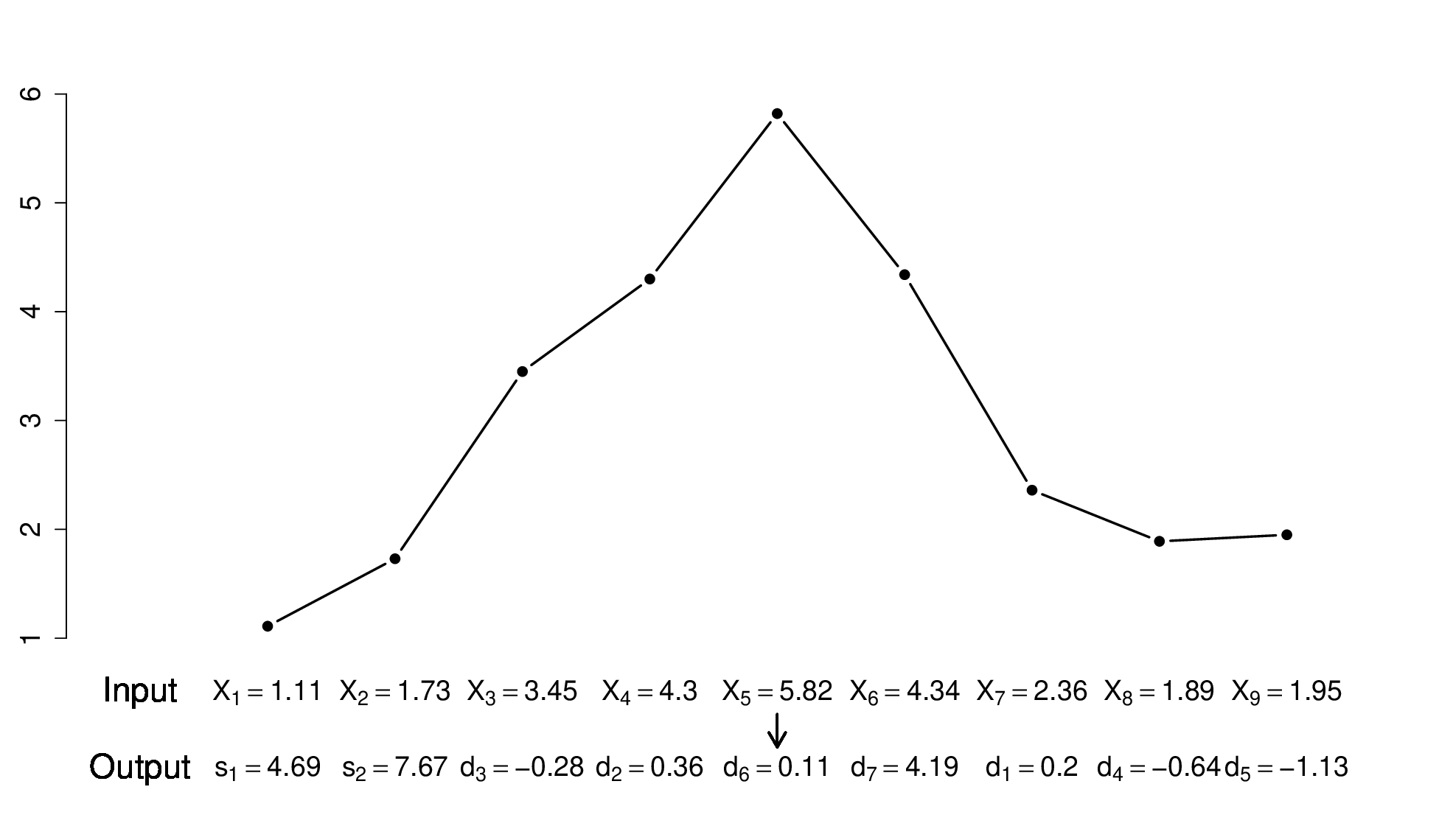}}
        \caption{One change-point with noise}
        \label{fig:tguw_ex4}
    \end{subfigure}
    
    \caption {Multiscale decomposition of the data through the TGUW transform when the data has no change-points ((a), (b)) or one change-point ((c), (d)). $s_1$ and $s_2$ are the smooth coefficients obtained through the TGUW transform and $d_k$ is the detail coefficient obtained in the $k^\text{th}$ merge. When the data has no noise ((a), (c)), $d_k = 0$ for all $k$ in (a) while two non-zero coefficients $d_6$ and $d_7$ encode the single change in (c). }
\label{fig:tguw_ex}
\end{figure}

\begin{figure}[ht!] 
     \centering
    \begin{subfigure}[t]{0.32\textwidth} 
        \raisebox{-\height}{\includegraphics[width=\textwidth]{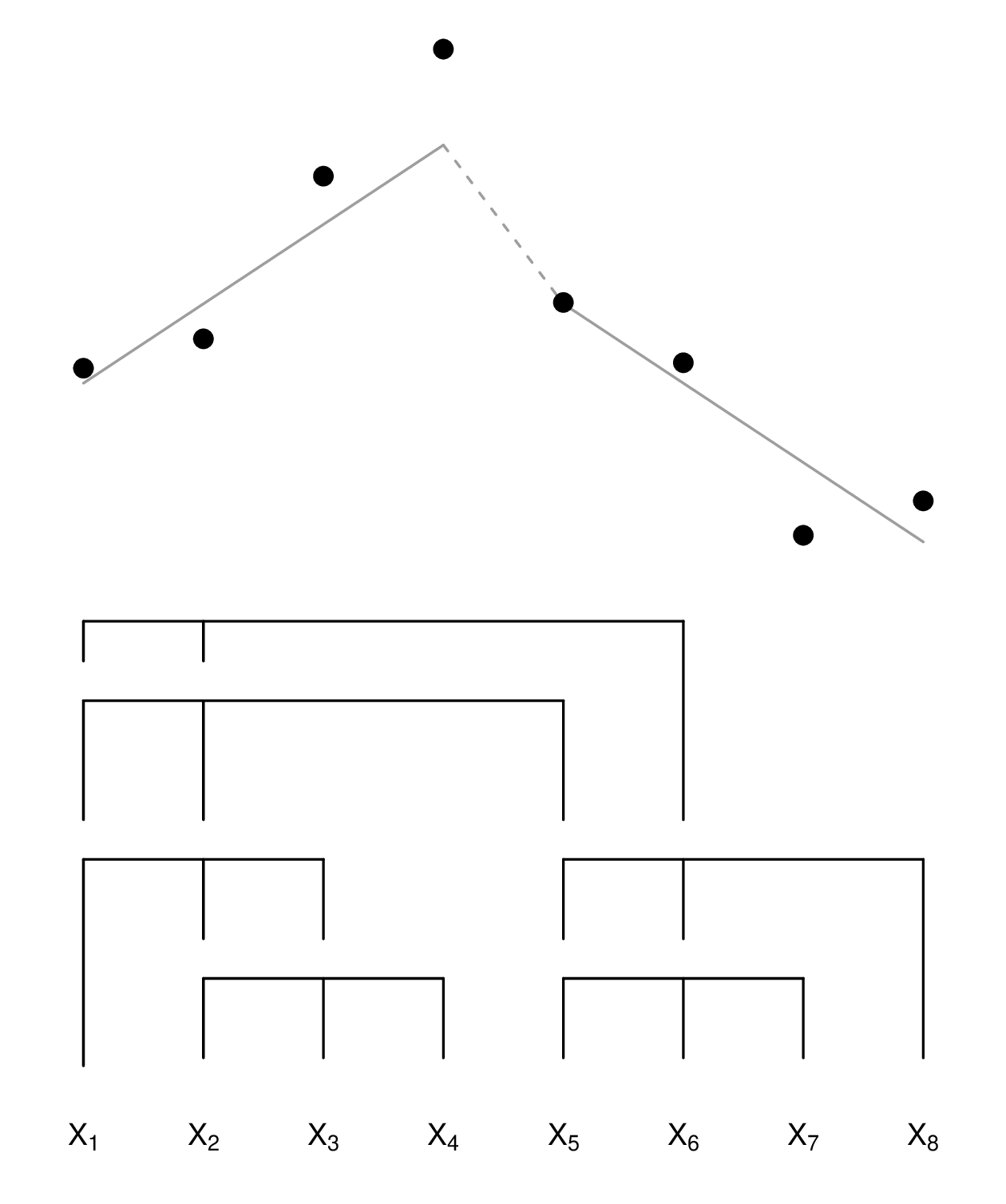}}
        \caption{TGUW transform}
        \label{fig:f1}
    \end{subfigure}
    \hfill
    \begin{subfigure}[t]{0.32\textwidth} 
        \raisebox{-\height}{\includegraphics[width=\textwidth]{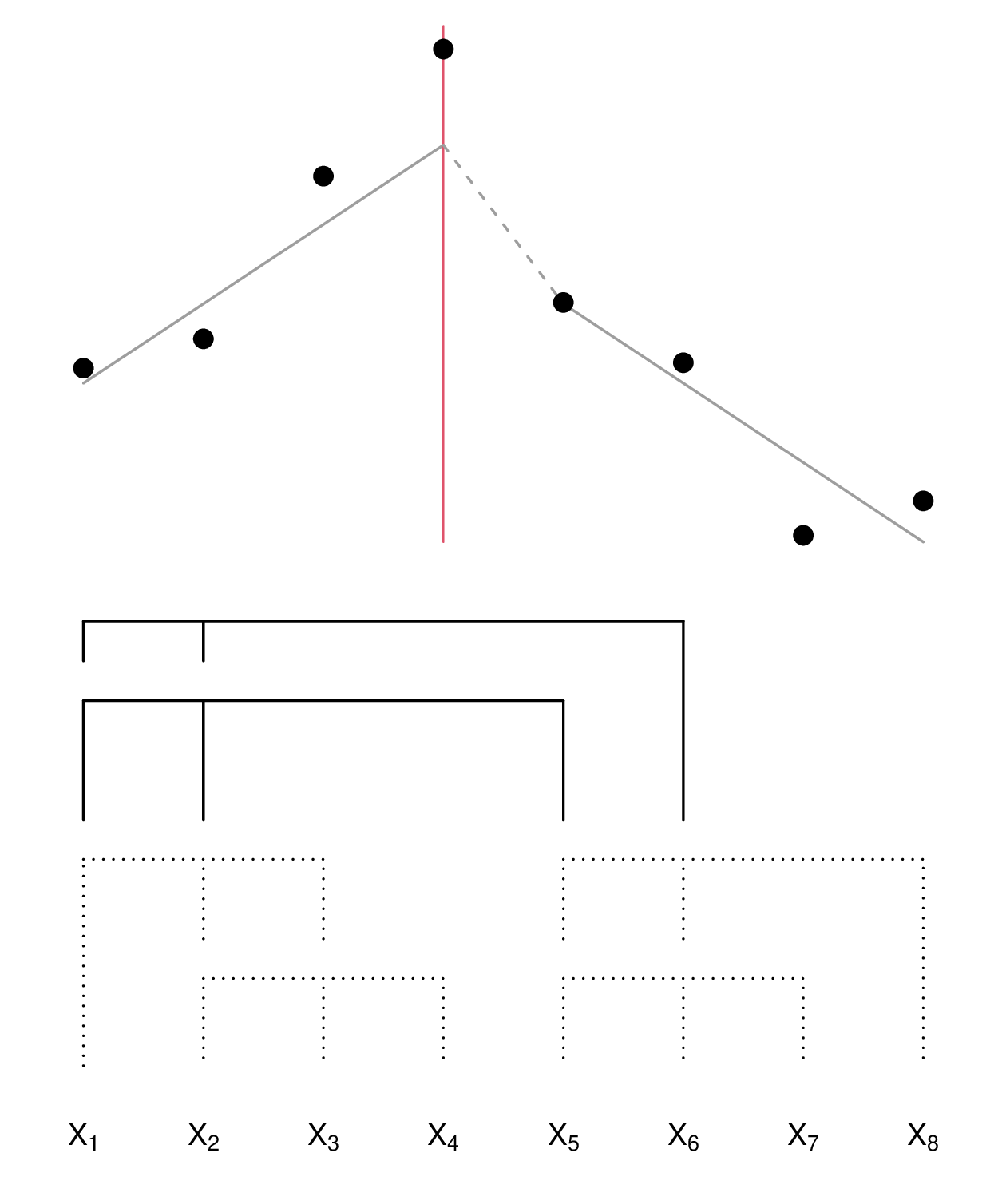}}
        \caption{Thresholding}
        \label{fig:f2}
    \end{subfigure}
    \hfill
    \begin{subfigure}[t]{0.32\textwidth} 
        \raisebox{-\height}{\includegraphics[width=\textwidth]{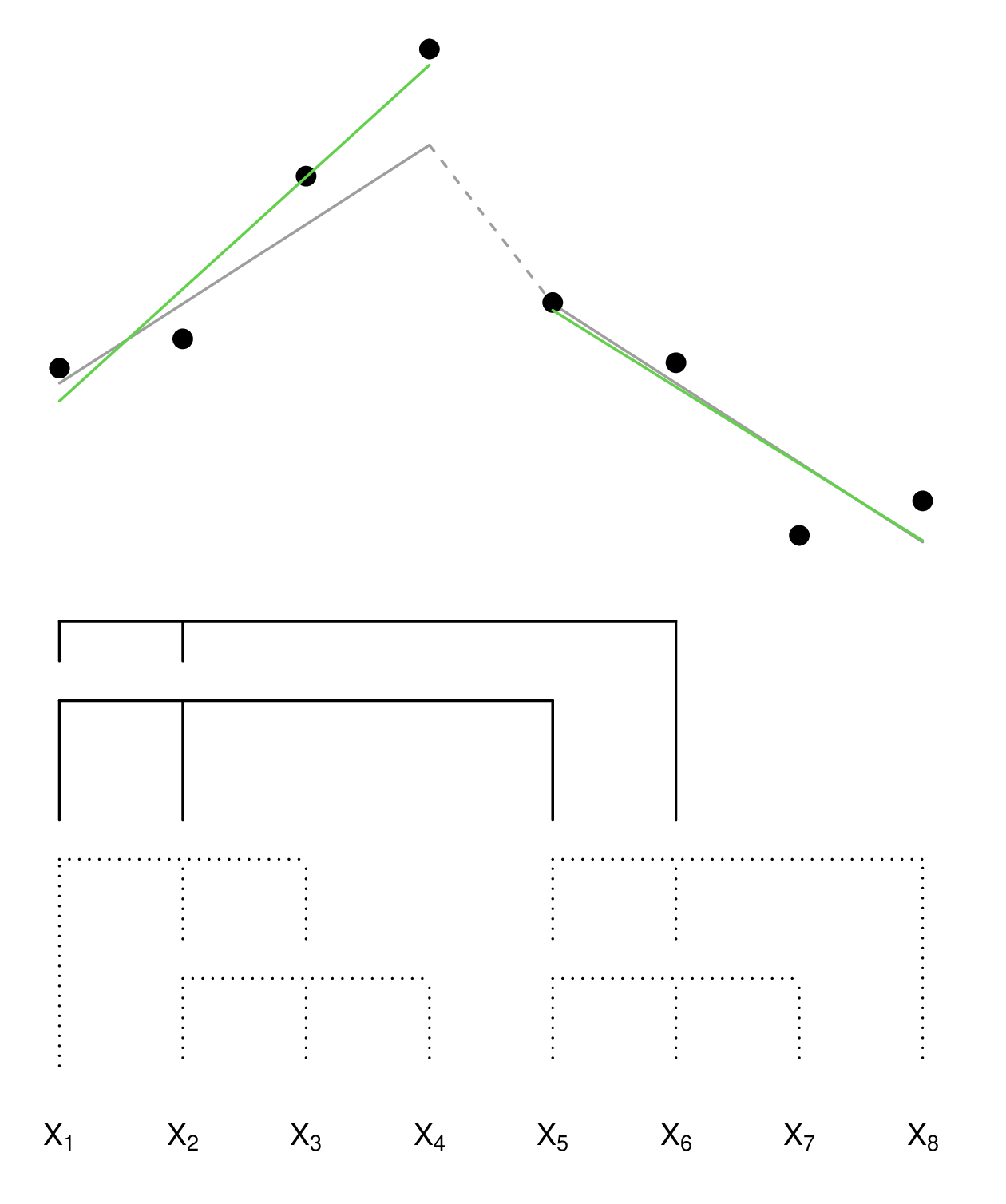}}
        \caption{Inverse TGUW transform}
        \label{fig:f3}
    \end{subfigure}
    
    \caption {Illustration of the first three steps of the TrendSegment procedure with the observed data $X_t$ (dots), the true signal $f_t$ (grey line) and the tree of mergings; (a) TGUW transform constructs a tree by merging neighbouring segments, (b) In thresholding, surviving coefficients (solid line in the tree) are chosen by a pre-specified threshold, which decides the location of the estimated change point (red), (c) Inverse TGUW transform gives the estimated signal (green) based on the estimated change points obtained in thresholding.}
\label{fig:tguw_flow}
\end{figure}

The TrendSegment procedure for estimating the number and the locations of change-points includes four steps. We give the broad picture first and outline details in later sections.
\begin{enumerate}[topsep=0pt, itemsep=-1ex, partopsep=1ex, parsep=1ex]
\item {\em TGUW transformation.} Perform the TGUW transform, a bottom-up unbalanced adaptive wavelet transformation of the input data $X_1, \ldots, X_T$, by recursively applying local conditionally orthonormal transformations. This produces a data-adaptive multiscale decomposition of the data with $T-2$ detail-type coefficients and 2 smooth coefficients.
The resulting conditionally orthonormal transform of the data hopes to encode most of the energy of the signal in only a few detail-type coefficients arising at coarse levels (see Figure \ref{fig:tguw_ex} for an example output). This representation sparsity justifies thresholding in the next step.
\item {\em Thresholding.} Set to zero those detail coefficients whose magnitude is smaller than a pre-specified threshold as long as all the non-zero detail coefficients are connected to each other in the tree structure. This step performs ``pruning'' as a way of deciding the significance of the sparse representation obtained in step 1. 
\item {\em Inverse TGUW transformation.} Obtain an initial estimate of  $f_t$ by carrying out the inverse TGUW transformation of the thresholded coefficient tree. The resulting estimator is discontinuous at the estimated change-points.  
It can be shown to be $l_2$-consistent, but not yet consistent for $N$ or $\eta_1, \ldots, \eta_N$.
\item {\em Post-processing.} Post-process the estimate from step 3 by removing some change-points perceived to be spurious, which enables us to achieve
estimation consistency for $N$ and $\eta_1, \ldots, \eta_N$.
\end{enumerate}
Figure \ref{fig:tguw_flow} illustrates the first three steps of the TrendSegment procedure. We devote the following four sections to describing each step above in order.

\subsection{TGUW transformation}\label{tguw}

\subsubsection{Key principles of the TGUW transform} \label{prcp}
In the initial stage, the data are considered smooth coefficients and the TGUW transform iteratively updates the sequence of smooth coefficients by merging the adjacent sections of the data which are the most likely to belong to the same segment.
The merging is done by performing an adaptively constructed orthonormal transformation to the chosen triplet of the smooth coefficients and in doing so, a data-adaptive unbalanced wavelet basis is established. 
The TGUW transform is completed after $T-2$ such orthonormal transformations and each merge is performed under the following principles.
\begin{enumerate}[topsep=0pt, itemsep=-1ex, partopsep=1ex, parsep=1ex, leftmargin=1.3em]
\item In each merge, three adjacent smooth coefficients are selected and the orthonormal transformation converts those three values into one detail and two (updated) smooth coefficients. 
The size of the detail coefficient gives information about the strength of the local linearity and the two updated smooth coefficients are associated with the estimated parameters (intercept and slope) of the local linear regression performed on the raw observations corresponding to the initially chosen three smooth coefficients.
\item {\em ``Two together'' rule.} The two smooth coefficients returned by the orthonormal transformation are paired in the sense that both contain information about one local linear regression fit.
Thus, we require that any such pair of smooth coefficients cannot be separated when choosing triplets in any subsequent merges. 
We refer to this recipe as the ``two together'' rule.
\item To decide which triplet of smooth coefficients should be merged next, we compare the corresponding detail coefficients as their magnitude represents the strength of the
corresponding local linear trend; the smaller the (absolute) size of the detail, the smaller the local deviation from linearity. 
Smooth coefficients corresponding to the smallest detail coefficients have priority in merging.
\end{enumerate}
\begin{table}[htbp]\caption{Notation. See Section \ref{algo} for formulae for the terms listed.}
\centering 
\begin{tabular}{l p{10.5cm} }
\toprule
$X_p$ & $p^\textnormal{th}$ element of the observation vector $\b{X}=\{X_1, X_2, \ldots, X_T\}^\top$. \\  
$s^0_{p, p}$ &  $p^\textnormal{th}$ initial smooth coefficient of the vector $\bs^0$  where $\b{X}=\bs^0$.\\
$d_{p, q, r}$  & detail coefficient obtained from $\{X_p, \ldots, X_r\}$ (merges of Types 1 or 2).\\
$s^{[1]}_{p, r}, s^{[2]}_{p, r}$ &  smooth coefficients obtained from $\{X_p, \ldots, X_r\}$, paired under the ``two together" rule.\\
$d^{[1]}_{p, q, r}, d^{[2]}_{p, q, r}$ &  paired detail coefficients obtained by merging two adjacent subintervals, $\{X_p, \ldots, X_q\}$ and $\{X_{q+1}, \ldots, X_r\}$, where $r > q+2$ and $q > p+1$ (merge of Type 3).\\
$\bs$ &  data sequence vector containing the (recursively updated) smooth and detail coefficients from the initial input $\bs^0$.\\
\bottomrule
\end{tabular}
\label{tab:nt}
\end{table}
As merging continues under the ``two together" rule, all mergings can be classified into one of three forms: 
\begin{itemize}
    \item Type 1: merging three initial smooth coefficients,
    \item Type 2: merging one initial and a paired smooth coefficient,
    \item Type 3: merging two sets of (paired) smooth coefficients,
\end{itemize}
where Type 3 is composed of two merges of triplets and more details are given in Section \ref{ex}.

\subsubsection{Example} \label{ex}
\begin{figure}[ht!] 
\centering
\includegraphics[width=12cm, height=7cm]{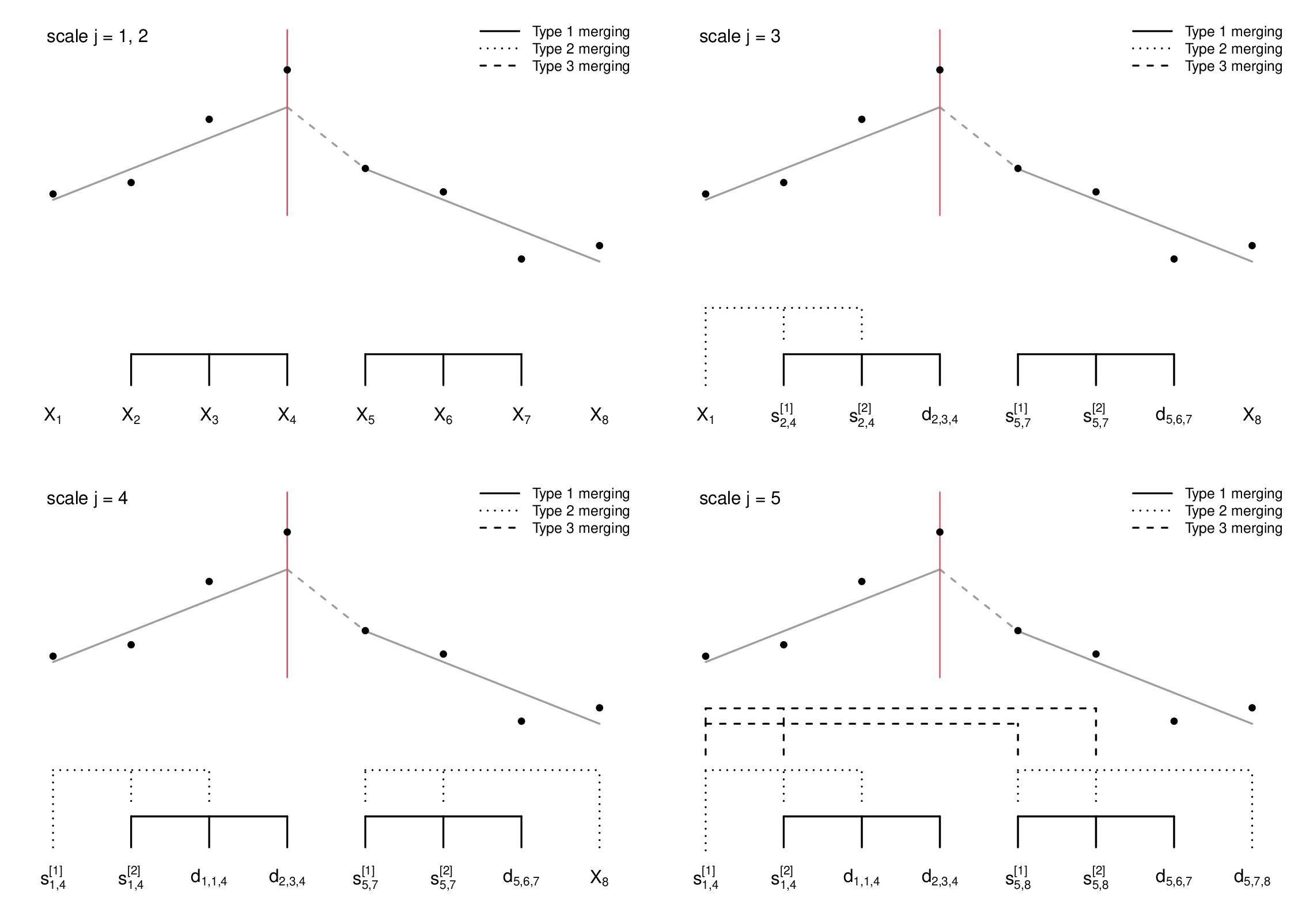}
 \caption {Construction of tree for the example in Section \ref{ex}; each diagram shows all merges performed up to the given scale with the data (dot), true signal (grey) and true change point (red).}
\label{fig:f0}
\end{figure}

We now provide a simple example of the TGUW transformation; the accompanying illustration is in Figure \ref{fig:f0}. 
The notation for this example and for the general algorithm introduced later is in Table \ref{tab:nt}. 
This example shows single merges at each pass through the data when the algorithm runs in a purely greedy way.
We will later generalise it to multiple passes through the data, which will speed up computation (this device is referred to as ``tail-greediness" as the algorithm merges those triplets corresponding to the lower tail of the distribution of local deviation from linearity in $\b{X}$).
We refer to $j^\text{th}$ pass through the data as scale $j$. 
Assume that we have the initial input $\bs^0 = (X_1, X_2, \ldots, X_8)$, so that the complete TGUW transform consists of 6 merges. 
We show 6 example merges one by one under the rules introduced in Section \ref{prcp}. 
This example demonstrates all three possible types of merges.

\textbf{Scale $j=1$}. From the initial input $\bs^0 = (X_1, \ldots, X_8)$, we consider 6 triplets $(X_1, X_2, X_3)$, $(X_2, X_3, X_4)$, $(X_3, X_4, X_5)$, $(X_4, X_5, X_6)$, $(X_5, X_6, X_7)$, $(X_6, X_7, X_8)$ and compute the size of the detail for each triplet, where the formula can be found in \eqref{e22}. 
Suppose that $(X_2, X_3, X_4)$ gives the smallest size of detail, $\vert d_{2, 3, 4}\vert$, then merge $(X_2, X_3, X_4)$ through the orthogonal transformation formulated in \eqref{e28} and update the data sequence into $\bs = (X_1, s^{[1]}_{2,4}, s^{[2]}_{2,4}, d_{2, 3, 4}, X_5, X_6, X_7, X_8)$. 
We categorise this transformation into Type 1 (merging three initial smooth coefficients). 

\textbf{Scale $j=2$}. From now on, the ``two together'' rule is applied. Ignoring any detail coefficients in $\bs$, the possible triplets for next merging are $(X_1, s^{[1]}_{2,4}, s^{[2]}_{2,4})$, $(s^{[1]}_{2,4}, s^{[2]}_{2,4}, X_5)$, $(X_5, X_6, X_7)$, $(X_6, X_7, X_8)$. 
We note that $(s^{[2]}_{2,4}, X_5, X_6)$ cannot be considered as a candidate for next merging under the ``two together'' rule as this triplet contains only one (not both) of the paired smooth coefficients returned by the previous merging. 
Assume that $(X_5, X_6, X_7)$ gives the smallest size of detail coefficient $\vert d_{5, 6, 7}\vert$ among the four candidates, then we merge them through the orthogonal transformation formulated in \eqref{e28} and now update the sequence into $\bs = (X_1, s^{[1]}_{2,4}, s^{[2]}_{2,4}, d_{2, 3, 4}, s^{[1]}_{5,7}, s^{[2]}_{5,7}$, $d_{5, 6, 7}, X_8)$. 
This transformation is also Type 1. 

\textbf{Scale $j=3$}. We now compare four candidates for merging, $(X_1, s^{[1]}_{2,4}, s^{[2]}_{2,4})$, $(s^{[1]}_{2,4}, s^{[2]}_{2,4}, s^{[1]}_{5,7})$, $(s^{[2]}_{2,4}, s^{[1]}_{5,7}, s^{[2]}_{5,7})$ and $(s^{[1]}_{5,7}, s^{[2]}_{5,7}, X_8)$. 
The two triplets in middle, $(s^{[1]}_{2,4}, s^{[2]}_{2,4}, s^{[1]}_{5,7})$ and $(s^{[2]}_{2,4}, s^{[1]}_{5,7}, s^{[2]}_{5,7})$, are paired together as they contain two sets of paired smooth coefficients, $(s^{[1]}_{2,4}, s^{[2]}_{2,4})$ and $(s^{[1]}_{5,7}, s^{[2]}_{5,7})$, and if we were to treat these two triplets separately, we would be violating the ``two together'' rule. 
The summary detail coefficient for this pair of triplets is obtained as $d_{2,4,7}=\max(\vert d^{[1]}_{2,4,7}\vert, \vert d^{[2]}_{2,4,7}\vert)$, which is compared with those of the other triplets.  
Now suppose that $(X_1, s^{[1]}_{2,4}, s^{[2]}_{2,4})$ has the smallest size of detail; we merge this triplet and update the data sequence into $\bs = (s^{[1]}_{1,4}, s^{[2]}_{1,4}, d_{1,1,4}, d_{2, 3, 4}, s^{[1]}_{5,7}, s^{[2]}_{5,7}, d_{5, 6, 7}, X_8)$. 
This transformation is of Type 2.

\textbf{Scale $j=4$}. We now have two pairs of paired coefficients: $(s^{[1]}_{1,4}, s^{[2]}_{1,4})$ and $(s^{[1]}_{5,7}, s^{[2]}_{5,7})$. 
Therefore, with the ``two together'' rule in mind, the only possible options for merging are: to merge the two pairs into $(s^{[1]}_{1,4}, s^{[2]}_{1,4}, s^{[1]}_{5,7}, s^{[2]}_{5,7})$, or to merge $(s^{[1]}_{5,7}, s^{[2]}_{5,7})$ with $X_8$. 
Suppose that the second merging is preferred. Then we perform Type 2 merge and update the data sequence into $\bs = (s^{[1]}_{1,4}, s^{[2]}_{1,4}, d_{1,1,4}, d_{2, 3, 4}, s^{[1]}_{5,8}, s^{[2]}_{5,8}, d_{5,6,7}, d_{5,7,8})$.

\textbf{Scale $j=5$}. The only remaining step is merging $(s^{[1]}_{1,4}, s^{[2]}_{1,4})$ and $(s^{[1]}_{5,8}, s^{[2]}_{5,8})$ into $(s^{[1]}_{1,4}, s^{[2]}_{1,4}, s^{[1]}_{5,8}, s^{[2]}_{5,8})$. This transformation is Type 3 and performed in two stages as follows. 
In the first stage, we merge $(s^{[1]}_{1, 4}, s^{[2]}_{1, 4}, s^{[1]}_{5, 8})$ and then update the sequence temporarily as $\bs = (s^{[1']}_{1,8}, s^{[2']}_{1,8}, d_{1,1,4}, d_{2, 3, 4}, d^{[1]}_{1,4,8}, s^{[2]}_{5,8}, d_{5, 6, 7}, d_{5,7,8})$. 
In the second stage, we merge $(s^{[1']}_{1,8}, s^{[2']}_{1,8}, s^{[2]}_{5, 8})$, which gives the updated sequence $\bs = (s^{[1]}_{1,8}, s^{[2]}_{1,8}, d_{1,1,4}, d_{2, 3, 4}, d^{[1]}_{1,4,8}, d^{[2]}_{1,4,8}, d_{5, 6, 7}, d_{5,7,8})$.
The transformation is now completed with the updated data sequence which contains $T-2=6$ detail and $2$ smooth coefficients.


\subsubsection{Some important features of TGUW transformation} \label{prop}
Before formulating the TGUW transformation in generality, we describe how it achieves sparse representation of the data. Sometimes, we will be referring to a detail coefficient $d_{p,q,r}^{\cdot}$ as $d_{p,q,r}^{(j, k)}$ or $d^{(j, k)}$, where $j = 1, \ldots, J$ is the scale of the transform (i.e. the consecutive pass through the data) at which $d_{p,q,r}^{\cdot}$ was computed, $k=1, \ldots, K(j)$ is the location index of $d_{p,q,r}^{\cdot}$ within all scale $j$ coefficients, and $d_{p,q,r}^{\cdot}$ is $d_{p,q,r}^{[1]}$ or $d_{p,q,r}^{[2]}$ or $d_{p,q,r}$, depending on the type of merge. 

The TGUW transform eventually converts the input data sequence $\b X$ of length $T$ into the sequence containing 2 smooth and $T-2$ detail coefficients through $T-2$ orthonormal transforms as follows,
\begin{equation} \label{e31}
\renewcommand{\arraystretch}{0.75}
\begin{pmatrix} 
 s^{[1]}_{1, T} \\  s^{[2]}_{1, T} \\ \begin{pmatrix} \\ {d^{(j, k)}}_{j=1, \ldots, J, k=1, \ldots, K(j)}  \\ \\ \end{pmatrix}
\end{pmatrix}  = \;
\begin{pmatrix} 
 \psi^{(0, 1)} \\  \psi^{(0, 2)} \\ \begin{pmatrix} \\ {\psi^{(j, k)}}_{j=1, \ldots, J, k=1, \ldots, K(j)}  \\ \\ \end{pmatrix}
\end{pmatrix}  
\begin{pmatrix} 
X_1 \\ X_2 \\ \vdots \\ X_{T} \\
\end{pmatrix} = \;
\Psi_{T \times T} \begin{pmatrix} 
X_1 \\ X_2 \\ \vdots \\ X_{T} \\\end{pmatrix},
\end{equation}
where $\Psi$ is a data-adaptively chosen orthonormal unbalanced wavelet basis for $\mathbb{R}^T$.
The detail coefficients $d^{(j, k)}$ can be regarded as scalar products between $\b X$ and a particular unbalanced wavelet basis $\psi^{(j, k)}$, where the formal representation is given as $\{d^{(j, k)} = \langle X, \psi^{(j, k)} \rangle,_{j=1, \ldots, J, k=1, \; \ldots, K(j)}\}$ for detail coefficients and $s^{[1]}_{1, T} = \langle X, \psi^{(0, 1)} \rangle$, $s^{[2]}_{1, T} = \langle X, \psi^{(0, 2)} \rangle$ for the two smooth coefficients. 

The TGUW transform is nonlinear, but it is also conditionally linear and orthonormal given the order in which the merges are performed. 
The orthonormality of the unbalanced wavelet basis, $\{\psi^{(j, k)}\}$, implies Parseval's identity:
\begin{equation} \label{ts.e32}
\sum_{t=1}^T X_t^2 = \sum_{j=1}^J \sum_{k=1}^{K(j)} (d^{(j, k)})^2 + (s^{[1]}_{1, T})^2 + (s^{[2]}_{1, T})^2.
\end{equation}
Furthermore, the filters $(\psi^{(0, 1)}, \psi^{(0, 2)})$ corresponding to the two smooth coefficients $s^{[1]}_{1, T}$ and $s^{[2]}_{1, T}$ form an orthonormal basis of the subspace $\{(x_1, x_2, \ldots, x_T) \; \vert \;  x_1-x_2=x_2-x_3=\cdots=x_{T-1}-x_{T} \}$ of $\mathbb{R}^{T}$; see Section E of the supplementary materials for further details.
This implies
\begin{equation} \label{ts.e32new}
\sum_{t=1}^T X_t^2 - (s^{[1]}_{1, T})^2 - (s^{[2]}_{1, T})^2 = \sum_{t=1}^T (X_t-\hat{X}_t)^2
\end{equation}
where $\hat{\b{X}} = s^{[1]}_{1, T}\psi^{(0, 1)} + s^{[2]}_{1, T}\psi^{(0, 2)}$ is the best linear regression fit to $\b X$ achieved by minimising the sum of squared errors. 
This, combined with the Parseval's identity above, implies
\begin{equation} \label{ts.e33}
\sum_{t=1}^T (X_t-\hat{X}_t)^2 = \sum_{j=1}^J \sum_{k=1}^{K(j)} (d^{(j, k)})^2.
\end{equation}

By construction, the detail coefficients $\vert d^{(j, k)}\vert$ obtained in the initial stages of the TGUW transform tend to be small in magnitude. 
Then the Parseval's identity in \eqref{ts.e32} implies that a large portion of $\sum_{t=1}^T (X_t-\hat{X}_t)^2$ is explained by only a few large $\vert d^{(j, k)}\vert$'s arising in the later stages of the transform;
in this sense, the TGUW transform provides sparsity of signal representation.

\subsubsection{TGUW transformation: general algorithm} \label{algo}
In this section, we formulate in generality the TGUW transformation illustrated in Section \ref{ex} by showing how an adaptive orthonormal unbalanced wavelet basis, $\Psi$ in \eqref{e31}, is constructed. 
One of the important principles is ``tail-greediness'' \citep{fryzlewicz2017tail} which enables us to reduce the computational complexity by performing multiple merges over non-overlapping regions in a single pass over the data. 
More specifically, it allows us to perform up to $\max\{2, \lceil \rho \alpha_j \rceil\}$ merges at each scale $j$, where $\alpha_j$ is the number of smooth coefficients in the data sequence $\bs$ and $\rho \in (0, 1)$ (the lower bound of 2 is essential to permit a Type 3 transformation, which consists of two merges). 


We now describe the TGUW algorithm. 

\begin{enumerate}[topsep=0pt, itemsep=-1ex, partopsep=1ex, parsep=1ex, leftmargin=1.3em]
\item At each scale $j$, find the set of triplets that are candidates for merging under the ``two together'' rule and compute the corresponding detail coefficients. 
Regardless of the type of merge, a detail coefficient $d_{p,q,r}^{\cdot}$ is, in general, obtained as
\begin{equation} \label{e22}
d_{p,q,r}^{\cdot} = a\bs_{p:r}^1 + b\bs_{p:r}^2 + c\bs_{p:r}^3,
\end{equation}
where $p \leq q < r$, $\bs_{p:r}^k$ is the $k^{\text{th}}$ smooth coefficient of the subvector $\bs_{p:r}$ with a length of $r-p+1$ and the constants $a, b, c$ are the elements of the detail filter $\b{h}=(a, b, c)^{\top}$. 
We note that $(a, b, c)$ also depends on $(p, q, r)$, but this is not reflected in the notation, for simplicity. 
The detail filter is a weight vector used in computing the weighted sum of a triplet of smooth coefficients which should satisfy the condition that the detail coefficient is zero if and only if the corresponding raw observations over the merged regions have a perfect linear trend. 
If $(X_{p}, \ldots, X_r)$ are the raw observations associated with the triplet of the smooth coefficients $(\bs_{p:r}^1, \bs_{p:r}^2, \bs_{p:r}^3)$ under consideration, then the detail filter $\b{h}$ is obtained in such a way as to produce zero detail coefficient only when $(X_{p}, \ldots, X_r)$ has a perfect linear trend, as the detail coefficient itself represents the extent of non-linearity in the corresponding region of data. 
This implies that the smaller the size of the detail coefficient, the closer the alignment of the corresponding data section with linearity. 

\item Summarise all $d_{p,q,r}^{\cdot}$ constructed in step 1 to a (equal length or shorter) sequence of $d_{p,q,r}$ by finding a summary detail coefficient $d_{p,q,r}=\max(\vert d^{[1]}_{p,q,r} \vert, \vert d^{[2]}_{p,q,r} \vert)$ for any pair of detail coefficients constructed by Type 3 merges.
\item Sort the size of the summarised detail coefficients $\vert d_{p,q,r}\vert$ obtained in step 2 in non-decreasing order.
\item Extract the (non-summarised) detail coefficient(s) $\lvert d^{\cdot}_{p,q,r} \rvert$ corresponding to the smallest (summarised) detail coefficient $\lvert d_{p,q,r} \rvert$ e.g. both $\lvert d^{[1]}_{p,q,r} \rvert$ and $\lvert d^{[2]}_{p,q,r} \rvert$ should be extracted only if $d_{p,q,r}=\max(\lvert d^{[1]}_{p,q,r} \rvert, \lvert d^{[2]}_{p,q,r} \rvert)$. 
Repeat the extraction until $\max\{2, \lceil \rho \alpha_j \rceil\}$ (or all possible, whichever is the smaller number) detail coefficients have been obtained, as long as the region of the data corresponding to each detail coefficient extracted does not overlap with the regions corresponding to the detail coefficients already drawn. 
\item For each $\lvert d_{p,q,r}^{\cdot} \rvert$ extracted in step 4, merge the corresponding smooth coefficients by updating the corresponding triplet in $\bs$ through the orthonormal transform as follows,
\begin{align} \label{e28}
\begin{pmatrix} 
s^{[1]}_{p, r} \\ 
s^{[2]}_{p, r} \\ 
d^\cdot_{p, q, r} 
\end{pmatrix} = 
\begin{pmatrix} 
& & \b{\ell}_1^\top & & \\ 
& & \b{\ell}_2^\top & &  \\
& & \b{h}^\top & & \\
\end{pmatrix}
\begin{pmatrix} 
\bs_{p:r}^1 \\
\bs_{p:r}^2\\
\bs_{p:r}^3
\end{pmatrix} =
\Lambda \begin{pmatrix} 
\bs_{p:r}^1 \\
\bs_{p:r}^2\\
\bs_{p:r}^3
\end{pmatrix}.
\end{align}
The key step is finding the  $3 \times 3$ orthonormal matrix, $\Lambda$, which is composed of one detail and two low-pass filter vectors in its rows. 
Firstly the detail filter $\b{h}^\top$ is determined to satisfy the condition mentioned in step 1, and then the two low-pass filters ($\b\ell_{1}^\top, \b\ell_{2}^\top$) are obtained by satisfying the orthonormality of $\Lambda$.
There is no uniqueness in the choice of ($\b\ell_{1}^\top, \b\ell_{2}^\top$), but this has no effect on the transformation itself.
The details of this mechanism can be found in Section E of the supplementary materials.
\item Go to step 1 and repeat at new scale $j=j+1$ as long as we have at least three smooth coefficients in the updated data sequence $\bs$. 
\end{enumerate}

More specifically, when Type 1 merge is performed in step 1 (i.e. $\bs_{p:r}$ in \eqref{e22} consists of three initial smoothing coefficients, which implies $r=p+2$), the corresponding detail filter $\b{h}$ is obtained as a unit normal vector to the plane $\{(x, y, z) \vert x - 2y + z = 0\}$, thus the detail coefficient $d$ presents the projection of three initial smoothing coefficients to the unit normal vector. In the same manner, due to the orthonormality of $\Lambda$ in \eqref{e28}, the two low-pass filters ($\b\ell_{1}^\top, \b\ell_{2}^\top$) form an arbitrary orthonormal basis of the plane $\{(x, y, z) \vert x - 2y + z = 0\}$. 
In practice, the detail filter $\b{h}$ in Step 1 is obtained by updating so-called weight vectors of constancy and linearity in which the initial inputs have a form of $(1,1, \ldots, 1)^\top$ and $(1, 2, \ldots, T)^\top$, respectively. The details can be found in Section F of the supplementary materials. 

We now comment briefly on the computational complexity of the TGUW transform.
Assume that $\alpha_j$ smooth coefficients are available in the data sequence $\bs$ at scale $j$ and we allow the algorithm to merge up to $\sceil{\rho \alpha_j}$ many triplets (unless their 
corresponding data regions overlap) where $\rho \in (0, 1)$ is a constant. 
This gives us at most $(1-\rho)^jT$ smooth coefficients remaining in $\bs$ after $j$ scales. 
Solving for $(1-\rho)^jT \leq 2$ gives the largest number of scales $J$ as $\ceil{\log(T) / \log\big((1-\rho)^{-1}\big) + \log(2) / \log(1-\rho)}$, at which point the TGUW transform terminates with two smooth coefficients remaining.
Considering that the most expensive step at each scale is sorting which takes $O(T \log(T))$ operations, the computational complexity of the TGUW transformation is $O(T \log^2(T))$.

\subsection{Thresholding} \label{dns}
Because at each stage, the TGUW transform constructs the smallest possible detail coefficients, but it is at the same time orthonormal and so preserves the $l_2$ energy of the input data,
the variability (= deviation from linearity) of the signal tends to be mainly encoded in only a few detail coefficients computed at the later stages of the transform. 
The resulting sparsity of representation of the input data in the domain of TGUW coefficients justifies thresholding as a way of deciding the significance of each detail coefficient (which measures the local deviation from linearity).

\begin{figure}[ht!] 
    \centering
    \begin{subfigure}[t]{0.99\textwidth} 
        \raisebox{-\height}{\includegraphics[width=\textwidth]{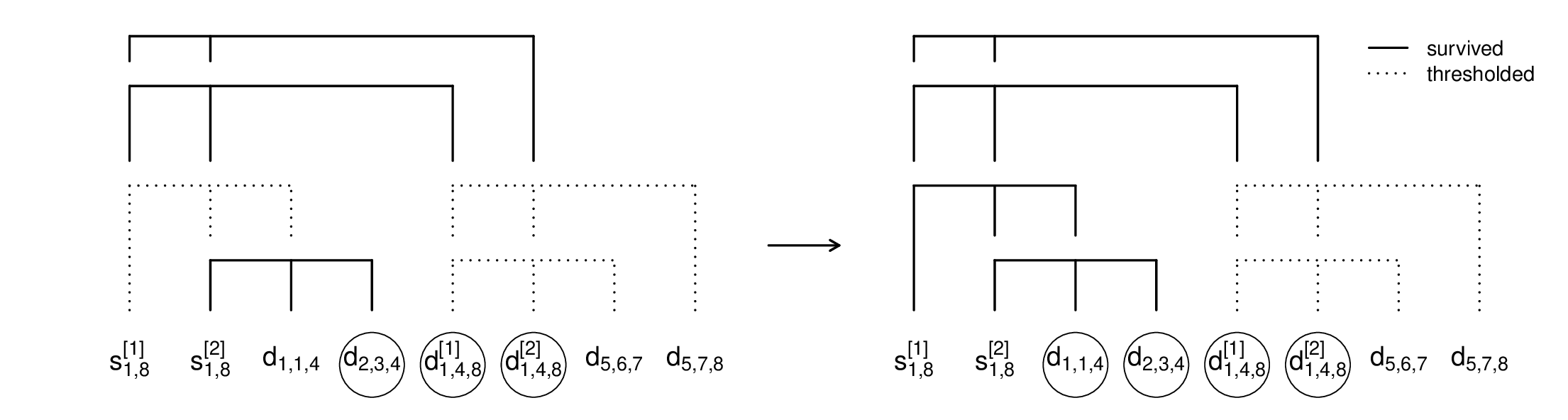}}
        \caption{survived detail coefficients before and after applying the ``connected'' rule}
        \label{fig:conn}
    \end{subfigure}
    \hfill
    \begin{subfigure}[t]{0.99\textwidth} 
        \raisebox{-\height}{\includegraphics[width=\textwidth]{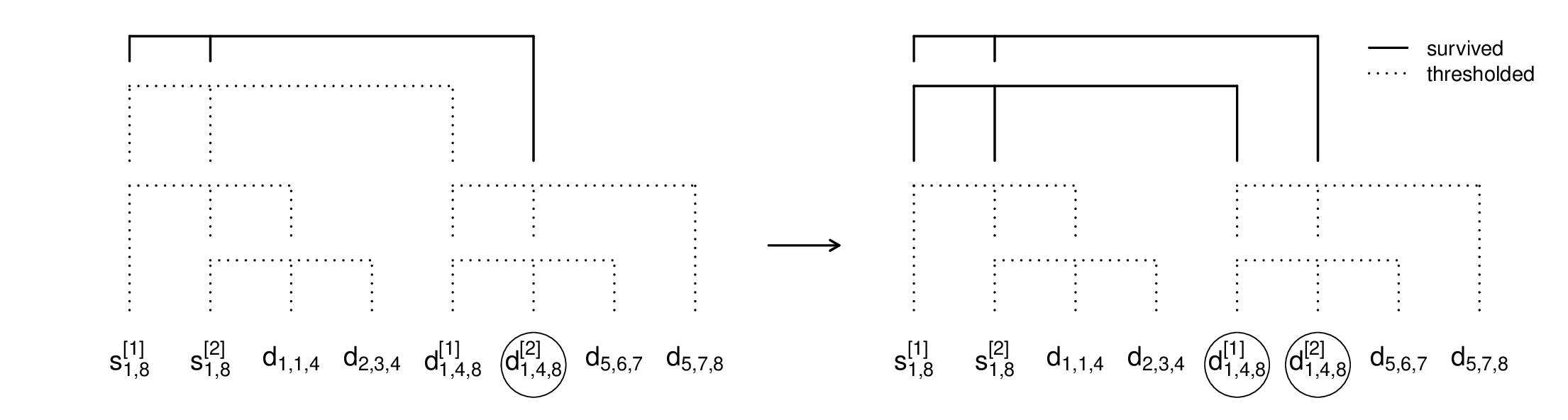}}
        \caption{survived detail coefficients before and after applying the ``two together'' rule}
        \label{fig:twoto}
    \end{subfigure}
    \hfill
    \caption {The tree of mergings in the example of Section \ref{ex}. Left-hand trees show the examples of tree obtained from initial hard thresholding, the right-hand trees come from processing the respective left-hand ones by applying (a) the ``connected'' rule and (b) the ``two together'' rule, respectively, described in Section \ref{dns}. The circled detail coefficients are the surviving ones.}
\label{fig:thd}
\end{figure}

We propose to threshold the TGUW detail coefficients under two important rules, which should simultaneously be satisfied; we refer to these as the ``connected'' rule and the ``two together'' rule. 
The ``two together'' rule in thresholding is similar to the one in the TGUW transformation except it targets pairs of detail rather than smooth coefficients, and only applies to pairs of detail coefficients arising from Type 3 merges. 
Figure \ref{fig:twoto} shows one such pair in the example of Section \ref{ex}, $(d_{1,4,8}^{[1]}, d_{1,4,8}^{[2]})$, and the ``two together'' rule means that both such detail coefficients should be kept if at least one survives the initial thresholding. 
This is a natural requirement as a pair of Type 3 detail coefficients effectively corresponds to a single merge of two adjacent regions. 

The ``connected'' rule which prunes the branches of the TGUW detail coefficients if and only if the detail coefficient itself and all of its children coefficients fall below a certain threshold in absolute value. 
This is illustrated in Figure \ref{fig:conn} along with the example of Section \ref{ex}; if both $d_{2,3,4}$ and $(d_{1,4,8}^{[1]}, d_{1,4,8}^{[2]})$ were to survive the initial thresholding, the ``connected'' rule would mean we also had to keep $d_{1,1,4}$, which is the child of $(d_{1,4,8}^{[1]}, d_{1,4,8}^{[2]})$ and the parent of $d_{2,3,4}$ in the TGUW coefficient tree. 

Through the thresholding, we wish to estimate the underlying signal $f$ in \eqref{e1} by estimating $\mu^{(j, k)} = \langle f, \psi^{(j, k)} \rangle$ where $\psi^{(j, k)}$ is an orthonormal unbalanced wavelet basis constructed in the TGUW transform from the data. 
Throughout the entire thresholding procedure, the ``connected'' and ``two together'' rules are applied in this order. 
We firstly threshold and apply the ``connected'' rule, which gives us $\hat{\mu}_0^{(j, k)}$, the initial estimator of $\mu^{(j, k)}$, as
\begin{equation} \label{e5}
\hat{\mu}_0^{(j, k)} = d^{(j, k)}_{p,q,r} \cdot \mathbb{I} \; \Big\{ \, \exists (j', k') \in \mathcal{C}_{j, k} \quad \big\vert d^{(j', k')}_{p',q',r'}\big\vert  > \lambda \, \Big\},
\end{equation}
where $\mathbb{I}$ is an indicator function and
\begin{equation} \label{e51}
\mathcal{C}_{j, k} = \{(j', k'), j' = 1, \ldots, j, k' = 1, \ldots, K(j'): d^{(j', k')}_{p',q',r'} \text{ is such that } [p', r'] \subseteq [p, r] \}.
\end{equation}

Now the ``two together'' rule is applied to the initial estimators $\hat{\mu}_0^{(j, k)}$ to obtain the final estimators $\hat{\mu}^{(j, k)}$. 
We firstly note that two detail coefficients, $d^{(j, k)}_{p,q,r}$ and $d^{(j', k+1)}_{p',q',r'}$ are called ``paired'' when they are formed by Type 3 mergings and when $(j, p, q, r)=(j', p',q',r')$. The ``two together'' rule is formulated as below,
\begin{numcases}{\hat{\mu}^{(j, k)}=}
  \hat{\mu}_0^{(j, k)}, & if $d_{p,q,r}^{(j, k)}$ is not paired, \nonumber \\
  \hat{\mu}_0^{(j, k)}, & if $d_{p,q,r}^{(j, k)}$ is paired with $d_{p,q,r}^{(j, k')}$ and both $\hat{\mu}_0^{(j, k)}$ and $\hat{\mu}_0^{(j, k')}$ are zero or nonzero, \nonumber \\
  d^{(j, k)}, & if $d_{p,q,r}^{(j, k)}$ is paired with $d_{p,q,r}^{(j, k')}$ and $\hat{\mu}_0^{(j, k')}\neq0$ and $\hat{\mu}_0^{(j, k)}=0$. \label{e52}
\end{numcases}

It is important to note that the application of the two rules ensures that $\tilde{f}$ is a piecewise-linear function composed of best linear fits (in the least-squares sense) for each interval of linearity.
As an aside, we note that the number of survived detail coefficients does not necessarily equal the number of change-points in $\tilde{f}$ as a pair of detail coefficients arising from a Type 3 merge are associated with a single change-point.

\subsection{Inverse TGUW transformation} \label{inv}
The estimator $\tilde{f}$ of the true signal $f$ in \eqref{e1} is obtained by inverting (= transposing) the orthonormal transformations in \eqref{e28} in reverse order to that in which they were originally performed.
This inverse TGUW transformation is referred to as $\text{TGUW}^{-1}$, and thus
\begin{equation} \label{e60}
\tilde{f} = \text{TGUW}^{-1} \big\{ \; \hat{\mu}^{(j, k)}, j=1, \ldots, J, \, k=1, \ldots, K(j) \; \| \; s^{[1]}_{1, T}, s^{[2]}_{1, T} \; \big\},
\end{equation}
where $\|$ denotes vector concatenation.

\subsection{Post processing for consistency of change-point detection} \label{pp}
As will be formalised in Theorem \ref{thm1} of Section \ref{sec3}, the piecewise-linear estimator $\tilde{f}$ in \eqref{e60} possibly overestimates the number of change-points. 
To remove the spurious estimated change-points and to achieve the consistency of the number and the locations of the estimated change-points, we adopt the post-processing framework of 
\citet{fryzlewicz2017tail}. 
\citet{lin2017sharp} show that we can usually post-process $l_2$-consistent estimators in this way as a fast enough $l_2$ error rate implies that each true change-point has an estimator nearby. 
The post-processing methodology includes two stages, i) execution of three steps, TGUW transform, thresholding and inverse TGUW transform, again to the estimator $\tilde{f}$ in \eqref{e60} and ii) examination of regions containing only one estimated change-point to check for its significance.
\paragraph{Stage 1.} \label{pp1}
We transform the estimated function $\tilde{f}$ in \eqref{e60} with change-points $(\tilde{\eta}_1, \tilde{\eta}_2, \ldots, \tilde{\eta}_{\tilde{N}})$ into a new estimator $\dbtilde{f}$ with corresponding change-points $(\dbtilde{\eta}_1, \dbtilde{\eta}_2, \ldots, \dbtilde{\eta}_{\dbtilde{N}})$. 
Using $\tilde{f}$ in \eqref{e60} as an input data sequence $\bs$, we perform the TGUW transform as presented in Section \ref{algo}, but in a greedy rather than tail-greedy way such that only one detail coefficient $d^{(j, 1)}$ is produced at each scale $j$, and thus $K(j)=1$ for all $j$. 
We repeat to produce detail coefficients until the first detail coefficient such that $\vert d^{(j, 1)}\vert  > \lambda$ is obtained where $\lambda$ is the parameter used in the thresholding procedure described in Section \ref{dns}. 
Once the condition, $\vert d^{(j, 1)}\vert  > \lambda$, is satisfied, we stop merging, relabel the surviving change-points as $(\dbtilde{\eta}_1, \dbtilde{\eta}_2, \ldots, \dbtilde{\eta}_{\dbtilde{N}})$ and construct the new estimator $\dbtilde{f}$ as
\begin{equation} \label{e62}
\dbtilde{f}_t = \hat{\theta}_{i, 1} +  \hat{\theta}_{i, 2} \; t \quad \text{for} \quad t \in \big[\dbtilde{\eta}_{i-1}+1, \dbtilde{\eta}_i\big], \quad i=1, \ldots, \dbtilde{N},  
\end{equation}
where $\dbtilde{\eta}_{0}= 0$, $\dbtilde{\eta}_{\dbtilde{N}+1}=T$ and ($\hat{\theta}_{i, 1}, \hat{\theta}_{i, 2}$) are the OLS intercept and slope coefficients, respectively, for the corresponding pairs $\{(t, X_t),\; t \in \big[\dbtilde{\eta}_{i-1}+1, \dbtilde{\eta}_i\big]\}$. The exception is when the region under consideration only contains a single data point $X_{t_0}$, in which case fitting a linear regression is impossible. We then set $\dbtilde{f}_{t_0} = X_{t_0}$.

\paragraph{Stage 2.} \label{pp2}
From the estimator $\dbtilde{f}_t$ in Stage 1, we obtain the final estimator $\hat{f}$ by pruning the change-points $(\dbtilde{\eta}_1, \dbtilde{\eta}_2, \ldots, \dbtilde{\eta}_{\dbtilde{N}})$ in $\dbtilde{f}_t$. 
For each $i=1, \ldots, \dbtilde{N}$, compute the corresponding detail coefficient $d_{p_i, q_i, r_i}$ as described in \eqref{e22}, where $p_i=\floor{\frac{\dbtilde{\eta}_{i-1} + \dbtilde{\eta}_i}{2}}+1$, $q_i=\dbtilde{\eta}_i$ and $r_i=\ceil{\frac{\dbtilde{\eta}_{i} + \dbtilde{\eta}_{i+1}}{2}}$. 
Now prune by finding the minimiser $i_0 = \argmin_i{\vert d_{p_i, q_i, r_i}\vert }$ and removing $\dbtilde{\eta}_{i_0}$ and setting $\dbtilde{N} := \dbtilde{N} - 1$ if $\vert d_{p_{i_0}, q_{i_0}, r_{i_0}}\vert   \leq \lambda$ where $\lambda$ is same as in Section \ref{dns}. 
Then relabel the change-points with the subscripts $i=1, \ldots, \dbtilde{N}$ under the convention $\dbtilde{\eta}_{0}=0$, $\dbtilde{\eta}_{\dbtilde{N}+1}=T$. 
Repeat the pruning while we can find $i_0$ which satisfies the condition $\big\vert d_{p_{i_0}, q_{i_0}, r_{i_0}}\big\vert  < \lambda$. 
Otherwise, stop, denote by $\hat{N}$ the number of detected change-points and by 
$\hat{\eta}_i$ -- the
change-points in increasing order for $i=0, \ldots, \hat{N}+1$ where $\hat{\eta}_{0}=0$ and $\hat{\eta}_{\hat{N}+1}=T$. 
The estimated function $\hat{f}$ is obtained by simple linear regression for each region determined by the final change-points $\hat{\eta}_1, \ldots, \hat{\eta}_{\hat{N}}$ as in \eqref{e62}, with the
exception for the case of single data point as described in Stage 1 above. 

Through these two stages of post processing, the estimation of the number and the locations of change-points become consistent, and further details can be found in Section \ref{sec3}.


\section{Theoretical results} \label{sec3}
We study the $l_2$ consistency of $\tilde{f}$ and $\dbtilde{f}$, and the change-point detection consistency of $\hat{f}$, where the estimators are
defined in Section \ref{sec2}. The $l_2$ risk of an estimator $\tilde{f}$ is defined as $\big\| \tilde{f} - f \big\|_T^2 = T^{-1}\sum_{i=1}^T(\tilde{f}_i-f_i)^2$, where $f$ is the underlying signal as in \eqref{e1}. 
We firstly investigate the $l_2$ behaviour of $\tilde{f}$. The proofs of Theorems \ref{thm1}-\ref{thm3} can be found in Appendix \ref{apx}. 

\begin{Thm} \label{thm1}
$X_t$ follows model \eqref{e1} with $\sigma=1$ and $\tilde{f}$ is the estimator in \eqref{e60}. 
If the threshold $\lambda = C_1\{2 \log (T) \}^{1/2}$ with a constant $C_1 \geq \sqrt{3}$, then we have 
\begin{equation} \label{e7}
\mathbb{P} \; \bigg(\;  \| \tilde{f}-f \|_T^2 \; \leq \; C_1^2 \; \frac{1}{T} \; \log(T) \; \Big\{ 4 + 8 N \; \lceil\; \log (T) / \log (1-\rho)^{-1} \; \rceil \; \Big\} \; \bigg) \; \rightarrow \; 1,
\end{equation}
as $T \rightarrow \infty$ and the piecewise-linear estimator $\tilde{f}$ contains $\tilde{N} \leq C N \log (T)$ change-points where $C$ is a constant. 
\end{Thm}
Thus, $\tilde{f}$ is $l_2$ consistent under the strong sparsity assumption (i.e. if $N$ is finite) or even under the relaxed condition that $N$ has the order of $\log T$. 
The crucial mechanism of $l_2$ consistency is the ``tail-greediness'' which allows up to $K(j) \geq 1$ smooth coefficients to be removed at each scale $j$. 
In other words, consistency is generally unachievable if we proceed in a greedy (as opposed to tail-greedy) way, i.e. if we only merge one triplet at each scale of the TGUW transformation.

We now move onto the estimator $\dbtilde{f}$ obtained in the first stage of post-processing. 

\begin{Thm} \label{thm2}
$X_t$ follows model \eqref{e1} with $\sigma=1$ and $\dbtilde{f}$ is the estimator in \eqref{e62}. 
Let the threshold $\lambda$ be as in Theorem \ref{thm1}.
Then we have $\big\| \dbtilde{f} - f \big\|_T^2 \; =  \; O\big(NT^{-1} \log^2(T)\big)$ with probability approaching $1$ as $T \rightarrow \infty$ and there exist at most two 
estimated change-points between each pair of true change-points $(\eta_i, \eta_{i+1})$ for $i=0, \ldots, N$, where $\eta_0=0$ and $\eta_{N+1}=T$.
Therefore $\dbtilde{N} \leq 2(N+1)$.
\end{Thm}
We see that $\dbtilde{f}$ is $l_2$ consistent, but inconsistent for the number of change-points.
Now we investigate the final estimators, $\hat{f}$ and $\hat{N}$.


\begin{Thm} \label{thm3}
$X_t$ follows model \eqref{e1} with $\sigma=1$ and ($\hat{f}$, $\hat{N}$) are the estimators obtained in Section \ref{pp2}. 
Let the threshold $\lambda$ be as in Theorem \ref{thm1} and suppose that the number of true change-points, 
$N$, has the order of $\log T$.
Let $\Delta_T = \min_{i=1, \ldots, N} \Big\{ \Big(\ubar{f}_T^i \Big)^{2/3} \cdot \delta_T^i \Big\}$ 
where $\ubar{f}_T^i = \min \Big( \vert f_{\eta_{i+1}}-2f_{\eta_{i}}+f_{\eta_{i-1}}\vert , \vert f_{\eta_{i+2}}-2f_{\eta_{i+1}}+f_{\eta_{i}}\vert  \Big)$ and 
$\delta_T^i = \min \Big( \vert \eta_i-\eta_{i-1}\vert , \vert \eta_{i+1}-\eta_{i}\vert  \Big)$.
Assume that $T^{1/3} R_T^{1/3} = o\Big(\Delta_T \Big)$ where $\big\| \dbtilde{f} - f \big\|_T^2 = O_p(R_T)$ is as in Theorem \ref{thm2}.
Then we have 
\begin{equation} 
\mathbb{P} \; \bigg( \hat{N}=N, \quad \max_{i=1, \ldots, N} \bigg\{ \vert \hat{\eta}_i-\eta_i\vert  \cdot \Big(\ubar{f}_T^i \Big)^{2/3} \bigg\} \leq C T^{1/3} R_T^{1/3} \bigg) \; \rightarrow \; 1,
\end{equation}
as $T \rightarrow \infty$ where $C$ is a constant. 
\end{Thm}

Our theory indicates that when $\min_i \ubar{f}_T^i \sim T^{-1}$, the change-point detection rate of the TrendSegment procedure is $O_p (T^{2/3} \log T)$. 
If the number of true change-points, $N$, is finite, then the detection accuracy becomes $O_p (T^{2/3} (\log T)^{2/3})$.
Comparing it with the rate of $O_p (T^{2/3} (\log T)^{1/3})$ derived by \citet{baranowski2019narrowest} and \citet{anastasiou2021detecting} and also with the rate of $O_p (T^{2/3})$ derived by \citet{raimondo1998minimax}, our detection accuracy is different by only a logarithmic factor.
In the case in which $\min_i \ubar{f}_T^i$ is bounded away from zero, the consistent estimation of the number and locations of change-point is achieved by assuming $T^{1/3} R_T^{1/3} = o(\delta_T)$ where $\delta_T=\min_{i=1, \ldots, N+1}\vert \eta_i-\eta_{i-1}\vert $ and $R_T = N T^{-1} \log^2(T)$.
In addition, when there exists a separate data segment containing only one data point, then the two consecutive change-points, $\eta_{k-1}$ and $\eta_k$, linked via $\eta_{k-1} = \eta_k -1$ under the definition of a change-point in \eqref{cp} can be detected exactly at their true locations only if the corresponding $\ubar{f}_T^i$s satisfy the condition $\min \Big( \ubar{f}_T^k, \ubar{f}_T^{k-1} \Big) \gtrsim \log(T)$.

In the supplementary material, the assumptions of the Gaussianity and the independence on $\varepsilon_t$ are relaxed and the corresponding Theorems B.1-B.3 are presented in a setting in which the noise is dependent and/or non-Gaussian.


\section{Simulation study} \label{sec4}

\subsection{Parameter choice and setting} \label{sec4.1}

\subsubsection{Post-processing}\label{sec4.1.1}
In what follows, we disable Stages 1 and 2 of post-processing by default: our empirical experience is that Stage 1 rarely makes a difference in practice but comes with an additional computational cost, and Stage 2 occasionally over-prunes change-point estimates.

\subsubsection{Choice of the ``tail-greediness" parameter}\label{sec4.1.2}
$\rho \in (0, 1)$ is a constant which controls the greediness level of the TGUW transformation in the sense that it decides how many merges are performed in a single pass over the data. A large $\rho$ can reduce the computational cost but it makes the procedure less adaptive, whereas a small $\rho$ gives the opposite effect. Based on our empirical experience, the best performance is stably achieved in the range $\rho \in (0, 0.05]$ and we use $\rho=0.04$ as a default in the simulation study and data analyses. 

\subsubsection{Choice of the minimum segment length}\label{sec4.1.3}
We can give a condition on the minimum segment length of the estimated signal returned by the TrendSegment algorithm. If it is set to $1$, two consecutive data-points can be detected as change-points. 
As theoretically shown in the supplementary material, the minimum length of the estimated segment should have an order of $\log(T)$ to achieve estimation consistency in the case of dependent and/or non-Gaussian errors.
To avoid too short segments, and to cover non iid Gaussian noise, we set the minimum segment length to $C \log(T)$ and use the default $C=0.9$ in the remainder of the paper, otherwise we are not able to detect those short segments in (M6). This constraint can be adjusted by users in the R package \code{trendsegmentR}.  


\subsubsection{Continuity at change-points} \label{sec4.1.4}
As described in Section \ref{sec2}, the TrendSegment algorithm works by detecting change-points first (in thresholding) and then estimating the best linear fit (in the least-squares sense) for each segment (in the inverse TGUW transform). These procedures normally ensure discontinuity at change-points, however our R package \code{trendsegmentR} has an option for ensuring continuous change-points by approximating $f$ using the linear spline fit with knots at detected change-points.

\subsubsection{Choice of threshold $\lambda$} \label{sec4.1.5}
Motivated by Theorem \ref{thm1}, we consider the simplest naïve threshold of the form 
\begin{equation}\label{thrnaive}
    \lambda^\text{Naïve} = C\sigma \sqrt{2 \log T},
\end{equation}
where $\sigma$ can be estimated in different ways depending on the type of noise. Under iid Gaussian noise, we can estimate $\sigma$ using the Median Absolute Deviation (MAD) estimator \citep{hampel1974influence} defined as  $\hat{\sigma}=\text{Median}(\vert X_1-2X_2+X_3\vert , \ldots, \vert X_{T-2}-2X_{T-1}+X_{T}\vert )/(\Phi^{-1}(3/4)\sqrt{6})$ where $\Phi^{-1}$ is the quantile function of the Gaussian distribution. We found that under iid Gaussian noise $C=1.3$ empirically leads to the best performance over a sequence of $C$, where the details and the relevant results for non-Gaussian and/or dependent errors can be found in Section C of the supplementary material. 
For completeness, we now present an algorithm for a threshold that works well in all circumstances. 
When the noise is not generated from iid Gaussian, it is reasonable to assume that the threshold is affected by the serial dependence structure and/or the extent of heavy-tailedness of noise, which motivates us to use threshold of the form:
\begin{equation}\label{newlambda}
    \lambda^\text{Robust} = C \mathcal{I} g(\mathcal{K}) \sqrt{2 \log T},
\end{equation}
where $\mathcal{I}$ is the long-run standard deviation, $\mathcal{K}$ is kurtosis and $g$ is a function. To estimate the unknown parameters in \eqref{newlambda}, we follow Algorithm \ref{algo_lambda}.

\begin{center}
\begin{algorithm}[H]
\setstretch{1.5}
\SetAlgoLined
\textbf{INPUT}: $\b{X}$, $\lambda^\text{Naïve}$, $C$, $\eta_{\max}$ 
\begin{enumerate}
    \item Pre-estimate the fit, $\hat{f}_t$, via the TrendSegment algorithm with $\eta_{\max}$, where $\eta_{\max}$ is a pre-specified maximum number of estimated change-points. \label{s1}
    \item Compute the empirical residuals, $\hat{\varepsilon}_t$, from the pre-fit obtained in 1. \label{s2}
    \item From $\hat{\varepsilon}_t$, compute the sample kurtosis ($\hat{\mathcal{K}}$) and the long-run standard deviation estimator ($\hat{\mathcal{I}}$) based on AR(1) model. See comments underneath the algorithm for details of this step. \label{s3}
    \item Plug in $\hat{\mathcal{I}}$ and $\hat{\mathcal{K}}$ into the threshold formula in \eqref{newlambda} with a pre-specified $C$. \label{s4}
    \item To estimate the function $g$, find a non-parametric regression fit with $X = \hat{\mathcal{K}}$ and $Y = \frac{\lambda^\text{Naïve}}{C \hat{\mathcal{I}} \sqrt{2 \log T}}$, where $\lambda^\text{Naïve}$ is chosen as the best performing threshold by repeating the simulations with a range of threshold constant $C$ over different types of noise.\label{s5}
    \item Obtain the threshold in \eqref{newlambda} based on the set of estimators, ($\hat{\mathcal{I}}$, $\hat{\mathcal{K}}$, $\hat{g}$). \label{s6} 
\end{enumerate}
\textbf{OUTPUT}: The robust threshold $\lambda^\text{Robust}$.
 \caption{Robust threshold selection}
 \label{algo_lambda}
\end{algorithm} 
\end{center}

We now describe the details of each step in Algorithm \ref{algo_lambda}.

\paragraph{Pre-estimated fit in Step 1.}
In \eqref{newlambda}, the heavy-tailedness and dependent structure of the noise are captured by $\mathcal{K}$ and $\mathcal{I}$, respectively. In practice, estimating $\mathcal{I}$ and $\mathcal{K}$ is challenging as the observation includes change-points in its underlying signal. One of the most straightforward way is pre-estimating the fit $\hat{f}_t$ via TrendSegment algorithm with a parameter $\eta_{\max}$, the maximum number estimated change-points. As long as $\eta_{\max}$ is not too large, some extent of overestimation would be acceptable, and we use $\eta_{\max} = \lceil 0.15 T \rceil$ as a default in practice, as it empirically led to the best performance and the simulation results do not vary by much over the range $\eta_{\max} \in [\lceil 0.1 T \rceil, \lceil 0.2 T \rceil]$. The pre-fitting gives us the estimated noise $\hat{\varepsilon}_t = X_t - \hat{f}_t$, from which we can estimate both $\mathcal{I}$ and $\mathcal{K}$.

\paragraph{Pre-specified constant $C$ in Step 4.}
We set $C=1.3$ as it empirically led to the best performance for iid Gaussian noise with the naive approach in \eqref{thrnaive}. Thus we hope to have both $\hat{\mathcal{I}}$ and $\hat{g}(\hat{\mathcal{K}})$ close to $1$ under iid Gaussian noise, but larger than $1$ when the noise has serial dependence and/or heavy-tailedness.

\paragraph{$\mathcal{I}$ and $\mathcal{K}$ in Step 4.}
$\mathcal{I}$ and $\mathcal{K}$ capture dependency and heavy-tailedness of noise, respectively. 
First, kurtosis is estimated from the estimated noise as follows:
\begin{align} \label{kts}
    \hat{\mathcal{K}} = \frac{\sum_{t=1}^T ( \hat{\varepsilon}_t - \bar{\hat{\varepsilon}} )^4}{T\hat{s}^4_{\hat{\varepsilon}}},
\end{align}
where $\bar{\hat{\varepsilon}}$ and $\hat{s}_{\hat{\varepsilon}}$ are sample mean and sample standard deviation of $\hat{\varepsilon}$, respectively.
For estimating $\mathcal{I}$, we consider the case when Gaussian noise has dependent structure. Then the dependencies increase the marginal variance of CUSUM statistic and one way of solving this issue is inflating the threshold by the following factor 
\begin{align} \label{ifl}
    \mathcal{I} = \sqrt{(1+\phi)/(1-\phi)},
\end{align}
where $\phi$ is the true parameter of a AR(1) process \citep{fearnhead2022detecting}. We can estimate $\phi$ by fitting AR(1) model to the estimated noise $\hat{\varepsilon}_t = X_t - \hat{f}_t$, and this gives us the estimated long-run standard deviation $\hat{\mathcal{I}}$. Although in theory the inflation factor in \eqref{ifl} is valid only for Gaussian noise, we use the estimator of \eqref{ifl} as an estimated long-run standard deviation even when the noise has both serial dependence and heavy-tailedness, hoping that the heavy-tailedness is captured reasonably well by $\mathcal{K}$. 

\paragraph{Kurtosis function $g$ in Step 5.}
We fit a non-parametric regression as described in step 5 of Algorithm \ref{algo_lambda} over different models and noise scenarios.  
We found that $g(\hat{\mathcal{K}})$ has no particular functional form in $\hat{\mathcal{K}}$, and is scattered between $0.9$ and $1.6$ over all noise scenarios and all simulations models considered in the paper. Therefore, the resulting non-parametric fit $\hat{g}(\hat{\mathcal{K}})$ also has a flat shape over a range of $\hat{\mathcal{K}}$, and we use this in finding the robust threshold in practice. This is due to the condition on the minimum segment length described earlier which helps the method to be robust to spikes. 

The detailed procedure of estimating $g$ is presented in Section C.2 of the supplementary material. Also, the simulation results using Algorithm \ref{algo_lambda} for dependent and/or heavy-tailed noise can be found in Tables C.1 - C.10 in Section C.1 of the supplementary material. The proposed robust threshold selection algorithm can also be applied to iid Gaussian noise without any knowledge on type of noise and the corresponding simulation results are given in Section \ref{sec4.3}.

\begin{figure}[ht!] 
    \centering
    \begin{subfigure}[t]{0.49\textwidth} 
        \raisebox{-\height}{\includegraphics[width=\textwidth]{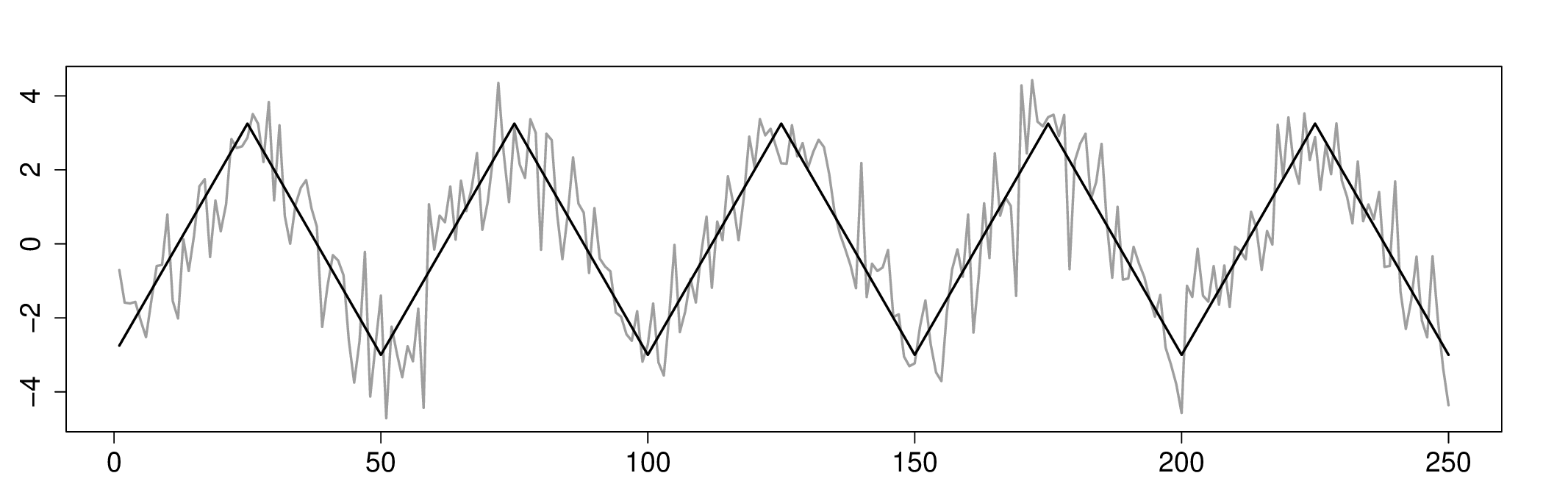}}
        \caption{(M1) wave1}
        \label{fig:sgn1}
    \end{subfigure}
    \hfill
    \begin{subfigure}[t]{0.49\textwidth} 
        \raisebox{-\height}{\includegraphics[width=\textwidth]{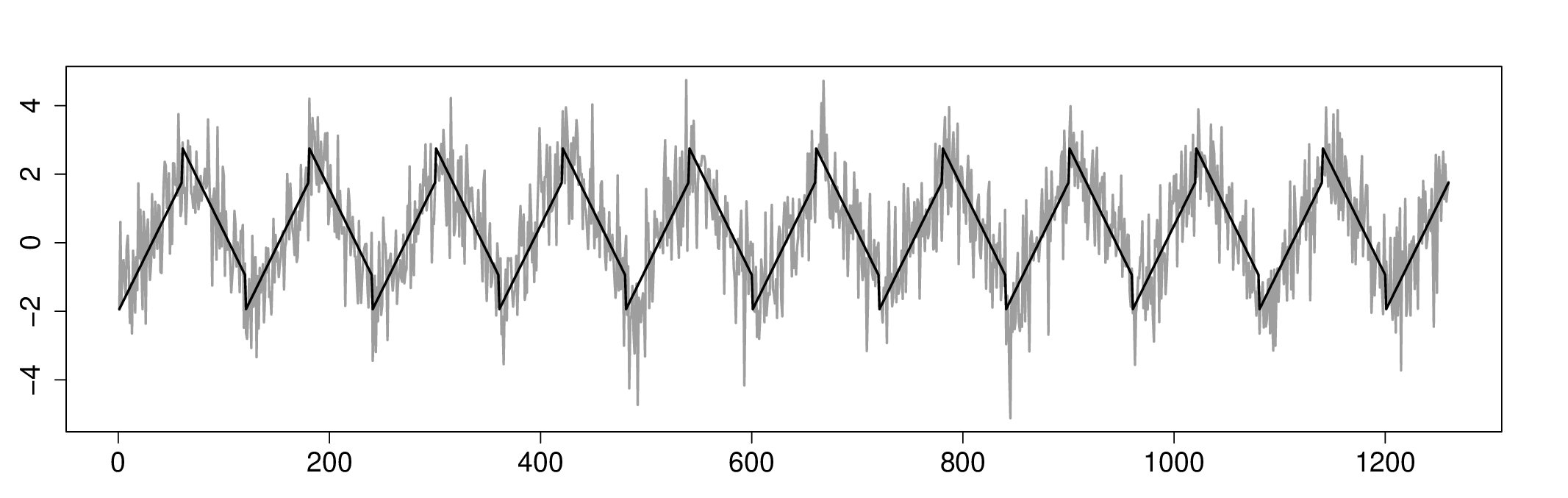}}
        \caption{(M2) wave2}
        \label{fig:sgn2}
    \end{subfigure}
    \begin{subfigure}[t]{0.49\textwidth} 
        \raisebox{-\height}{\includegraphics[width=\textwidth]{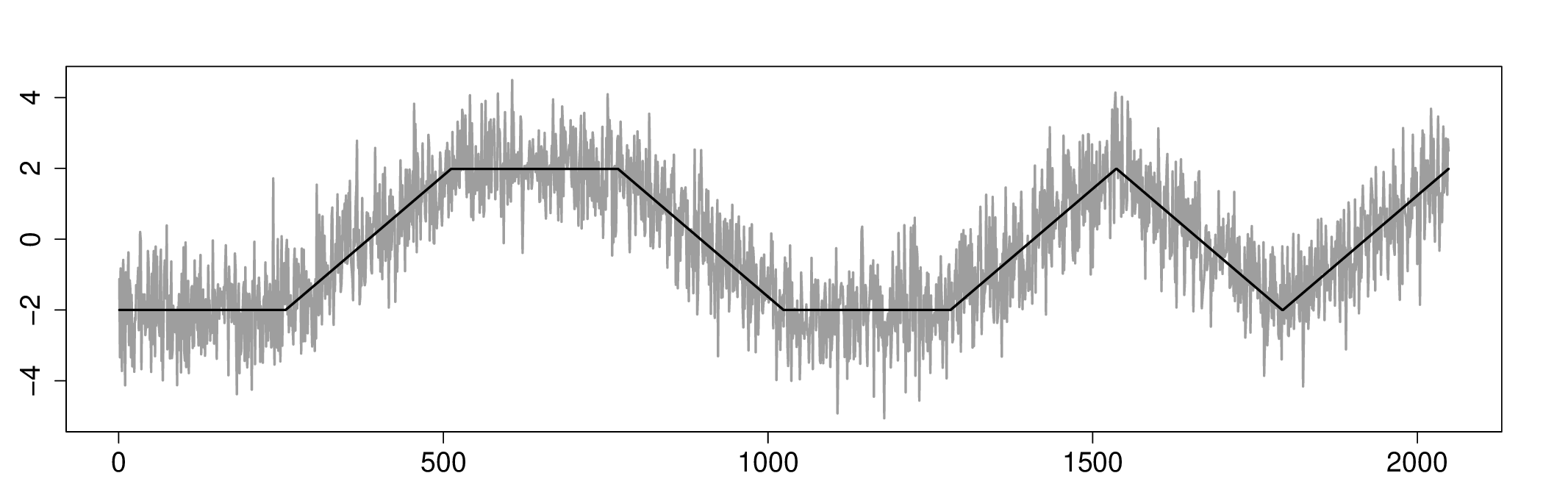}}
        \caption{(M3) mix1}
        \label{fig:sgn3}
    \end{subfigure}
    \hfill
    \begin{subfigure}[t]{0.49\textwidth}
        \raisebox{-\height}{\includegraphics[width=\textwidth]{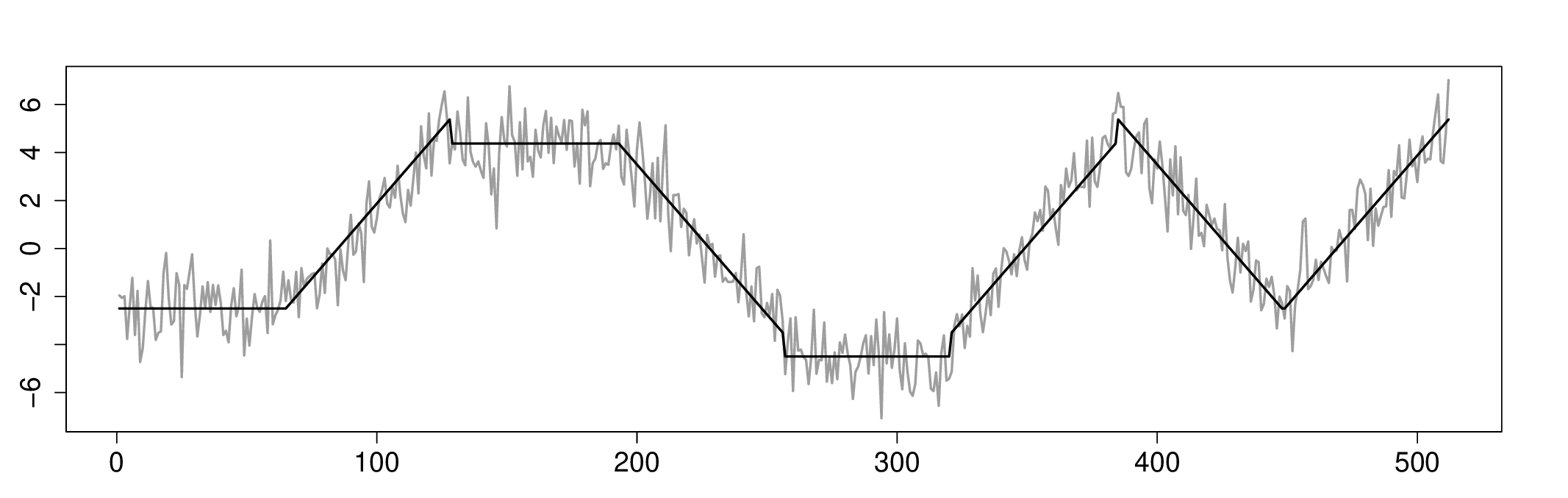}}
        \caption{(M4) mix2}  
    		\label{fig:sgn4}
    \end{subfigure}
    \begin{subfigure}[t]{0.49\textwidth} 
        \raisebox{-\height}{\includegraphics[width=\textwidth]{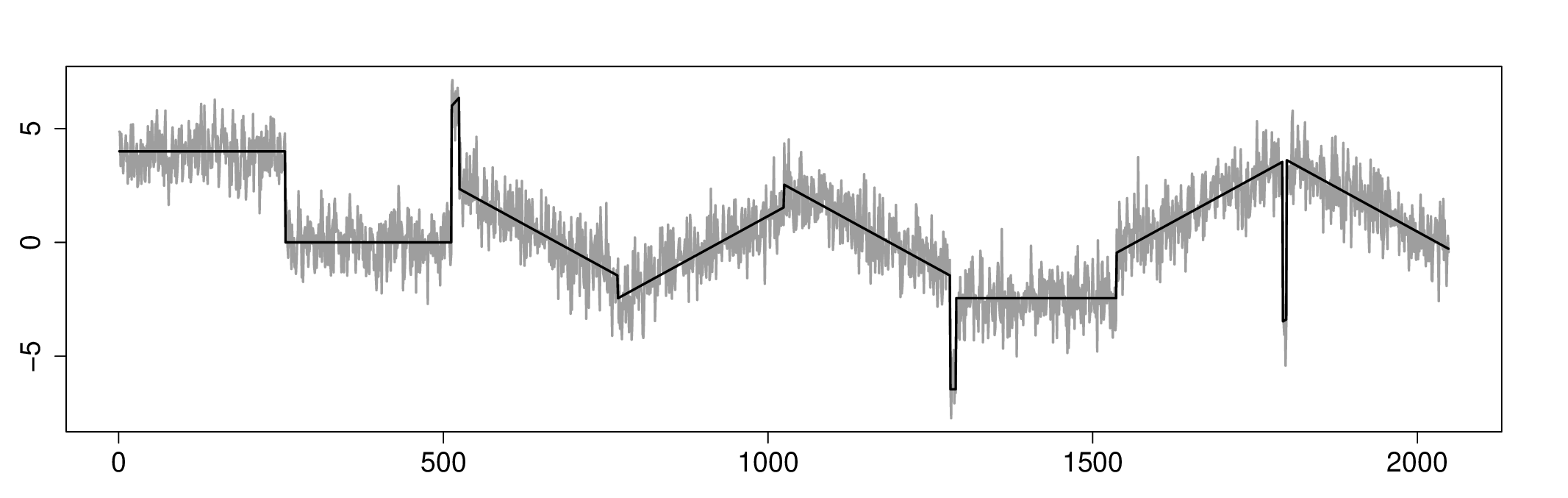}}
        \caption{(M5) mix3}
        \label{fig:sgn5}
    \end{subfigure}
    \hfill
    \begin{subfigure}[t]{0.49\textwidth}
        \raisebox{-\height}{\includegraphics[width=\textwidth]{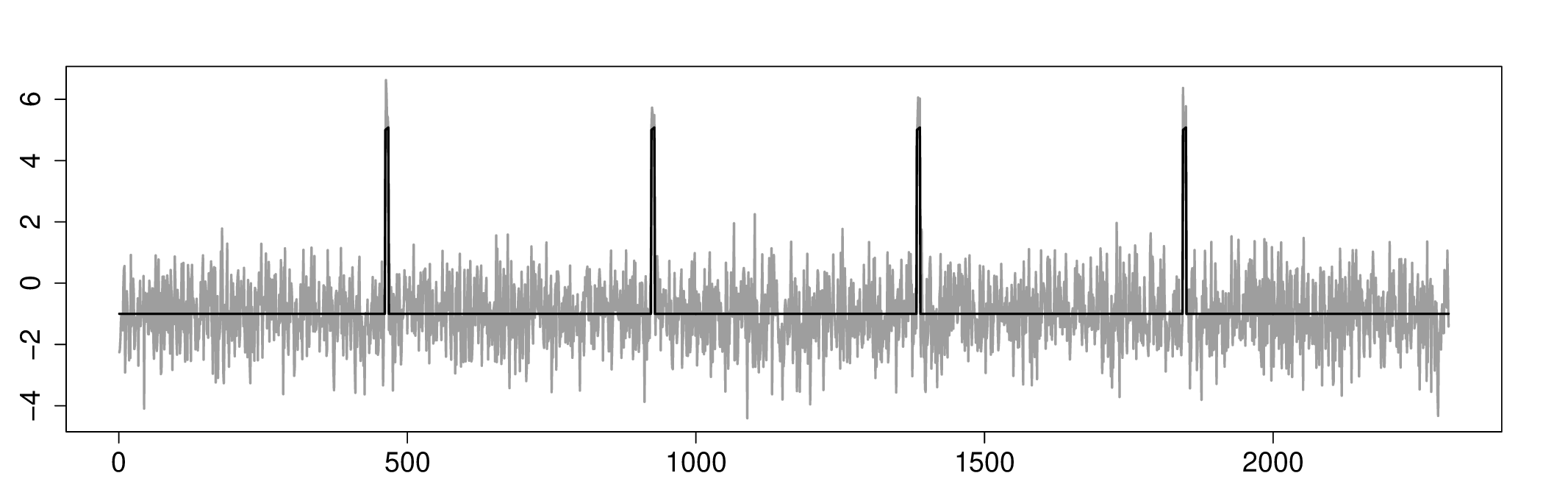}}
        \caption{(M6) lin.sgmts} 
        \label{fig:sgn6}
    \end{subfigure}
    \begin{subfigure}[t]{0.49\textwidth}
        \raisebox{-\height}{\includegraphics[width=\textwidth]{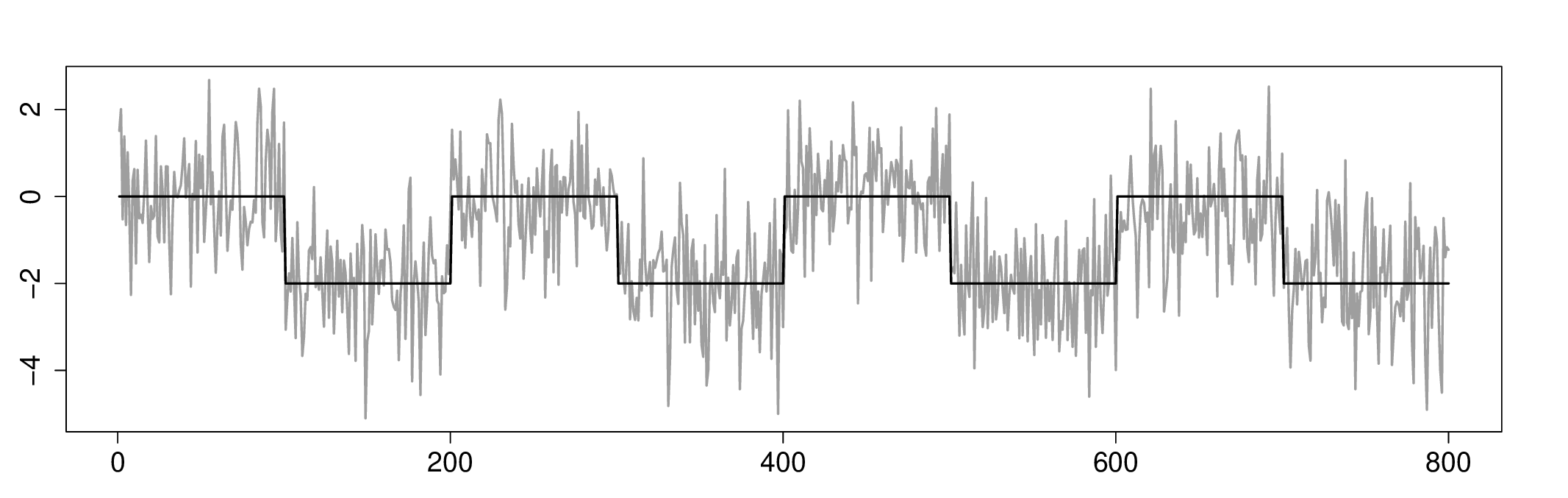}}
        \caption{(M7) teeth} 
        \label{fig:sgn9}
    \end{subfigure}
    \hfill
    \begin{subfigure}[t]{0.49\textwidth}
        \raisebox{-\height}{\includegraphics[width=\textwidth]{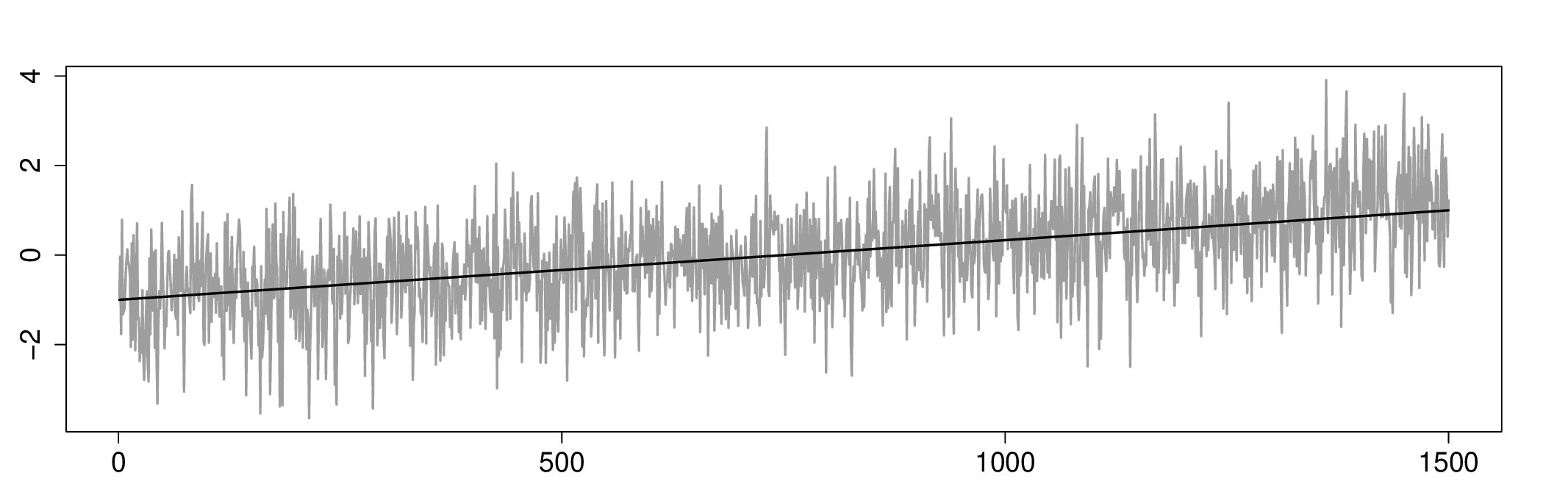}}
        \caption{(M8) lin} 
        \label{fig:sgn10}
    \end{subfigure}
    \hfill
    \caption {Examples of data with its underlying signal studied in Section \ref{sec4}. (a)-(h) data series $X_t$ (light grey) and true signal $f_t$ (black).}
\label{fig:signal}
\end{figure}

We consider iid Gaussian noise and simulate data from model \eqref{e1} using 8 signals, (M1) wave1, (M2) wave2, (M3) mix1, (M4) mix2, (M5) mix3, (M6) lin.sgmts, (M7) teeth and (M8) lin, shown in Figure \ref{fig:signal}. (M1) is continuous at change-points, while (M2) has discontinuities. (M3) contains both constant and linear segments and is continuous at change-points, whereas (M4) is of the similar type but has a mix of continuous and discontinuous change-points. (M5) has three particularly short segments containing 12, 9 and 6 data points, respectively and (M6) has isolated spike-type short segments containing 6 data points each. (M7) is piecewise-constant, and (M8) is a linear signal without change-points. The signals and R code for all simulations can be downloaded from our GitHub repository \citep{maeng2021tsgithub} and the simulation results under dependent or heavy-tailed errors can be found in Section C of the supplementary materials. 

\subsection{Competing methods and estimators} \label{sec4.2}
We perform the TrendSegment procedure based on the parameter choice in Section \ref{sec4.1} and compare the performance with that of the following competitors: Narrowest-Over-Threshold detection ($\textbf{NOT}$, \citet{baranowski2019narrowest}) implemented in the R package \code{not} from CRAN, Isolate-Detect ($\textbf{ID}$, \citet{anastasiou2021detecting}) available in the R package \code{IDetect}, trend filtering ($\textbf{TF}$, \citet{kim2009ell_1}) available from \url{https://github.com/glmgen/genlasso}, Continuous-piecewise-linear Pruned Optimal Partitioning ($\textbf{CPOP}$, \citet{fearnhead2019detecting}) available from \url{https://www.maths.lancs.ac.uk/~fearnhea/Publications.html} and a bottom-up algorithm based on the residual sum of squares (RSS) from a linear fit ($\textbf{BUP}$, \citet{keogh2004segmenting}). 
The TrendSegment methodology is implemented in the R package \code{trendsegmentR}.
 
As BUP requires a pre-specified number of change-points (or a well-chosen stopping criterion which can vary depending on the data), we include it in the simulation study (with the stopping criterion optimised for the best performance using the knowledge of the truth) but not in data applications. We do not include the methods of \citet{spiriti2013knot} and \citet{bai2003computation} implemented in the R packages \code{freeknotsplines} and \code{strucchange} as we have found them to be particularly slow. For instance, the minimum segment size in \code{strucchange} can be adjusted to be small as long as it is greater than or equal to 3 for detecting linear trend changes. This is suitable for detecting very short segments (e.g in (M6) lin.sgmts), however this setting is accompanied by extremely heavy computation: with this minimum segment size in place, a single signal simulated from (M6) took us over three hours to process on a standard PC.

Out of the competing methods tested, ID, TF and CPOP return continuous change-points, while the estimated signals of Trendsegment and BUP is in principle discontinuous at change-points. For NOT, we use the contrast function for not necessarily continuous piecewise-linear signals. Regarding the tuning parameters for the competing methods, we follow the recommendation of each respective paper or the corresponding R package.


\subsection{Results} \label{sec4.3}
The summary of the results for all models and methods can be found in Tables \ref{M15} and \ref{M67}. 
We run 100 simulations and as a measure of accuracy of estimators, we use Monte-Carlo estimates of the Mean Squared Error of the estimated signal defined as MSE=$\mathbb{E}\{(1/T)\sum_{t=1}^T (f_t-\hat{f}_t)^2\}$. 
The empirical distribution of $\hat{N}-N$ is also reported where $\hat{N}$ is the estimated number of change-points and $N$ is the true one. 
In addition to this, for comparing the accuracy of the locations of the estimated change-points $\hat{\eta}_i$, we show estimates of the scaled Hausdorff distance given by
\begin{equation} \label{e42}
d_H = \frac{1}{T} \mathbb{E} \max \bigg\{ \max_i \min_j \big\vert \eta_i-\hat{\eta}_j \big\vert , \quad \max_j \min_i  \big\vert \hat{\eta}_j-\eta_i \big\vert  \bigg\}
\end{equation}
where $i=0, \ldots, N+1$ and $j=0, \ldots, \hat{N}+1$ with the convention $\eta_0=\hat{\eta}_0=0, \eta_{N+1}=\hat{\eta}_{N+1}=T$ and $\hat{\eta}$ and $\eta$ denote estimated and true locations of the change-points. 
The smaller the Hausdorff distance, the better the estimation of the change-point locations. For each method, the average computation time in seconds is shown.

\begin{table}[ht!]
\centering
\caption {Distribution of $\hat{N}-N$ for models (M1)-(M4) and all methods listed in Section \ref{sec4.1} and \ref{sec4.2} over 100 simulation runs. Also the average MSE (Mean Squared Error) of the estimated signal $\hat{f}_t$ defined in Section \ref{sec4.3}, the average Hausdorff distance $d_H$ given by \eqref{e42} and the average computational time in seconds using an Intel Core i5 2.9 GHz CPU with 8 GB of RAM, all over 100 simulations. Bold: methods within 10\% of the highest empirical frequency of $\hat{N}-N=0$ or within 10\% of the lowest empirical average $d_H(\times 10^2)$. Note that TrendSegment is shortened to TS.}
\renewcommand{\arraystretch}{1} 
\begin{tabular}{ccrrrrrrrrrr}
  \hline
    & & & & & $\hat{N}-N$ & & & \\
  \cline{3-9}
Model & Method & $\leq$-3 & -2 & -1 & 0 & 1 & 2 & $\geq$3 & MSE & $d_H(\times 10^2)$ & time \\ 
  \hline
   \multirow{7}*{(M1)} & TS($\lambda^\text{Naïve}$) & 0 & 0 & 2 & \textbf{98} & 0 & 0 & 0 & 0.23 &  2.96 & 0.22 \\ 
  & TS($\lambda^\text{Robust}$) & 0 & 0 & 2 & \textbf{97} & 1 & 0 & 0 & 0.23 & 2.97 & 0.09 \\ 
  & NOT & 0 & 0 & 0 & \textbf{98} & 2 & 0 & 0 & 0.19 &  2.28 & 0.22 \\ 
  & ID & 0 & 0 & 0 & \textbf{97} & 3 & 0 & 0 & 0.14 &  \textbf{1.52} & 0.02 \\ 
  & TF & 0 & 0 & 0 & 0 & 0 & 0 & 100 & 0.11 &  4.50 & 3.18 \\ 
  & CPOP & 0 & 0 & 0 & \textbf{97} & 2 & 1 & 0 & 0.09 &  \textbf{1.09} & 0.05 \\ 
  & BUP & 100 & 0 & 0 & 0 & 0 & 0 & 0 & 2.65 & 10.75 & 0.35 \\ 
   \hline
\multirow{7}*{(M2)} & TS($\lambda^\text{Naïve}$) & 0 & 0 & 2 & \textbf{98} & 0 & 0 & 0 & 0.11 & 1.90 & 0.50 \\ 
   & TS($\lambda^\text{Robust}$) & 0 & 0 & 4 & \textbf{96} & 0 & 0 & 0 & 0.11 & 1.91 & 0.24 \\ 
   & NOT & 0 & 0 & 2 & \textbf{98} & 0 & 0 & 0 & 0.09 & 1.56 &  0.35 \\ 
   & ID & 0 & 0 & 0 & \textbf{94} & 6 & 0 & 0 & 0.09 & \textbf{1.44} &  0.23 \\ 
   & TF & 0 & 0 & 0 & 0 & 0 & 0 & 100 & 0.06 & 2.31 & 31.34 \\ 
   & CPOP & 0 & 0 & 0 & \textbf{93} & 7 & 0 & 0 & 0.06 & \textbf{1.15} &  2.09 \\ 
   & BUP & 100 & 0 & 0 & 0 & 0 & 0 & 0 & 0.75 & 4.69 &  2.21 \\ 
   \hline
\multirow{7}*{(M3)} & TS($\lambda^\text{Naïve}$) & 0 & 0 & 0 & \textbf{99} & 1 & 0 & 0 & 0.03 & 3.33 &  0.61 \\ 
   & TS($\lambda^\text{Robust}$) & 0 & 0 & 0 & \textbf{100} & 0 & 0 & 0 & 0.03 & 3.33 & 0.29 \\ 
   & NOT & 0 & 0 & 0 & \textbf{100} & 0 & 0 & 0 & 0.02 & 2.70 &  0.33 \\ 
   & ID & 0 & 0 & 0 & \textbf{100} & 0 & 0 & 0 & 0.02 & \textbf{1.86} &  0.02 \\ 
   & TF & 0 & 0 & 0 & 0 & 0 & 0 & 100 & 0.01 & 5.41 & 28.89 \\ 
   & CPOP & 0 & 0 & 0 & \textbf{100} & 0 & 0 & 0 & 0.01 & \textbf{1.02} & 17.38 \\ 
   & BUP & 0 & 0 & 0 & 2 & 22 & 48 & 28 & 0.03 & 5.46 &  2.20  \\ 
   \hline
\multirow{7}*{(M4)} & TS($\lambda^\text{Naïve}$) & 0 & 0 & 0 & \textbf{100} & 0 & 0 & 0 & 0.09 &  3.24 & 0.31 \\ 
  & TS($\lambda^\text{Robust}$) & 0 & 0 & 0 & \textbf{100} & 0 & 0 & 0 & 0.09 & 3.24 & 0.09 \\ 
  & NOT & 0 & 0 & 0 & \textbf{99} & 1 & 0 & 0 & 0.08 &  \textbf{2.71} & 0.23 \\ 
  & ID & 0 & 0 & 0 & \textbf{97} & 3 & 0 & 0 & 0.07 &  \textbf{2.04} & 0.02 \\ 
  & TF & 0 & 0 & 0 & 0 & 0 & 0 & 100 & 0.05 &  5.47 & 8.50 \\ 
  & CPOP & 0 & 0 & 0 & \textbf{97} & 3 & 0 & 0 & 0.04 &  \textbf{1.83} & 0.39 \\ 
  & BUP & 7 & 64 & 27 & 2 & 0 & 0 & 0 & 0.52 & 10.66 & 0.56 \\ 
   \hline
\end{tabular}\\
 \label{M15}
\end{table}

\begin{table}[ht!]
\centering
\caption {Distribution of $\hat{N}-N$ for models (M5)-(M8) and all methods listed in Section \ref{sec4.1} and \ref{sec4.2} over 100 simulation runs. Also the average MSE (Mean Squared Error) of the estimated signal $\hat{f}_t$ defined in Section \ref{sec4.3}, the average Hausdorff distance $d_H$ given by \eqref{e42} and the average computational time in seconds using an Intel Core i5 2.9 GHz CPU with 8 GB of RAM, all over 100 simulations. Bold: methods within 10\% of the highest empirical frequency of $\hat{N}-N=0$ or within 10\% of the lowest empirical average $d_H(\times 10^2)$. Note that TrendSegment is shortened to TS.}
\renewcommand{\arraystretch}{1} 
\begin{tabular}{ccrrrrrrrrrr}
  \hline
    & & & & & $\hat{N}-N$ & & & \\
  \cline{3-9}
Model & Method & $\leq$-3 & -2 & -1 & 0 & 1 & 2 & $\geq$3 & MSE & $d_H(\times 10^2)$ & time \\ 
  \hline
 \multirow{7}*{(M5)} 
   & TS($\lambda^\text{Naïve}$) & 0 & 0 & 0 & \textbf{90} & 10 & 0 & 0 & 0.03 & \textbf{1.40} & 1.30 \\ 
   & TS($\lambda^\text{Robust}$) & 0 & 0 & 0 & \textbf{89} & 11 & 0 & 0 & 0.03 & \textbf{1.41} & 0.32 \\
   & NOT & 0 & 12 & 9 & 75 & 3 & 0 & 1 & 0.05 & \textbf{0.73} & 0.25 \\
   & ID & 0 & 0 & 0 & 1 & 5 & 25 & 69 & 0.29 & 8.09 & 0.03 \\ 
   & TF & 0 & 0 & 0 & 0 & 0 & 0 & 100 & 0.14 & 6.15 & 28.53 \\ 
   & CPOP & 0 & 0 & 0 & 8 & 27 & 31 & 34 & 0.03 & \textbf{1.42} & 3.50 \\  
   & BUP & 0 & 0 & 0 & 41 & 44 & 13 & 2 & 0.10 & 4.72 & 2.25 \\ 
   \hline
\multirow{7}*{(M6)} & TS($\lambda^\text{Naïve}$) & 0 & 0 & 0 & \textbf{99} & 1 & 0 & 0 & 0.01 & \textbf{0.05} & 0.90 \\ 
   & TS($\lambda^\text{Robust}$) & 0 & 3 & 1 & \textbf{96} & 0 & 0 & 0 & 0.02 & 0.64 & 0.34 \\ 
   & NOT & 2 & 13 & 37 & 45 & 2 & 1 & 0 & 0.07 & 1.74 & 0.25 \\ 
   & ID & 0 & 0 & 0 & 0 & 0 & 1 & 99 & 0.07 & \textbf{0.17} & 0.04 \\ 
   & TF & 0 & 0 & 0 & 0 & 0 & 0 & 100 & 0.13 & 9.87 & 30.72 \\ 
   & CPOP & 0 & 0 & 0 & 21 & 28 & 40 & 11 & 0.03 & \textbf{0.22} & 3.02 \\  
   & BUP & 0 & 0 & 0 & 0 & 0 & 0 & 100 & 0.12 & 9.29 & 2.70 \\ 
   \hline
\multirow{7}*{(M7)} & TS($\lambda^\text{Naïve}$) & 0 & 5 & 21 & 40 & 28 & 6 & 0 & 0.10 &  7.02 & 0.31 \\ 
   & TS($\lambda^\text{Robust}$) & 1 & 10 & 38 & 31 & 16 & 4 & 0 & 0.13 & 8.64 & 0.13 \\ 
   & NOT & 1 & 1 & 8 & \textbf{56} & 31 & 3 & 0 & 0.06 & \textbf{2.62} &  0.25 \\ 
   & ID & 3 & 0 & 16 & 14 & 26 & 13 & 28 & 0.32 & 10.87 &  0.12 \\ 
   & TF & 0 & 0 & 0 & 0 & 0 & 0 & 100 & 0.10 &  6.11 & 23.19 \\ 
   & CPOP & 0 & 0 & 1 & 1 & 3 & 17 & 78 & 0.05 & \textbf{3.37} &  1.19 \\ 
   & BUP & 70 & 25 & 5 & 0 & 0 & 0 & 0 & 0.28 & 11.89 &  1.58 \\ 
   \hline
\multirow{7}*{(M8)} & TS($\lambda^\text{Naïve}$) & 0 & 0 & 0 & \textbf{100} & 0 & 0 & 0 & 0.00 & \textbf{0.00} &  0.43 \\ 
   & TS($\lambda^\text{Robust}$) & 0 & 0 & 0 & \textbf{100} & 0 & 0 & 0 & 0.00 & \textbf{0.00} & 0.19 \\ 
   & NOT & 0 & 0 & 0 & \textbf{100} & 0 & 0 & 0 & 0.00 & \textbf{ 0.00} &  0.17 \\ 
   & ID & 0 & 0 & 0 & \textbf{100} & 0 & 0 & 0 & 0.00 & \textbf{ 0.00} &  0.59 \\ 
   & TF & 0 & 0 & 0 & 78 & 5 & 2 & 15 & 0.00 &  9.08 & 35.79 \\ 
   & CPOP & 0 & 0 & 0 & \textbf{100} & 0 & 0 & 0 & 0.00 & \textbf{ 0.00} & 12.96 \\ 
   & BUP & 0 & 0 & 0 & 0 & 0 & 0 & 100 & 0.01 & 46.34 &  2.63 \\ 
   \hline
\end{tabular} \label{M67}
\end{table}

We first emphasise that the results with both the naïve and the robust thresholds ($\lambda^\text{Naïve}$ in \eqref{thrnaive} and $\lambda^\text{Robust}$ in \eqref{newlambda}) are reported for TrendSegment, and the performances are nearly the same except (M7). For simplicity, we call both methods as TrendSegment in the remainder of this section. 

The results for (M1) and (M2) are similar. TrendSegment shows comparable performance to NOT, ID and CPOP  in terms of the estimation of the number of change-points while it is less attractive in terms of the estimated locations of change-points. TF tends to overestimate the number of change-points throughout all models. When the signal is a mix of constant and linear trends as in (M3) and (M4), TrendSegment, NOT and ID still perform well in terms of the estimation of the number of change-points. CPOP tends to overestimate in (M4) when there exists discontinuity at change-points, however it shows the best performs in terms of localisation (i.e. the smallest mean of Hausdorff distance) as it tends to estimate more than one (and somewhat frequent) change-points at discontinuous change-points. 
As TrendSegment and NOT deal with the piecewise-linear signals that is not necessarily continuous at change-points, they performs better than others in (M2) and (M4).

We see that TrendSegment has a particular advantage over the other methods especially in (M5) and (M6), when frequent change-points composed of the isolated spike-type short segments of length 6 exist. This is due to the bottom-up nature of TrendSegment which focuses on local features in the early stage of merges and enables TrendSegment to detect those short segments. TrendSegment shows its relative robustness in estimating the number and the location of change-points while NOT, ID and CPOP significantly underperform. 

For the estimation of the piecewise-constant signal (M7), no methods show good performances and NOT, ID and TrendSegment tend to underestimate the number of change-points while CPOP and TF overestimate. In the case of the no-change-point signal (M8), all methods estimate well except TF and BUP. In summary, TrendSegment is never among the worst methods, is almost always among the best ones, and is particularly attractive for signals containing frequent change-points with short segments. With respect to computation time, NOT and ID are very fast in all cases, TrendSegment is slower than these two but is faster than TF, CPOP and BUP, especially when the length of the time series is larger than 2000.

\section{Data applications} \label{sec5}

\subsection{Average January temperatures in Iceland} \label{iceland}
\begin{figure}[ht!] 
     \centering
    \begin{subfigure}[t]{0.49\textwidth} 
        \raisebox{-\height}{\includegraphics[width=\textwidth]{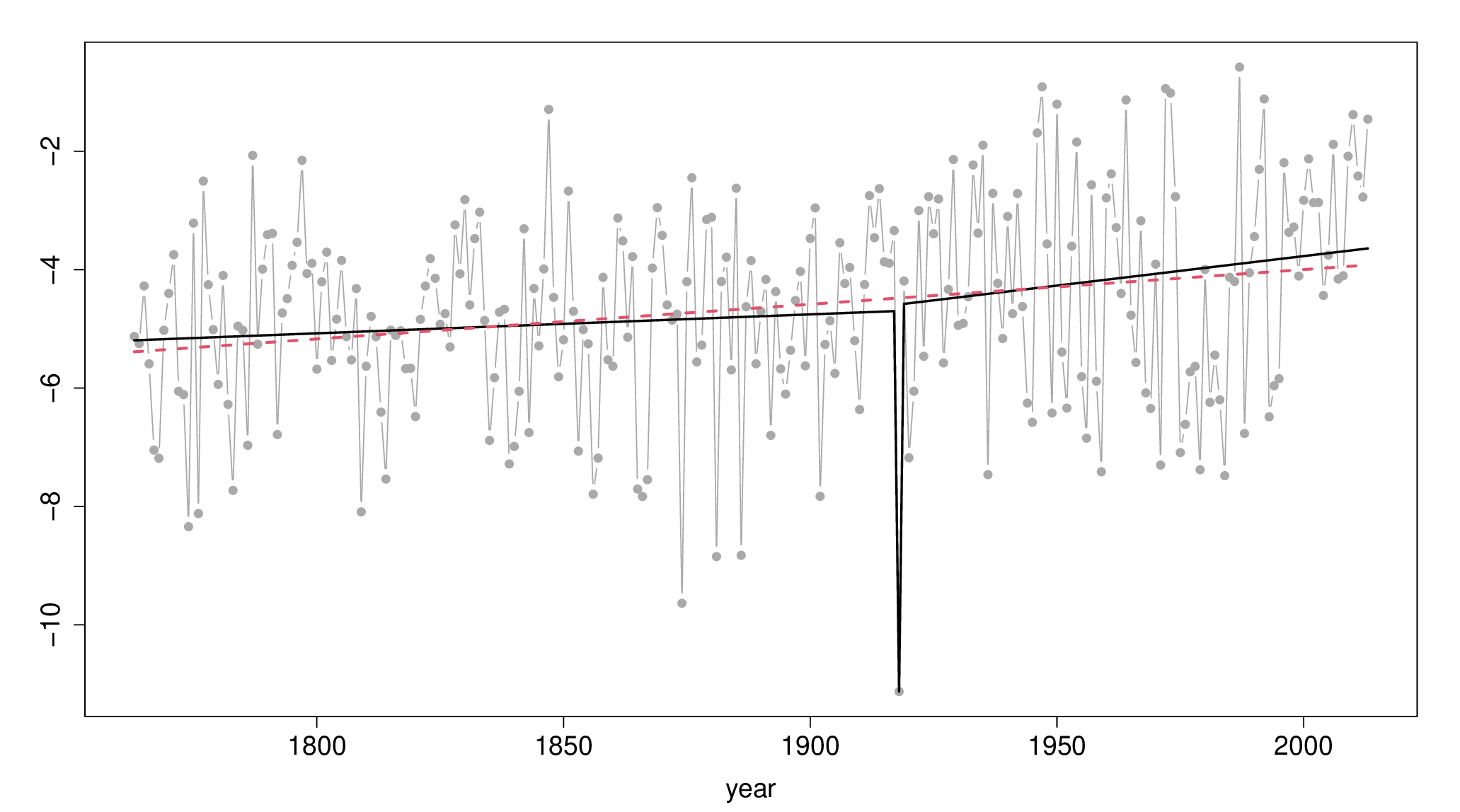}}
        \caption{TrendSegment with $\lambda^\text{Naïve}$}
        \label{fig:ice1}
    \end{subfigure}
    \hfill
    \begin{subfigure}[t]{0.49\textwidth} 
        \raisebox{-\height}{\includegraphics[width=\textwidth]{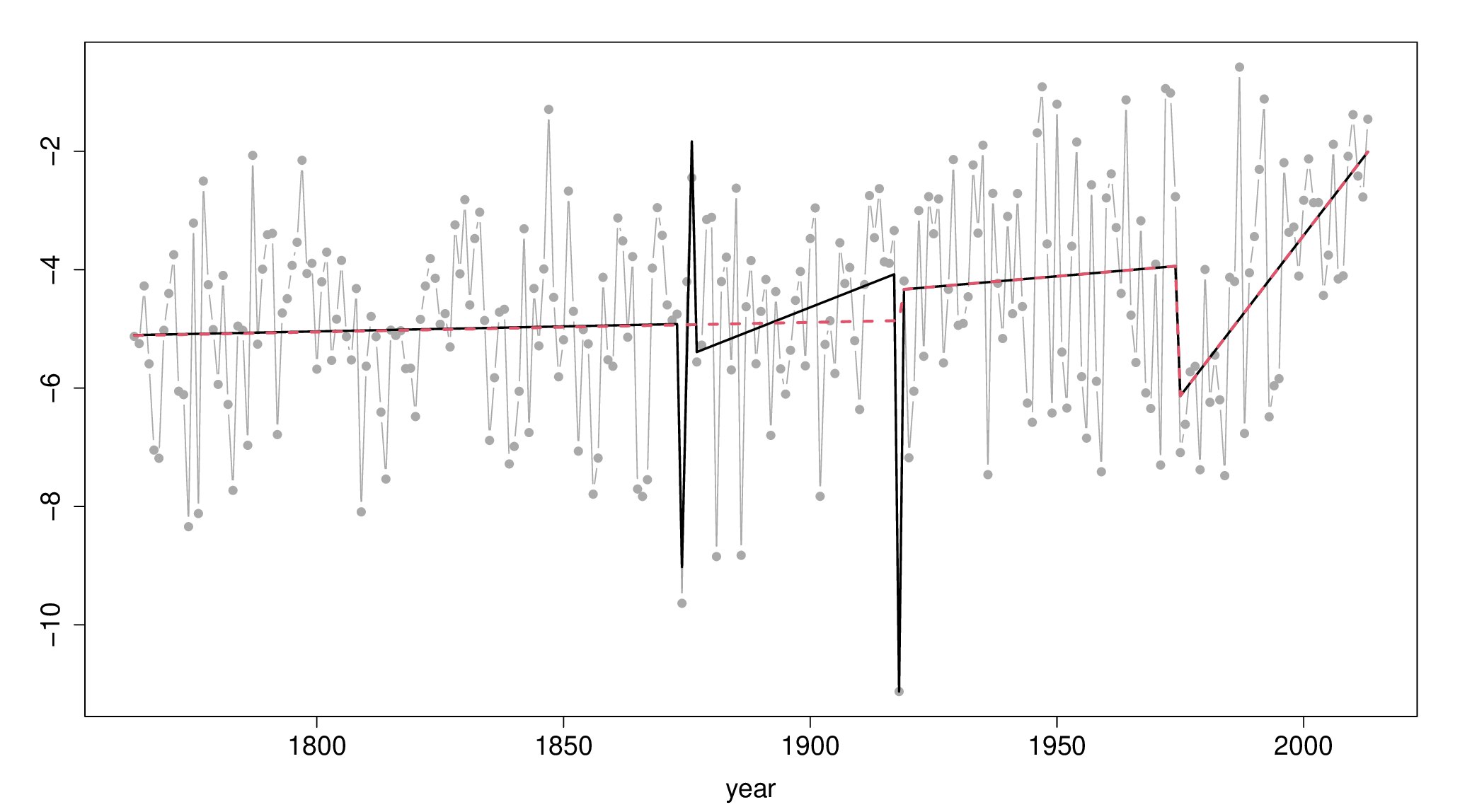}}
        \caption{TrendSegment with $\lambda^\text{Robust}$}
        \label{fig:ice2}
    \end{subfigure}
    \begin{subfigure}[t]{0.49\textwidth} 
        \raisebox{-\height}{\includegraphics[width=\textwidth]{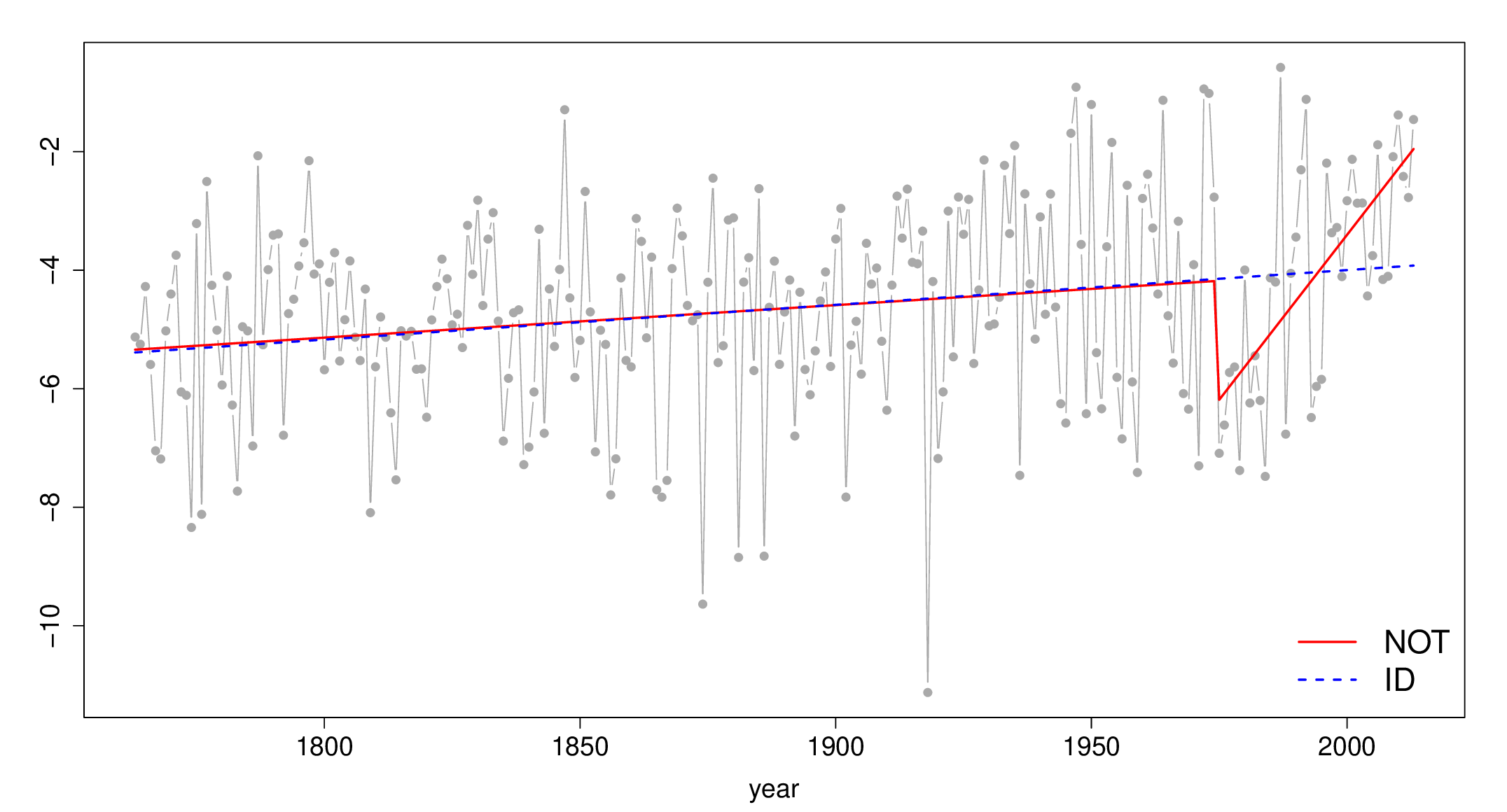}}
        \caption{NOT and ID}
        \label{fig:ice3}
    \end{subfigure}
    \hfill
    \begin{subfigure}[t]{0.49\textwidth} 
        \raisebox{-\height}{\includegraphics[width=\textwidth]{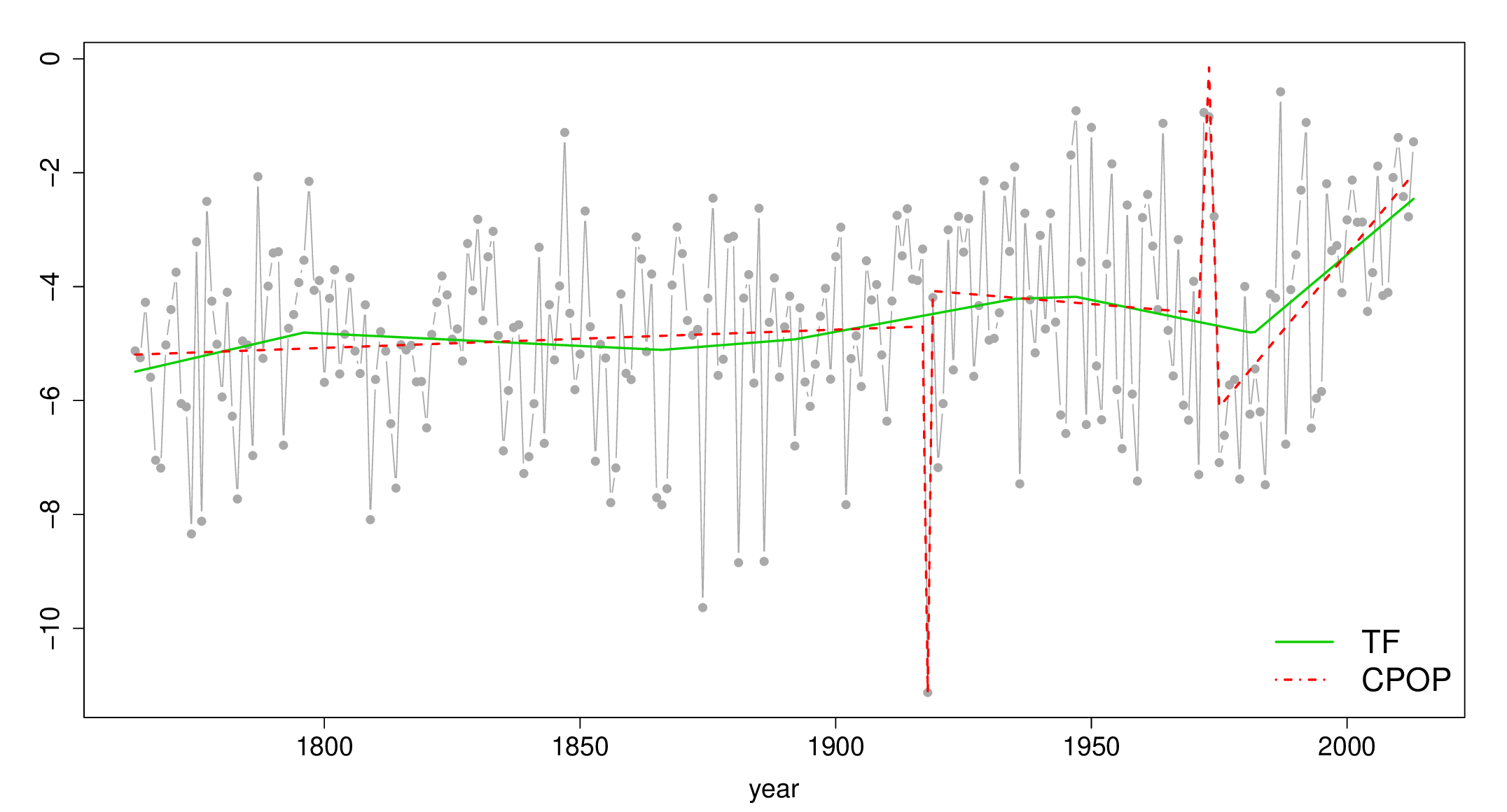}}
    		\caption{TF and CPOP} 
    		\label{fig:ice4}
    \end{subfigure}
    \caption {Change-point analysis for January average temperature in Reykjavik from 1763 to 2013 in Section \ref{iceland}. The data series (grey dots) and estimated signal with change-points returned by (a) TrendSegment using $\lambda^\text{Naïve}$ in \eqref{thrnaive} with minimum segment length equals to 1 (black solid) and equals to $\lfloor 0.9 \log(T) \rfloor$ (red dashed), (b) TrendSegment using $\lambda^\text{Robust}$ in \eqref{newlambda} with minimum segment length equals to 1 (black solid) and equals to $\lfloor 0.9 \log(T) \rfloor$ (red dashed), (c) NOT and ID, (d) TF and CPOP.}
\label{fig:janiceland}
\end{figure}

We analyse a land temperature dataset available from \url{https://www.kaggle.com/berkeleyearth/climate-change-earth-surface-temperature-data}, consisting of average temperatures in January recorded in Reykjavik recorded from 1763 to 2013. Figure \ref{fig:janiceland} shows the data; the point corresponding to 1918 appears to be an anomalous point. This is sometimes called point anomaly which can be viewed as a separate data segment containing only one datapoint. 
Regarding the 1918 observation, \citet{moore2017iceland} report that  ``[t]he winter of 1917/1918 is referred to as the Great Frost Winter in Iceland. 
It was the coldest winter in the region during the twentieth century. 
It was remarkable for the presence of sea ice in Reykjavik Harbour as well as for the unusually large number of polar bear sightings in northern Iceland.'' 

Out of the competing methods tested, ID, TF and CPOP are in principle able to classify two consecutive time point as change-points, and therefore they are able to detect separate data segments containing only one data point each. NOT and BUP are not designed to detect two consecutive time point as change-points as their minimum distance between two consecutive change-points is restricted to be at least two.
In the TrendSegment algorithm, the minimum segment length can flexibly set by the users as described in Section \ref{sec4}. Figures \ref{fig:ice1} and \ref{fig:ice2} show that the change-point estimators depend on the type of threshold we use ($\lambda^\text{Naïve}$ or $\lambda^\text{Robust}$) and also vary over conditions on the minimum segment length. Regardless of the minimum segment length, the robust threshold selection tends to detect more change-points than the naïve threshold. 
When the minimum segment length is set to $1$, with both naïve and robust thresholds, TrendSegment commonly identifies change-points in 1917 and 1918, where the temperature in 1918 is fitted as a single point. 
As shown in Figure \ref{fig:ice4}, out of the competing methods, only CPOP detects the temperature in 1918 as an anomalous point.
Figures \ref{fig:ice2}, \ref{fig:ice3} and \ref{fig:ice4} show that TrendSegment with $\lambda^\text{Robust}$, NOT and CPOP detect the change of slope in 1974, ID returns an increasing function with no change-points and TF reports 6 points with the most recent one in 1981, but none of them detect the point in 1918 as a separate data segment. When setting the minimum segment length equals to the default ($\lfloor 0.9 \log(T) \rfloor$) in TrendSegment with $\lambda^\text{Naïve}$ in Figure \ref{fig:ice1}, it returns no change-points as ID does. This example illustrates the flexibility of the TrendSegment as it detects not only change-points in linear trend but it can identify a separate data segment at the same time, which the competing methods do not achieve.

\subsection{Monthly average sea ice extent of Arctic and Antarctic} \label{seaice}
We analyse the average sea ice extent of the Arctic and the Antarctic available from \url{https://nsidc.org} to estimate the change-points in its trend. 
As mentioned in \citet{serreze2018arctic}, sea ice extent is the most common measure for assessing the feature of high-latitude oceans and it is defined as the area covered with an ice concentration of at least 15$\%$. 
Here we use the average ice extent in February and September as it is known that the Arctic has the maximum ice extent typically in February while the minimum occurs in September and the Antarctic does the opposite. 

\citet{serreze2018arctic} indicate that the clear decreasing trend of sea ice extent of the Arctic in September is one of the most important indicator of climate change. 
In contrast to the Arctic, the sea ice extent of the Antarctic has been known to be stable in the sense that it shows a weak increasing trend in the decades preceding 2016 \citep{comiso2017positive, serreze2018arctic}. 
However, \citet{rintoul2018choosing} warn of a possible collapse of the past stability by citing a significant decline of the sea ice extent in 2016. 
We now use the most up-to-date records (to 2020) and re-examine the concerns expressed in 
\citet{rintoul2018choosing} with the help of our change-point detection methodology.

\begin{figure}[ht!] 
     \centering
    \begin{subfigure}[t]{0.49\textwidth} 
        \raisebox{-\height}{\includegraphics[width=\textwidth]{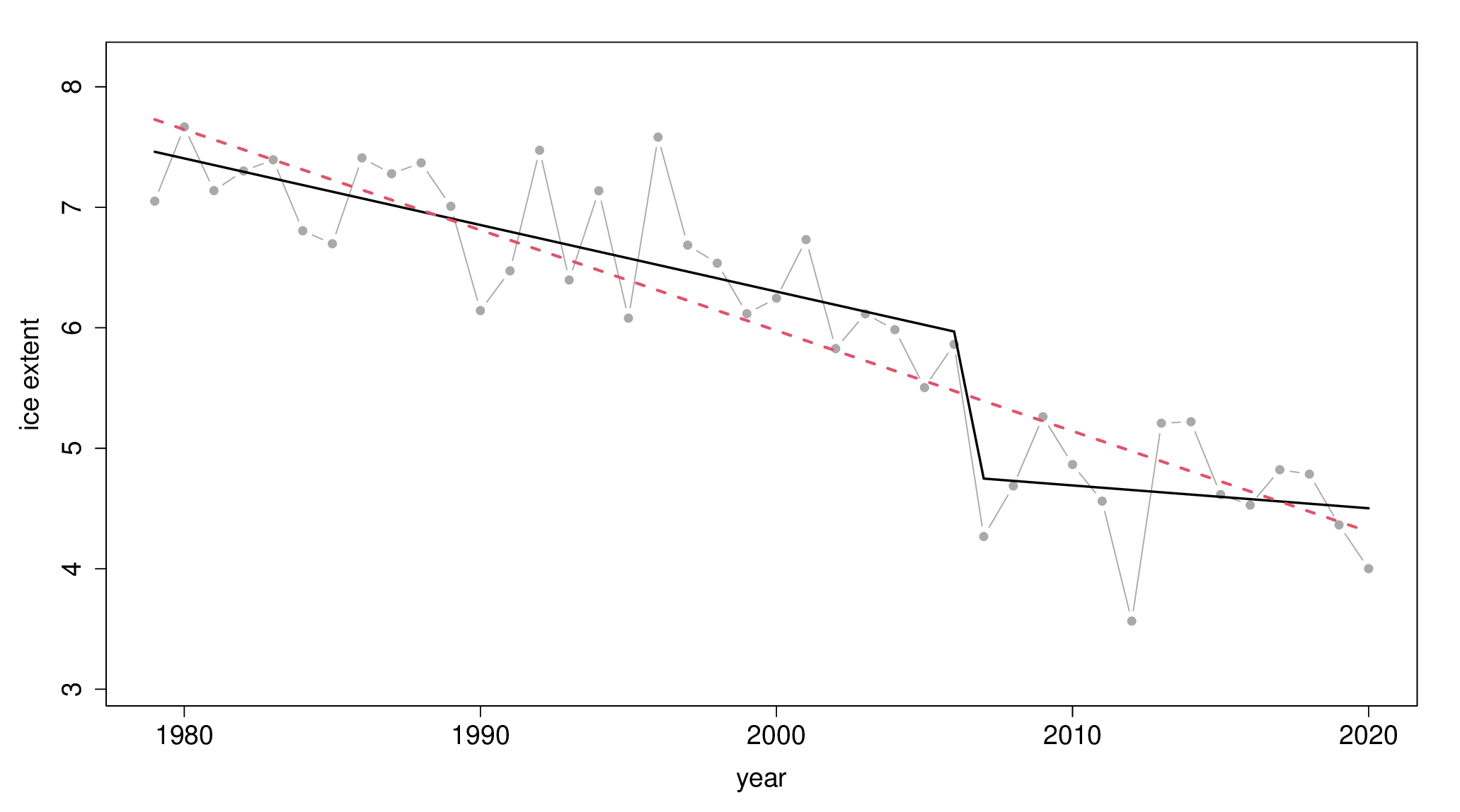}}
        \caption{Arctic in February}
        \label{fig:seaice1}
    \end{subfigure}
    \hfill
    \begin{subfigure}[t]{0.49\textwidth} 
        \raisebox{-\height}{\includegraphics[width=\textwidth]{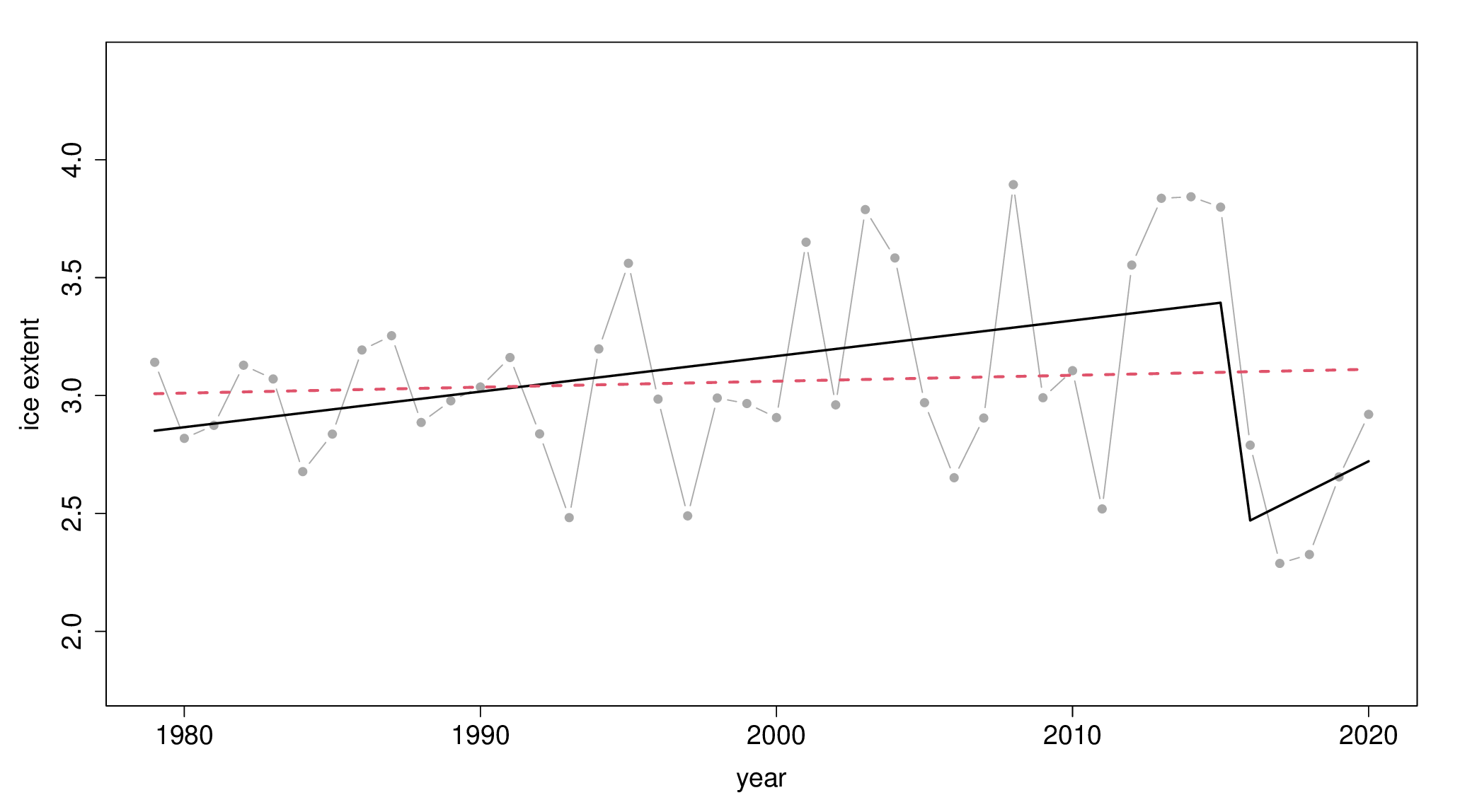}}
        \caption{Antarctic in February}
        \label{fig:seaice2}
    \end{subfigure}
    \begin{subfigure}[t]{0.49\textwidth} 
        \raisebox{-\height}{\includegraphics[width=\textwidth]{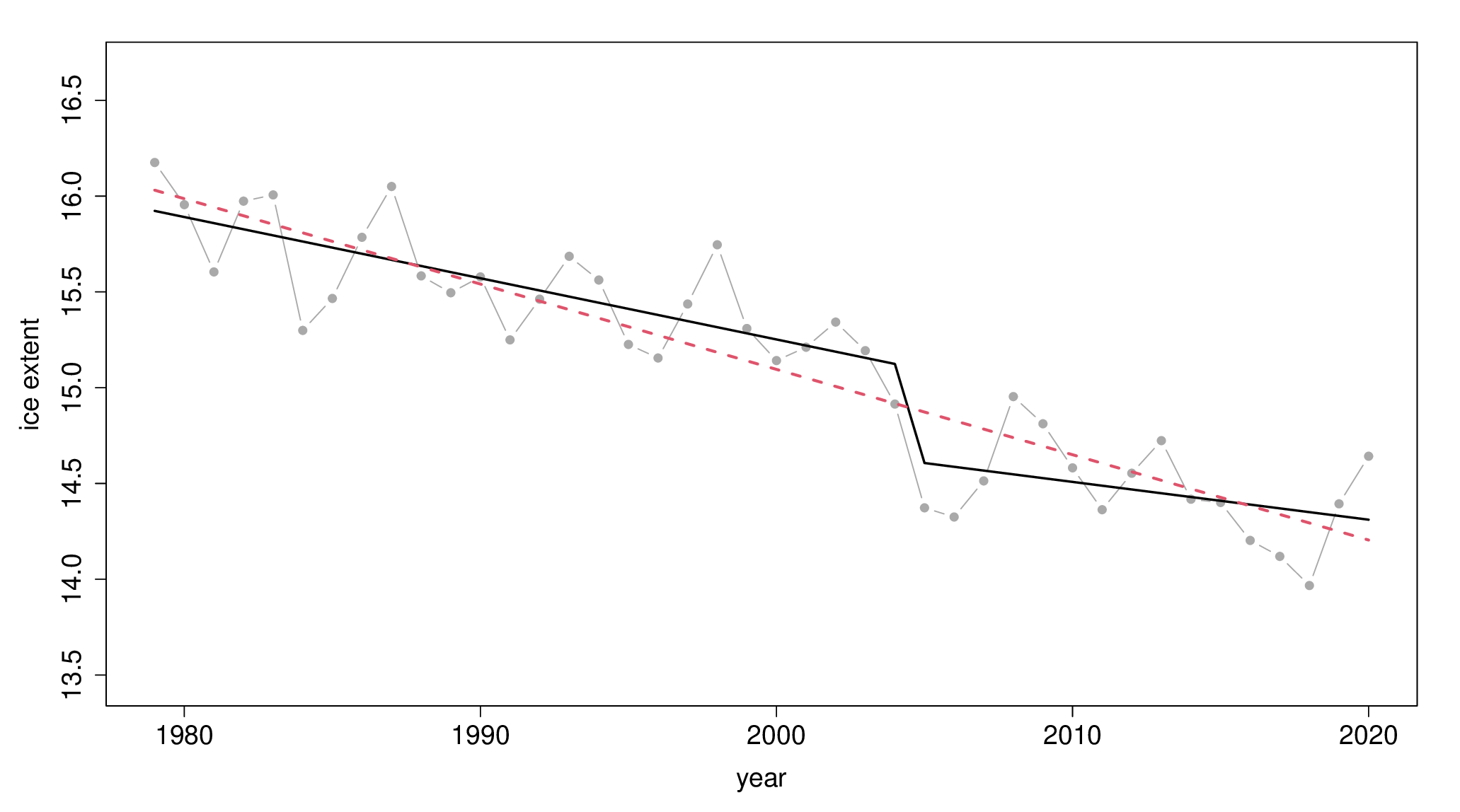}}
        \caption{Arctic in September}
        \label{fig:seaice3}
    \end{subfigure}
    \hfill
    \begin{subfigure}[t]{0.49\textwidth} 
        \raisebox{-\height}{\includegraphics[width=\textwidth]{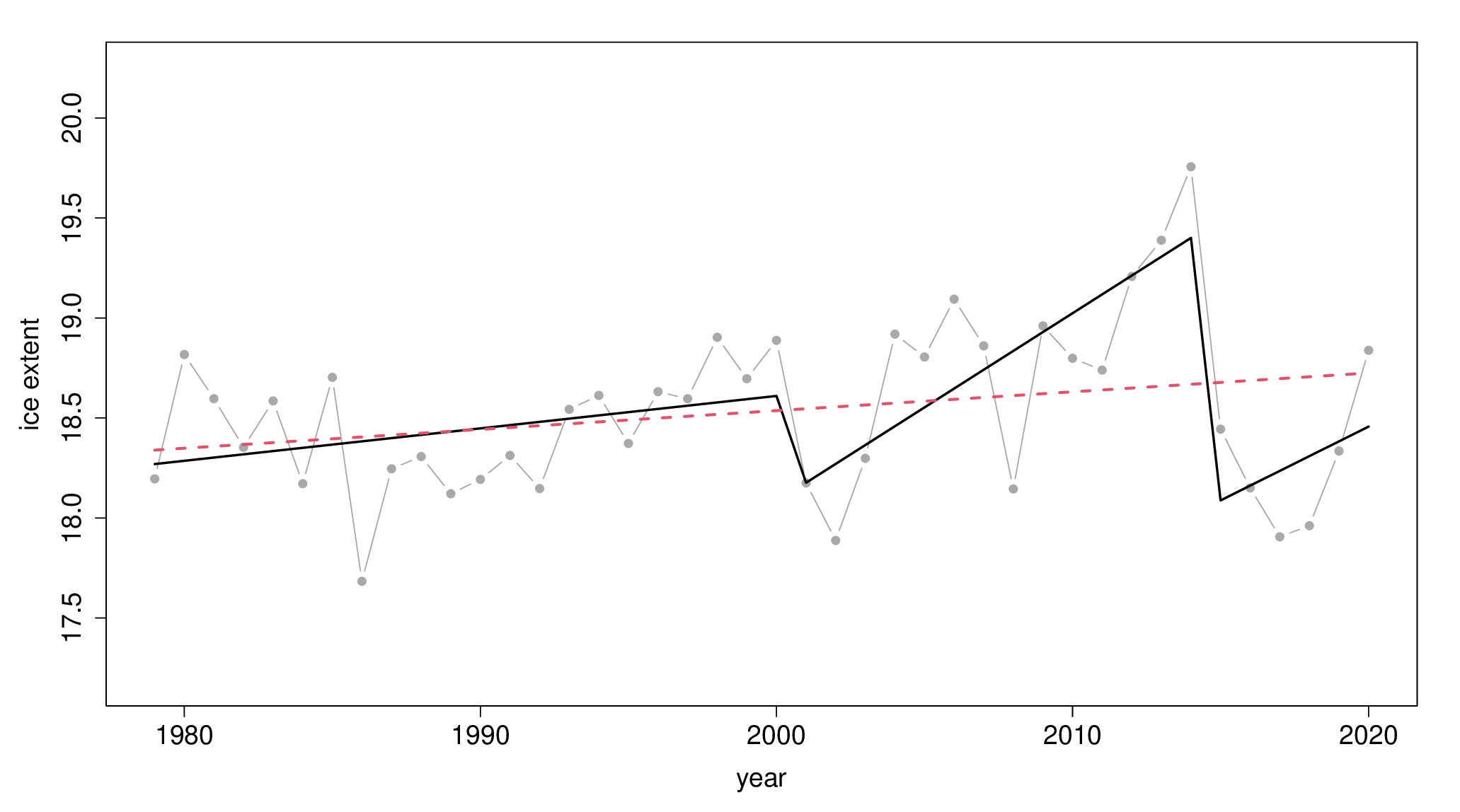}}
    		\caption{Antarctic in September} 
    		\label{fig:seaice4}
    \end{subfigure}
    \caption {The TrendSegment estimate of piecewise-linear trend for the monthly average sea ice extent from 1979 to 2020 in Section \ref{seaice}. (a) the data series (grey dots); the TrendSegment estimate using $\lambda^\text{Naïve}$ in \eqref{thrnaive} (solid black) and TrendSegment estimate using $\lambda^\text{Robust}$ (red dashed) for average sea ice extent of the Arctic in February, (b) Antarctic in February, (c) Arctic in September, (d) Antarctic in September.}
\label{fig:seaice}
\end{figure}

In this example, the condition on the minimum segment length does not affect the change-point estimation results, thus Figure \ref{fig:seaice} shows the results obtained from the default minimum segment length. Also, as shown in Figure \ref{fig:seaice}, TrendSegment estimate with $\lambda^\text{Robust}$ identifies no change-point over all four datasets, thus we focus on giving interpretations for the TrendSegment estimate with $\lambda^\text{Naïve}$ in the following.

Figures \ref{fig:seaice1} and \ref{fig:seaice3} show the well-known decreasing trend of the average sea ice extent in the Arctic both in its winter (February) and summer (September). 
In Figure \ref{fig:seaice1}, the TrendSegment estimate identifies change-points in 2005 and detects a sudden drop during 2003-2005. One change-point in 2007 is identified in Figure \ref{fig:seaice3}, which differentiates the decreasing speed of winter ice extent in the Arctic before and after 2007. 
As observed in the above-mentioned literature, the sea ice extent of the Antarctic shows a modest increasing trend up until recently (Figures \ref{fig:seaice2} and \ref{fig:seaice4}); however, TrendSegment procedure estimates change-point in 2016 and detects a sudden drop during 2015-2017 for the Antarctic summer (February) and similarly detects two sudden drops by the estimated change-points in 2001 and 2015 for the Antarctic winter (September), which is in line with the message of \citet{rintoul2018choosing}. 
The results of other competing methods can be found in Section D.1 of the supplementary materials. 


\section{Extension to non-Gaussian and/or dependent noise} \label{sec6}
Our TrendSegment algorithm can be extended to more realistic settings e.g. when the noise $\ve_t$ is possibly dependent and/or non-Gaussian. The extension is performed by slightly altering the estimators $\tilde{f}, \dbtilde{f}$ and $\hat{f}$ and keeping the rate of threshold the same as the one used in Theorems \ref{thm1}-\ref{thm3} (i.e. $\lambda = O((\log T)^{1/2})$) that is established under the iid Gaussian noise. We add an additional step to ensure that only the detail coefficients $d^{(j, k)}_{p, q, r}$ corresponding to a long enough interval $[p, r]$ survive, as this step enables us to apply strong asymptotic normality of $\sum_{t=p}^r \ve_t$. Under dependent or non-Gaussian noise, Theorems \ref{thm1}-\ref{thm3} presented in Section \ref{sec3} still hold with a larger rate that is different by only a logarithmic factor, where the corresponding theories and proofs can be found in Section B of the supplementary material. 

In Algorithm \ref{algo_lambda} in Section \ref{sec4.1.5}, we propose a robust way of threshold selection that works well in all circumstances including iid Gaussian noise.
To demonstrate the robustness of our threshold selection in case the noise has serial dependence and/or heavy-tailedness, additional simulations are performed for five distributions of the noise; (a) $\varepsilon_t \sim $ i.i.d. scaled $t_5$ distribution with unit-variance, (b) $\varepsilon_t$ follows a stationary AR(1) process with $\phi = 0.3$ and Gaussian innovation, (c) the same setting with (b) but with $\phi = 0.6$, (d) $\varepsilon_t$ follows a stationary AR(1) process with $\phi = 0.3$ and $t_5$ innovation and (e) the same setting with (d) but with $\phi = 0.6$, where the results are summarised in Tables C.1-C.10 in Section C.1 of the supplementary material. Lastly, in Section D.2 of the supplementary material, we demonstrate that our TrendSegment algorithm shows a good performance on London air quality data that possibly has some non-negligible autocorrelation.



\section*{Appendix A  \hspace{10pt} Technical proofs} \label{apx}
The proof of Theorems \ref{thm1}-\ref{thm3} are given below and Lemmas 1 and 2 can be found in Section A of the supplementary materials.

\vspace{10pt}

\noindent {\bf Proof of Theorem \ref{thm1}.}\label{pf1}
\normalfont Let $\mathcal{S}^1_j$ and $\mathcal{S}^0_j$ as in Lemma 2.
From the conditional orthonormality of the unbalanced wavelet transform, on the set $A_T$ defined in Lemma 1, we have
{\footnotesize
\begin{align*} 
\| \tilde{f}-f \|_T^2 \;  = \; & \frac{1}{T} \sum_{j=1}^J \sum_{k=1}^{K(j)} \Big( d^{(j, k)} \cdot \mathbb{I} \big\{ \, \exists (j', k') \in \mathcal{C}_{j, k} \quad \vert d^{(j', k')}\vert > \lambda \, \big\} - \mu^{(j, k)} \Big)^2 + \; T^{-1} (s^{[1]}_{1, T} - \mu^{(0, 1)})^2 + \; T^{-1} (s^{[2]}_{1, T} - \mu^{(0, 2)})^2 \\
\; \leq \; & \frac{1}{T} \sum_{j=1}^J \Bigg( \sum_{k \in \mathcal{S}^0_j} + \sum_{k \in \mathcal{S}^1_j} \Bigg) \Big( d^{(j, k)} \cdot \mathbb{I} \big\{ \, \exists (j', k') \in \mathcal{C}_{j, k} \quad \vert d^{(j', k')} \vert > \lambda \, \big\} - \mu^{(j, k)} \Big)^2 + \; 4C_1^2T^{-1} \log T \\
\; =: & \; \mathit{I} + \mathit{II} + 4C_1^2T^{-1} \log T \numberthis \label{e92},
\end{align*}
}
where $\mu^{(0, 1)} = \langle {f}, \psi^{(0, 1)} \rangle$ and $\mu^{(0, 2)} = \langle {f}, \psi^{(0, 2)} \rangle$. 
We note that $\big(s^{[1]}_{1, T} - \mu^{(0, 1)}\big)^2 \leq 2C_1^2\log T$ is simply obtained by combining Lemma 2 and the fact that $s^{[1]}_{1, T} - \mu^{(0, 1)}= \langle \bve, \psi^{(0, 1)} \rangle$, which can also be applied to obtain $\big(s^{[2]}_{1, T} - \mu^{(0, 2)}\big)^2 \leq 2C_1^2\log T$.
By Lemma 2, $\mathbb{I} \big\{ \, \exists (j', k') \in \mathcal{C}_{j, k} \quad \vert d^{(j', k')} \vert > \lambda \big\} = 0 $ for $k \in \mathcal{S}^0_j$ and also by the fact that $\mu^{(j, k)} = 0$ for $j=1, \ldots, J, k \in \mathcal{S}^0_j$, we have $\mathit{I} = 0$. For $\mathit{II}$, we denote $\mathcal{B} =  \big\{ \, \exists (j', k') \in \mathcal{C}_{j, k} \quad \vert d^{(j', k')} \vert > \lambda \, \big\}$ and have
{\footnotesize
\begin{align*} 
\big( d^{(j, k)} \cdot \mathbb{I} \big\{ \mathcal{B} \big\} - \mu^{(j, k)} \big)^2 \numberthis  \; = \; & \big( d^{(j, k)} \cdot \mathbb{I} \big\{ \mathcal{B} \big\} - d^{(j, k)} + d^{(j, k)} - \mu^{(j, k)} \big)^2 \\
\;  \leq \; & \big( d^{(j, k)} \big)^2 \mathbb{I}\big( \vert d^{(j', k')} \vert \leq \lambda \big) + 2\vert d^{(j, k)}\vert \; \mathbb{I}\big( \vert d^{(j', k')}\vert \leq \lambda \big) \; \vert d^{(j, k)} - \mu^{(j, k)}\vert +  \; \big( d^{(j, k)} - \mu^{(j, k)}  \big)^2 \\  
\; \leq \;  & \lambda^2 + 2\lambda C_1 \{2 \log T \}^{1/2} + 2 C_1^2 \log T.
\end{align*}}
Combining with the upper bound of $J$, $ \lceil \log(T) / \log((1-\rho)^{-1}) + \log(2) / \log(1-\rho) \rceil$, and the fact that $\vert \mathcal{S}^1_j \vert  \leq N$, we have 
$\mathit{II} \leq 8 C_1^2 N T^{-1} \lceil \log(T) / \log((1-\rho)^{-1}) + \log(2) / \log(1-\rho) \rceil \log T$, and therefore $\| \tilde{f}-f \|_T^2 \; \leq \; C_1^2 \; T^{-1} \; \log(T) \; \Big\{ 4 + 8 N \; \lceil \log(T) / \log((1-\rho)^{-1}) + \log(2) / \log(1-\rho) \rceil \; \Big\}$.
As the estimated change-points are obtained through those detail coefficients, thus at each scale, up to $N$ estimated change-points are added. Combining it with the largest scale $J$ whose order is $\log T$, the number of change-points in $\tilde{f}$ returned from the inverse TGUW transformation is up to $CN\log T$ where $C$ is a constant. 

\vspace{10pt}

\noindent {\bf Proof of Theorem \ref{thm2}.}\label{pf2}
\normalfont Let $\tilde{B}$ and $\dbtilde{B}$ the unbalanced wavelet basis corresponding to $\tilde{f}$ and $\dbtilde{f}$, respectively. As the change-points in $\dbtilde{f}$ are a subset of those in $\tilde{f}$, establishing $\dbtilde{f}$ can be considered as applying the TGUW transform again to $\tilde{f}$ which is just a repetition of procedure done in estimating $\tilde{f}$ in the greediest way. Thus $\dbtilde{B}$ is classified into two categories, 1) all basis vectors $\psi^{(j, k)} \in \tilde{B}$ such that $\psi^{(j, k)}$ is not associated with the change-points in $\tilde{f}$ and  $\vert \langle \b{X}, \psi^{(j, k)} \rangle\vert  = \vert d^{(j, k)}\vert  < \lambda$ and 2) all vectors $\psi^{(j, 1)}$ produced in Stage 1 of post-processing. 

We now investigate how many scales are used for this particular transform. First, the detail coefficients $d^{(j, k)}$ corresponding to the basis vectors $\psi^{(j, k)} \in \tilde{B}$ live on no more than $J=O(\log T)$ scales and we have $\vert \mathcal{S}^1_j\vert  \leq N$ by the argument used in the proof of Theorem \ref{thm1}. In addition, the vectors $\psi^{(j, 1)}$ in the second category correspond to different change-points in $\tilde{f}$ and there exist at most $\tilde{N}=O(N\log T)$ change-points in $\tilde{f}$  which we examine one at once (i.e. $\vert \mathcal{S}^1_j\vert  \leq 1$), thus at most $\tilde{N}$ scales are required for $d^{(j, 1)}$. 
Combining the results of two categories, the equivalent of quantity $\mathit{II}$ in the proof of Theorem \ref{thm1} for $\dbtilde{f}$ is bounded by $\mathit{II} \leq C_3 N T^{-1} \log^2 T$ and this completes the proof of the $l_2$ result, $\big\| \dbtilde{f} - {f} \big\|_T^2 \; =  \; O\big(NT^{-1} \log^2(T)\big)$ where $C_3$ is a positive constant large enough. 

Finally, we show that there exist at most two change-points in $\dbtilde{f}$ between true change-points $(\eta_\ell, \eta_{\ell+1})$ for $\ell=0, \ldots, N$ where $\eta_0=0$ and $\eta_{N+1}=T$. Consider the case where three change-point for instance ($\dbtilde{\eta}_l, \dbtilde{\eta}_{l+1}, \dbtilde{\eta}_{l+2}$) lie between a pair of true change-point, $(\eta_\ell, \eta_{\ell+1})$. In this case, by Lemma 2, the maximum magnitude of two detail coefficients computed from the adjacent intervals, $[\dbtilde{\eta}_l+1, \dbtilde{\eta}_{l+1}]$ and $[\dbtilde{\eta}_{l+1}+1, \dbtilde{\eta}_{l+2}]$, is less than $\lambda$ and $ \dbtilde{\eta}_{l+1}$ would be get removed from the set of estimated change-points. This satisfies $\dbtilde{N} \leq 2(N+1)$. 


\vspace{10pt}

\noindent {\bf Proof of Theorem \ref{thm3}.}\label{pf3}
\normalfont From the assumptions of Theorem 3, the followings hold. 
\begin{itemize}
\item Given any $\epsilon>0$ and $C>0$, for some $T_1$ and all $T>T_1$, it holds that \\ $\mathbb{P} \Big( \big\| \dbtilde{f} - {f} \big\|_T^2 > \frac{C^3}{4} R_T \Big) \leq \epsilon$ where $\dbtilde{f}$ is the estimated signal specified in Theorem \ref{thm2}. 
\item For some $T_2$, and all $T>T_2$, it holds that $C^{1/3}T^{1/3} R_T^{1/3} (\ubar{f}_T^\ell)^{-2/3} < \delta_T^\ell$ for all $\ell=1, \ldots, N$. 
\end{itemize}
Following the argument used in the proof of Theorem 19 in \citet{lin2016approximate}, we take $T \geq T^*$ where $T^* = \max\{T_1, T_2\}$ and let $r_{\ell, T}=\lfloor C^{1/3} T^{1/3} R_T^{1/3} (\ubar{f}_T^\ell)^{-2/3} \rfloor$ for $\ell=1, \ldots, N$.
Suppose that there exist at least one $\eta_\ell$ whose closest estimated change-point is not within the distance of $r_{\ell, T}$. 
Then there are no estimated change-points in $\dbtilde{f}$ within $r_{\ell, T}$ of $\eta_\ell$ which means that $\dbtilde{f}_j$ displays a linear trend over the entire segment $j \in \{\eta_\ell-r_{\ell, T}, \ldots, \eta_\ell+r_{\ell, T}\}$. Hence 
\begin{equation} 
\frac{1}{T} \sum_{j=\eta_\ell-r_{\ell, T}}^{\eta_\ell+r_{\ell, T}} \big(\dbtilde{f}_j - f_j\big)^2 \geq \frac{13 r_{\ell, T}^3}{24T} \big(\ubar{f}_T^\ell\big)^2 > \frac{C^3}{4} R_T.
\end{equation}
The first inequality holds by Lemma 20 of \citet{lin2016approximate}, and the second one holds by the definition of $r_{\ell, T}$. 
Assuming that at least one $\eta_\ell$ does not have an estimated change-point within the distance of $r_{\ell, T}$ implies that the estimation error exceeds $\frac{C^3}{4} R_T$ which is a contradiction as it is an event that we know occurs with probability at most $\epsilon$. Therefore, there must exist at least one estimated change-point within the distance of $r_{\ell, T}$ from each true change point $\eta_\ell$.

Throughout Stage 2 of post-processing, $\dbtilde{\eta}_{\ell_0}$ is either the closest estimated change-point of any $\eta_\ell$ or not. If $\dbtilde{\eta}_{\ell_0}$ is not the closest estimated change-point to the nearest true change-point on either its left or its right, by the construction of detail coefficients in Stage 2 of post-processing, Lemma 2 guarantees that the corresponding detail coefficient has the magnitude less than $\lambda$ and $\dbtilde{\eta}_{\ell_0}$ gets removed. Suppose $\dbtilde{\eta}_{\ell_0}$ is the closest estimated change-point of a true change-point $\eta_\ell$ and it is within the distance of $C T^{1/3} R_T^{1/3} \big(\ubar{f}_T^\ell \big)^{-2/3}$ from $\eta_\ell$. If the corresponding detail coefficient has the magnitude less than $\lambda$ and $\dbtilde{\eta}_{\ell_0}$ is removed, there must exist another $\dbtilde{\eta}_\ell$ within the distance of $C T^{1/3} R_T^{1/3} \big(\ubar{f}_T^\ell \big)^{-2/3}$ from $\eta_\ell$. 
If there are no such $\dbtilde{\eta}_\ell$, then by the construction of the detail coefficient, the order of magnitude of $\big\vert d_{p_{\ell_0}, q_{\ell_0}, r_{\ell_0}}\big\vert $ would be such that $\big\vert d_{p_{\ell_0}, q_{\ell_0}, r_{\ell_0}}\big\vert  > \lambda$ thus $\dbtilde{\eta}_{\ell_0}$ would not get removed. 
Therefore, after Stage 2 of post-processing is finished, each true change-point $\eta_\ell$ has its unique estimator within the distance of $C T^{1/3} R_T^{1/3} \big(\ubar{f}_T^\ell \big)^{-2/3}$.

\pagebreak
\bigskip
\begin{center}
{\Large\bf Supplementary materials for ``Detecting linear trend changes in data sequences''} \vspace{.5cm}\\
{\large{Hyeyoung Maeng and Piotr Fryzlewicz}} \vspace{.3cm} \\
\end{center}

\newcommand{\Appendix}{\appendix
\def\thesection{\Alph{section}.}
\def\thesubsection{\Alph{section}.\arabic{subsection}}
\def\thetable{\Alph{section}.\arabic{table}}
\def\thefigure{\Alph{section}.\arabic{figure}}
}

\begin{appendix} 
\Appendix
\renewcommand{\theequation}{S.\arabic{equation}}
\setcounter{equation}{0}
\setcounter{table}{0}
\setcounter{figure}{0}
\baselineskip=18pt

\bigskip


\noindent This document includes the following sections:\\ \space 
{\bf A.} \; Proofs \\
{\bf B.} \; Extension to dependent non-Gaussian noise \\
{\bf C.} \; Additional simulation results  \\
{\bf D.} \; Additional data application results \\
{\bf E.} \; Shape of the unbalanced wavelet basis \\
{\bf F.} \; A practical way to implement the TGUW transformation \\
{\bf G.} \; Extension to piecewise-quadratic signal


\setcounter{equation}{0}
\baselineskip=18pt

\section{Proofs} \label{pf}
\subsection{Some useful lemmas for Theorems 1-3 of the main article}

\begin{Lem} \label{lm1}
\normalfont Let the distribution of $\varepsilon_t$ in model (1) of the main article be iid standard Gaussian. Let $\psi^{(j, k)} = \sum_{i=1}^{I^{(j, k)}} \phi_i^{(j, k)} g_i^{(j, k)}$ where $\phi_i^{(j, k)}$ are constants and $g_i^{(j, k)}$ are vectors of equal length with $\psi^{(j, k)}$ where $I^{(j, k)} \in \{3, 4\}, j=1, \ldots, J, \; k=1, \ldots, K(j)$.
If we define the set $G = \{g_l\}$ where there is a unique correspondence between $\big\{{g_i^{(j, k)}}_{i=1, \ldots, I^{(j, k)}, j=1, \ldots, J, \, k=1, \ldots, K(j)}\big\}$ and $\{g_l\}$, 
we then have $P(A_T) \geq 1-C_2T^{-1}$ where 
\begin{equation} \label{at}
A_T=\bigg\{\max_{g_l\in G} \; |g_l^\top \bve| \leq \lambda \bigg\}, 
\end{equation}
$\lambda$ is as in Theorem 1 and $C_2$ is a positive constant.
\end{Lem}

\textbf{Proof.} 
We firstly show that for any fixed ${(j, k)}$, $g_i^{(j, k)}$ and $\phi_i^{(j, k)}$ satisfy the conditions, $\big(g_i^{(j, k)}\big)^\top g_i^{(j, k)}=1$, $\big(g_i^{(j, k)}\big)^\top g_{i^\prime}^{(j, k)}=0$ and $\sum_i \big(\phi_i^{(j, k)}\big)^2=1$, where $\psi^{(j, k)} = \sum_{i=1}^{I^{(j, k)}} \phi_i^{(j, k)} g_i^{(j, k)}$.
Depending on the type of merge, $\psi^{(j, k)}$ fall into one of the followings,
\begingroup
\allowdisplaybreaks
\begin{align*} 
\text{Type 1: }\psi^{(j, k)}_{p, q, r} & = \alpha_1e_p + \alpha_2e_{p+1} + \alpha_3e_{p+2}, \label{e91} \numberthis \\
\text{Type 2: }\psi^{(j, k)}_{p, q, r} & = \beta_1e_p + \beta_2 (\underbrace{0, \ldots, 0}_{p\times 1}, \b\ell_{1, p+1, r}^\top, \underbrace{0, \ldots, 0}_{(T-r)\times 1}) + \beta_3 (\underbrace{0, \ldots, 0}_{p\times 1},  \b\ell_{2, p+1, r}^\top, \underbrace{0, \ldots, 0}_{(T-r)\times 1}), \label{et2} \numberthis \\ 
\psi^{(j, k)}_{p, q, r} & = \beta_4 (\underbrace{0, \ldots, 0}_{(p-1)\times 1}, \b\ell_{1, p, r-1}^\top, \underbrace{0, \ldots, 0}_{(T-r+1)\times 1}) + \beta_5 (\underbrace{0, \ldots, 0}_{(p-1)\times 1},  \b\ell_{2, p, r-1}^\top, \underbrace{0, \ldots, 0}_{(T-r+1)\times 1}) + \beta_6e_r,\\ 
\text{Type 3: } \psi^{(j, k)}_{p, q, r} & = \gamma_1 (\underbrace{0, \ldots, 0}_{(p-1)\times 1}, \b\ell_{1, p, q}^\top, \underbrace{0, \ldots, 0}_{(T-q)\times 1}) + \gamma_2 (\underbrace{0, \ldots, 0}_{(p-1)\times 1},  \b\ell_{2, p, q}^\top, \underbrace{0, \ldots, 0}_{(T-q)\times 1}) \label{et3} \numberthis \\
& + \gamma_3 (\underbrace{0, \ldots, 0}_{q\times 1}, \b\ell_{1, q+1, r}^\top, \underbrace{0, \ldots, 0}_{(T-r)\times 1}) + \gamma_4 (\underbrace{0, \ldots, 0}_{q\times 1},  \b\ell_{2, q+1, r}^\top, \underbrace{0, \ldots, 0}_{(T-r)\times 1}), 
\end{align*}
\endgroup
where $e_i$ is a vector of length $T$ having $1$ only at $i^{th}$ element and zero for the others. 
As will be shown in Section \ref{geo}, $\b\ell_{1, i, j}$ and $\b\ell_{2, i, j}$ are an arbitrary orthonormal basis of the subspace $\{(x_1, x_2, \ldots, x_{j-i+1})\;| \; x_1-x_2=x_2-x_3= \cdots =x_{j-i}-x_{j-i+1} \}$ of $\mathbb{R}^{j-i+1}$. 

In any case, we can obtain the representation $\psi^{(j, k)} = \sum_{i=1}^{I^{(j, k)}} \phi_i^{(j, k)} g_i^{(j, k)}$ from \eqref{e91} if the constants $\phi_i^{(j, k)}$ correspond to $\{\alpha_i\}_{i=1}^3$ in Type 1, $\{\beta_i\}_{i=1}^3$ or $\{\beta_i\}_{i=4}^6$ in Type 2 and $\{\gamma_i\}_{i=1}^4$ in Type 3 and $g_i^{(j, k)}$ is the corresponding vector. 
From the orthonormality of the basis ($\b\ell_{1, m, n}, \b\ell_{2, m, n}$) for any $(m, n)$, we see that the conditions, $\big(g_i^{(j, k)}\big)^\top g_i^{(j, k)}=1$ and $\big(g_i^{(j, k)}\big)^\top g_{i^\prime}^{(j, k)}=0$, are satisfied for any $(i, i^\prime, j, k)$ where $i \neq i^\prime$. 
In addition, as $\psi^{(j, k)}$ keep orthonormality, we can argue that $\phi_i^{(j, k)}$ is bounded by the condition $\sum_i \big(\phi_i^{(j, k)}\big)^2=1$ for any ${(i, j, k)}$ which implies $\sum_{i=1}^3 \alpha_i^2=\sum_{i=1}^3 \beta_i^2=\sum_{i=4}^6 \beta_i^2=\sum_{i=1}^4 \gamma_i^2=1$ in \eqref{e91}.

If we predefine the pairs ($\b\ell_{1, m, n}, \b\ell_{2, m, n}$) for any $(m, n)$ by choosing an orthonormal basis of the subspace $\{(x_1, x_2, \ldots, x_{n-m+1})\;| \; x_1-x_2=x_2-x_3=\cdots=x_{n-m}-x_{n-m+1} \}$ of $\mathbb{R}^{n-m+1}$, then there exist at most $T^2$ vectors $g_l$ in the set $G$. This is because $m$ and $n$ can be randomly chosen from $\{1, 2, \ldots, T\}$ with replacement and if $m \neq n$, the two drawn pairs, $(m, n)$ and $(n, m)$, correspond to the same basis vectors, ($\b\ell_{1, m, n}, \b\ell_{2, m, n}$), while $(m, m)$ correspond to one vector $e_m$. 
Now we are in position to show that $P(A_T) \geq 1-C_2T^{-1}$. Using a simple Bonferroni inequality, we have
\begin{equation} 
1-P(A_T) \leq \sum_G P(|Z|>\lambda) \leq 2T^2 \frac{\phi_Z(\lambda)}{\lambda} = \frac{1}{C_1 \sqrt{\pi} T^{C_1^2-2} \sqrt{\log T}} \leq \frac{C_2}{T}
\end{equation}
where $\phi_Z$ is the p.d.f. of a standard normal $Z$. This completes the proof.

\vspace{0.3cm}


\begin{Lem} \label{lm2}
\normalfont Let $\mathcal{S}^1_j = \{1 \leq k \leq K(j) : d^{(j, k)}$ is $d_{p, q, r}$ such that $p < \eta_i + 1/2 < r$ for some $i=1, \ldots, N$ $\}$, and $\mathcal{S}^0_j = \{1, \ldots, K(j)\} \setminus \mathcal{S}^1_j$. On the set $A_T$ in \eqref{at} which satisfies $P(A_T) \rightarrow 1$ as $T \rightarrow \infty$, we have 
\begin{equation} \label{bt}
\max_{\substack{j=1, \ldots, J, \\ k \in \mathcal{S}^0_j}} \big|d^{(j, k)}\big| \leq \lambda, 
\end{equation}
where $\lambda$ is as in Theorem 1. 
\end{Lem}

\textbf{Proof.} 
On the set $A_T$, the following holds for $j=1, \ldots, J, k \in \mathcal{S}^0_j$,
\begin{align*}
\big|d^{(j, k)}\big| = \; & \big|(\psi^{(j, k)})^\top  \boldsymbol{\ve} \big| \\
= \; &  \Big|\phi_1^{(j, k)} \big(g_1^{(j, k)}\big)^\top  \bve + \phi_2^{(j, k)} \big(g_2^{(j, k)}\big)^\top  \bve + \phi_3^{(j, k)} \big(g_3^{(j, k)}\big)^\top  \bve + \phi_4^{(j, k)} \big(g_4^{(j, k)}\big)^\top  \bve \Big| \\
\leq \; & \max_{j, \; k}\Big(\big|\phi_1^{(j, k)}\big|+\big|\phi_2^{(j, k)}\big|+\big|\phi_3^{(j, k)}\big|+\big|\phi_4^{(j, k)}\big|\Big)\cdot \Big(\max_{l:\; g_l\in G} \; \big|g_l^\top  \bve \big|\Big), 
\end{align*}
where $\boldsymbol{\ve} = (\ve_1, \ldots, \ve_T)^\top$ and $\psi^{(j, k)}_{p, q, r}$ are as in \eqref{e91}.
The condition, $\sum_i \big(\phi_i^{(j, k)}\big)^2=1$ for any fixed ${(j, k)}$, given in the proof of Lemma \ref{lm1} implies that $\max_i \big|\phi_{i}^{(j, k)} \big| \leq 1$ for any ${(j, k)}$, thus we have \eqref{bt} when the constant $C_1$ for $\lambda$ in \eqref{bt} is larger than or equal to $4$ times $C_1$ used in \eqref{at}.


\section{Extension to dependent non-Gaussian noise} \label{dependent}
In this section, we extend the TGUW methodology to more realistic settings when the noise $\varepsilon_t$ is possibly dependent and/or non-Gaussian. We borrow the idea proposed in the supplementary material of \citet{fryzlewicz2017tail} in the sense that the extension is performed in a way of altering the estimators $\tilde{f}, \dbtilde{f}$ and $\hat{f}$ and keeping the rate of threshold, $O((\log T)^{1/2})$, used in Theorems 1-3 of the main article established under the iid Gaussian noise. 
However, our technique is distinguished from \citet{fryzlewicz2017tail} in that we put an additional step which ensures that only the detail coefficients $d^{(j, k)}_{p, q, r}$ corresponding to a long enough interval $[p, r]$ are survived, while \citet{fryzlewicz2017tail} gives a condition that both the left ($[p, q]$) and the right ($[q+1, r]$) segments should be long enough. 
This enables us to use the same size of threshold, $O((\log T)^{1/2})$, used in the iid Gaussian model without any further procedure such as basis rearrangement proposed in \citet{fryzlewicz2017tail}.

We now define the sets of short-segment and long-segment coefficients at each scale $j$ as follows:
\begin{align*}
    &\mathcal{W}_j^S(a) = \{1 \leq k \leq K(j) : d^{(j, k)}_{p, q, r} \text{ is such that} \; r-p \leq a \},\\ \numberthis \label{wl}
    &\mathcal{W}_j^L(a) = \{1 \leq k \leq K(j)\} \;\CS \;\mathcal{W}_j^S(a),
\end{align*}
where $a$ will be specified later this section. Those detail coefficients obtained from short segments are set to zero in the construction of the new estimators $\tilde{f}^L, \dbtilde{f}^L$ and $\hat{f}^L$, where $L$ in $f^L$ stands for ``Long-segment''. The initial estimator $\tilde{f}^L$ is obtained from the estimator of $\mu^{(j, k)}$ for $j \geq 1$ by applying the ``connected'' rule that is modified from the original one in Section 2.3 of the main article to satisfy the condition that the minimum segment length is longer than $a$:
\begin{equation} \label{muhat}
\hat{\mu}^{(j, k)} = d^{(j, k)}_{p,q,r} \cdot \mathbb{I} \; \Big\{ \, \exists (j', k') \in \mathcal{C}_{j, k} \quad \big|d^{(j', k')}_{p',q',r'}\big| > \lambda \quad \text{and} \quad  k' \in \mathcal{W}_{j'}^L(a) \Big\},
\end{equation}
where $\mathbb{I}$ is an indicator function and
\begin{equation*}
\mathcal{C}_{j, k} = \{(j', k'), j' = 1, \ldots, j, k' = 1, \ldots, K(j'): d^{(j', k')}_{p',q',r'} \text{ is such that } [p', r'] \subseteq [p, r] \}.
\end{equation*}
We then apply the ``two together'' rule to \eqref{muhat} in which both of the paired detail coefficients (formed by Type 3 mergings) should be survived if at least one is survived as done in thresholding of the main article. 
Compared to the estimator $\hat{\mu}^{(j, k)}$ obtained under the iid Gaussian setting, the only added step is setting all short-segment coefficient $d^{(j, k)}_{p,q,r}$ to zero.

\subsection{Preparatory lemmas} \label{lemma_nonG_error}


\begin{Lem} \label{lm3}
\normalfont Let the distribution of $\varepsilon_t$ in model (1) of the main article as in Theorem \ref{thm1}. Then for a constant $C_3 >0$ and $\lambda$ as in Theorem \ref{thm1}, we have $P(A_T) \geq 1-C_3T^{-1}$ for a constant $C_3 >0$, where
\begin{equation} \label{atl}
    A_T^L = \Bigg\{ \forall 1 \leq t_1 \leq t_2 \leq T, \; \forall k \in \{1, 2\} \quad s.t. \quad t_2-t_1 \geq C_1 \log T \quad \Bigg|\sum_{t=t_1}^{t_2} \b\ell_{k, t_1, t_2}^t \varepsilon_t \Bigg| \leq \lambda \Bigg \},
\end{equation}
and $\b\ell_{k, t_1, t_2}^t$ is the $t$-th element of the vector $\b\ell_{k, t_1, t_2}$ of length $t_2-t_1+1$ and the pairs ($\b\ell_{1, t_1, t_2}, \b\ell_{2, t_1, t_2}$) are predetermined for any $(t_1, t_2)$ by choosing an orthonormal basis of the subspace $\{(x_1, x_2, \ldots, x_{t_2-t_1+1})\;| \; x_1-x_2=x_2-x_3=\cdots=x_{t_2-t_1}-x_{t_2-t_1+1} \}$ of $\mathbb{R}^{t_2-t_1+1}$.
\end{Lem}

\textbf{Proof.} 
In the following, we consider the single sum $\sum_{t=1}^a w_t \varepsilon_t$ from the interval $[1, a]$ where $w_t = \b\ell_{k, 1, a}^t$ for a fixed $k \in \{1, 2\}$.  
The results can in principle be applied to an interval with different ends given that the length of the interval is at least $a$. Since $\varepsilon_t$ is $m$-dependent, we have $\alpha(l)=0$ for $l > m$ where $\alpha(\cdot)$ is the $\alpha$-mixing coefficients of $\varepsilon_t$.

From Theorem 1.4 in \citet{bosq1998nonparametric}, if $m_2^2 < \infty$, for each $\epsilon > 0$ and for a constant $c > 0$, we obtain
\begin{equation}\label{bosq}
    P\Bigg(\sqrt{a}\Bigg|\sum_{t=1}^a w_t \varepsilon_t\Bigg| > a\epsilon \Bigg) \leq a_1 \exp{\Bigg(-\frac{q\epsilon^2}{25m_2^2 + 5c\epsilon}\Bigg)} + a_2(k)\alpha\Bigg(\Bigg[\frac{a}{q+1}\Bigg]\Bigg)^{\frac{2k}{2k+1}},    
\end{equation}
where 
\begin{align*}
    & a_1 = 2\frac{a}{q} + 2\Bigg(1+\frac{\varepsilon^2}{25m_2^2 + 5c\epsilon }\Bigg), \quad \text{with} \quad m_2^2 = \max_{1 \leq t \leq a} E\Big[\Big(\sqrt{a} w_t \varepsilon_t\Big)^2\Big], \\
    & a_2(k) = 11n\Bigg(1+\Bigg(\frac{5m_k}{\epsilon}\Bigg)^{\frac{2k}{2k+1}}\Bigg), \quad \text{with} \quad m_k = \max_{1 \leq t \leq a} \big\|\sqrt{a} w_t \varepsilon_t \big\|_k.
\end{align*}
The assumption $m_2^2 < \infty$ is reasonably achievable as we can show $m_2^2  = a \max_t(w^2_t)$ is bounded by a constant from two conditions given on $w_t$, 1)$\{w_1-w_2= \cdots =w_{a-1}-w_a\}$ and 2) $\sum_{t=1}^a w_t^2 = 1$.

By setting $\epsilon=\lambda/\sqrt{a}$, $a=C \log(T)$ and $\lambda=C_1 \log^{1/2} T$ for  large enough $C>0$ and $C_1>0$, and setting $q=[c_1a]$ with a small $c_1$ (which gives $\Big[\frac{a}{q+1}\Big] \geq m+1$), we have that $a_1$ is bounded by a constant and $\alpha([a/(q+1)])=0$, thus \eqref{bosq} can be bounded as
\begin{equation*}
    P\Bigg(\Bigg|\sum_{t=1}^a w_t \varepsilon_t\Bigg| > \lambda \Bigg) \leq  \exp{\Bigg(-\frac{q\frac{\lambda^2}{a}}{25m_2^2 + 5c\frac{\lambda}{\sqrt{a}}}\Bigg)} \leq \exp \{-C_2 \log T \} = T^{-C_2},
\end{equation*}
where $C_2>0$ is suitably large. Since there exist at most $T^2$ sub-intervals $[t_1, t_2]$, applying a simple Bonferroni inequality, we have
\begin{align*}
    P\Bigg( \forall 1 \leq t_1 \leq t_2 \leq T, \; \forall k \in \{1, 2\} \quad s.t. \quad t_2-t_1 \geq C \log T \quad \Bigg|\sum_{t=t_1}^{t_2} \b\ell_{k, t_1, t_2}^t \varepsilon_t \Bigg| \leq \lambda \Bigg) \geq 1-\frac{C_3}{T}
\end{align*}
as $T \rightarrow \infty$ for a large enough $C>0$ and a certain constant $C_3>0$.


\begin{Lem} \label{lm4}
\normalfont  Let $\mathcal{S}^1_j$ and $\mathcal{S}^0_j$ as in Lemma \ref{lm2}. On the set $A_T^L$ in \eqref{atl} that satisfies $P(A_T^L) \rightarrow 1$ as $T \rightarrow \infty$, we have
\begin{equation*} 
\max_{\substack{j=1, \ldots, J, \\ k \in \mathcal{S}^0_j}} \big|d^{(j, k)}\big| \leq \lambda, 
\end{equation*}
where $\lambda$ is as in Theorem \ref{thm1}. 
\end{Lem}

\textbf{Proof.} 
The argument follows the proof of Lemma \ref{lm2}.

\subsection{Theoretical results of the length-lowerbounded-basis estimators} \label{nonG_error}
We now describe the behaviour of the initial estimator $\tilde{f}^L$ that is built from the basis vectors whose non-zero elements have length larger than $a$.

\begin{Thm} \label{thm10}
\normalfont  Let the distribution of $\varepsilon_t$ in model (1) of the main article as follows:
\begin{enumerate}[label=(\alph*)]
    \item $\varepsilon_t$ has mean zero and satisfies Cramer's conditions that
\begin{equation*}
    E|\varepsilon_t|^k \leq c^{k-2} k! E(\varepsilon_t^2) < \infty, \quad t=1, \ldots, T, \quad k=3, 4, \ldots,
\end{equation*}
where $c>0$.
\item $\{\varepsilon_t\}_t$ is the stationary sequence and $m$-dependent i.e. $\sigma(\varepsilon_s, s \leq t)$ and $\sigma(\varepsilon_s, s \geq t+k)$ are independent for $k > m$. 
\end{enumerate}
Let $\bar{f} = \max_t f_t - \min_t f_t$ be bounded and let the estimator $\tilde{f}^L$ is obtained from the estimator $\hat{\mu}^{(j, k)}$ in \eqref{muhat}, with $a=C \log(T)$ and the threshold $\lambda = C_1 \log^{1/2}(T)$, for large enough $C$ and $C_1$. 
Then on the set $A_T^L$ in \eqref{atl}, we have 
\begin{equation*}
  \| {\tilde{f}}^L-{f} \|_T^2 \; \leq \; \tilde{C} \; \frac{1}{T} \; N \; \log^2(T) \; \lceil\; \log (T) / \log (1-\rho)^{-1} \; \rceil,
\end{equation*}
for a constant $\tilde{C}>0$.
\end{Thm}

\textbf{Proof.} 
Let $\mathcal{S}^1_j$ and $\mathcal{S}^0_j$ as in Lemma \ref{lm2}.
From the conditional orthonormality of the unbalanced wavelet transform, on the set $A_T^L$ in \eqref{atl}, we have
\begin{align*} 
\| {\tilde{f}}^L-{f} \|_T^2 \;  = \; & \frac{1}{T} \sum_{j=1}^J \sum_{k=1}^{K(j)} \Big( d^{(j, k)} \cdot \mathbb{I} \big\{ \, \exists (j', k') \in \mathcal{C}_{j, k} \quad |d^{(j', k')}| > \lambda \;\; \text{and} \;\;  k' \in \mathcal{W}_{j'}^L(a) \, \big\} - \mu^{(j, k)} \Big)^2 \\
& + \; T^{-1} \big(s^{[1]}_{1, T} - \mu^{(0, 1)}\big)^2 + \; T^{-1} \big(s^{[2]}_{1, T} - \mu^{(0, 2)}\big)^2 \\
\; \leq \; & \frac{1}{T} \sum_{j=1}^J \Bigg( \sum_{k \in \mathcal{S}^0_j} + \sum_{k \in \mathcal{S}^1_j \cap \mathcal{W}_j^S(a)} + \sum_{k \in \mathcal{S}^1_j \cap \mathcal{W}_j^L(a)} \Bigg) \\
\; & \Big( d^{(j, k)} \cdot \mathbb{I} \big\{ \, \exists (j', k') \in \mathcal{C}_{j, k} \quad |d^{(j', k')}| > \lambda \;\; \text{and} \;\;  k' \in \mathcal{W}_{j'}^L(a) \, \big\} - \mu^{(j, k)} \Big)^2 \\
& + \; 4C_1^2T^{-1} \log T \\
\; =: & \; \mathit{I} + \mathit{II} + \mathit{III} + 4C_1^2T^{-1} \log T \numberthis \label{e92_1},
\end{align*}
where $\mu^{(0, 1)} = \langle {f}, \psi^{(0, 1)} \rangle$, $\mu^{(0, 2)} = \langle {f}, \psi^{(0, 2)} \rangle$ and $\mathcal{W}_{j'}^S(a)$ and $\mathcal{W}_{j'}^L(a)$ are as in \eqref{wl}. 
We note that $\big(s^{[1]}_{1, T} - \mu^{(0, 1)}\big)^2 \leq 2C_1^2\log T$ is simply obtained by combining Lemma \ref{lm4} and the fact that $s^{[1]}_{1, T} - \mu^{(0, 1)}= \langle \bve, \psi^{(0, 1)} \rangle$, which can also be applied to obtain $\big(s^{[2]}_{1, T} - \mu^{(0, 2)}\big)^2 \leq 2C_1^2\log T$.
We now examine the terms $\mathit{I}, \mathit{II}$ and $\mathit{III}$ in \eqref{e92_1}.

\textbf{Term $\mathit{I}$}: By Lemma \ref{lm4}, on the set $A_T^L$, $\mathbb{I} \big\{ \, \exists (j', k') \in \mathcal{C}_{j, k} \quad |d^{(j', k')}| > \lambda \big\} = 0 $ for $k \in \mathcal{S}^0_j$ if $k' \in \mathcal{W}_{j'}^L(a)$. Also by the fact that $\mu^{(j, k)} = 0$ for $j=1, \ldots, J, k \in \mathcal{S}^0_j$, we obtain $\mathit{I} = 0$. 

\textbf{Term $\mathit{II}$}: As there is no short-segment parent coefficient whose children is from long-segment due to the principle of bottom-up merging, the indicator function returns zero and the term $\mathit{II}$ is simplified to $\frac{1}{T} \sum_{j=1}^J \sum_{k \in \mathcal{S}^1_j \cap \mathcal{W}_j^S(a)} \Big(\mu^{(j, k)} \Big)^2$. 

We now examine the bound of individual $\mu^{(j, k)}_{p, q, r}$. Note that only Type 2 and Type 3 basis vectors are considered due to the minimum length constraint given on the set $A_T^L$. Borrowing the generalised form of $\psi^{(j, k)}_{p, q, r}$ in \eqref{e91}, for Type 3 basis vector, we obtain
\begin{align*}
    \mu^{(j, k)}_{p, q, r} = \langle {f}, \psi^{(j, k)}_{p, q, r} \rangle &= \gamma_1 \b\ell_{1, p, q}^\top {f}_{p:q} + \gamma_2 \b\ell_{2, p, q}^\top {f}_{p:q} + \gamma_3 \b\ell_{1, q+1, r}^\top {f}_{q+1:r} + \gamma_4 \b\ell_{2, q+1, r}^\top {f}_{q+1:r} \\ \numberthis \label{mujk}
    & \leq \gamma_1 \|{f}_{p:q}\| + \gamma_2 \|{f}_{p:q}\| + \gamma_3 \|{f}_{q+1:r}\| + \gamma_4 \|{f}_{q+1:r}\|,
\end{align*}
where ${f}_{p:q}$ is the subvector of ${f}$ containing $q-p+1$ elements. 
The inequality \eqref{mujk} is obtained from the orthonormality of $\b\ell_{1, p, q}, \b\ell_{2, p, q}, \b\ell_{1, q+1, r}, \b\ell_{2, q+1, r}$ and the definition of inner product $a\cdot b = \|a\|\cdot\|b\|\cdot \cos(\theta)$, where $\theta$ is the angle between $a$ and $b$. 
Note that if ${f}_{p:q}$ does not contain a change point, the corresponding  $\cos(\theta)=0$ as ${f}_{p:q}$ has a perfect linear trend and in the case when ${f}_{p:q}$ includes a change point, the size of angle is bounded as $|\cos(\theta)|\leq 1$. 
As $\bar{f} = \max_t f_t - \min_t f_t$ is assumed to be bounded and $\|{f}_{p:q}\|^2 \leq C [(q-p+1)^2 + \bar{f}^2]$ regardless of whether there exists a true change point in $[p, q]$, we have
\begin{align*}
    (\mu^{(j, k)}_{p, q, r})^2 \leq & (\gamma_1 + \gamma_2)^2 \cdot \|{f}_{p:q}\|^2 + (\gamma_3 + \gamma_4)^2 \cdot \|{f}_{q+1:r}\|^2 + 2\cdot(\gamma_1 + \gamma_2)\cdot(\gamma_3 + \gamma_4)\cdot \|{f}_{p:q}\|\cdot\|{f}_{q+1:r}\| \\
    \leq & c_1 [(q-p+1)^2 + (f_q-f_p)^2] + c_2 [(r-q)^2 + (f_r-f_{q+1})^2]  \\
    & + c_3  \sqrt{(q-p+1)^2 + (f_q-f_p)^2} \cdot \sqrt{(r-q)^2 + (f_r-f_{q+1})^2}\\
    \leq &  c_4 (r-p+1)^2 + c_5 \bar{f}^2 \\
    \leq & C (r-q+1)^2
\end{align*}
where $c_i > 0$ and $C>0$.
Without loss of generality, we assume $r-q+1 \leq a$, then using $a=O(\log(T))$ and applying the same upper bounds, $J \leq \lceil \log(T) / \log((1-\rho)^{-1}) + \log(2) / \log(1-\rho) \rceil$ and $|\mathcal{S}^1_j| \leq N$, used in the proof of Theorem 1 in the main article, we obtain
\begin{align*}
    \mathit{II} \leq \frac{1}{T}N \lceil \log(T) / \log((1-\rho)^{-1}) + \log(2) / \log(1-\rho) \rceil (\log(T))^2
\end{align*}

\textbf{Term $\mathit{III}$}:
Denote $\mathcal{B} =  \big\{ \, \exists (j', k') \in \mathcal{C}_{j, k} \quad |d^{(j', k')}| > \lambda \, \;\; \text{and} \;\;  k' \in \mathcal{W}_{j'}^L(a) \big\}$ and on the set $A_T^L$ in \eqref{atl} we have
\begin{align*} 
\big( d^{(j, k)} \cdot \mathbb{I} \{ \mathcal{B} \} - \mu^{(j, k)} \big)^2  \; = \; & \big( d^{(j, k)} \cdot \mathbb{I} \{ \mathcal{B} \} - d^{(j, k)} + d^{(j, k)} - \mu^{(j, k)} \big)^2 \\
\;  \leq \; & \big( d^{(j, k)} \big)^2 \mathbb{I}\Big( \big|d^{(j', k')}\big| \leq \lambda \;\; \text{or} \;\; k' \in \mathcal{W}_{j'}^S(a) \Big) +  \; \big( d^{(j, k)} - \mu^{(j, k)}  \big)^2 \\
& + \; 2\big|d^{(j, k)}\big| \; \mathbb{I}\Big( \big|d^{(j', k')}\big| \leq \lambda \;\; \text{or} \;\; k' \in \mathcal{W}_{j'}^S(a) \Big) \; \big|d^{(j, k)} - \mu^{(j, k)}\big| \\  \numberthis \label{e93}
\; \leq \;  & \lambda^2 + 2 C_1^2 \log T + 2\lambda C_1 \{2 \log T \}^{1/2}.
\end{align*}
Following the same argument used in the proof of Theorem 1 in the main article, we have
\begin{align*}
    \mathit{III} \leq \frac{1}{T}N \log(T) \lceil \log(T) / \log((1-\rho)^{-1}) + \log(2) / \log(1-\rho) \rceil.
\end{align*}

To complete the proof, considering all terms in \eqref{e92_1}, we finally obtain 
\begin{align} \label{new_rate}
    \| {\tilde{f}}^L-{f} \|_T^2 \; \leq \; \tilde{C} \; T^{-1} \; N \; (\log(T))^2  \; \lceil (\log(T) / \log((1-\rho)^{-1}) + \log(2) / \log(1-\rho) \rceil,
\end{align}
where $\tilde{C} > 0$. Comparing it with Theorem 1 of the main article that is presented under the iid Gaussian noise assumption, the $\ell_2$ rate in \eqref{new_rate} is different by only a logarithmic factor.

\begin{Thm} \label{thm20}
\normalfont  $X_t$ follows model (1) with $\sigma=1$. Let the distribution of $\varepsilon_t$ and the threshold $\lambda$ be as in Theorem \ref{thm10}. 
Further, let $\bar{f} = \max_t f_t - \min f_t$ be bounded. 
Then we have $\big\| \dbtilde{f}^L - f \big\|_T^2 \; =  \; O\big(NT^{-1} \log^3(T)\big)$ with probability approaching $1$ as $T \rightarrow \infty$, where $\dbtilde{f}^L$ is the estimator constructed from ${\tilde{f}}^L$ through Stage 1 of the post-processing described in Section 2.5 of the main article.   
And there exist at most two estimated change-points between each pair of true change-points $(\eta_i, \eta_{i+1})$ for $i=0, \ldots, N$, where $\eta_0=0$ and $\eta_{N+1}=T$.
Therefore $\dbtilde{N} \leq 2(N+1)$, where $\dbtilde{N}$ is the number of estimated change points in $\dbtilde{f}^L$.
\end{Thm}

\textbf{Proof.} 
The proof proceeds the same as the proof of Theorem 2 of the main article.

\begin{Thm} \label{thm30}
\normalfont  
$X_t$ follows model (1) with $\sigma=1$. Let the distribution of $\varepsilon_t$ and the threshold $\lambda$ be as in Theorem \ref{thm10}.
Further, let the number of true change-points, $N$, have the order of $§log T$ and let $\bar{f} = \max_t f_t - \min f_t$ be bounded. 
Let the estimators $\hat{f}^L$, $\hat{N}$ and $(\hat{\eta}_1, \ldots, \hat{\eta}_{\hat{N}})$ are constructed through Stage 2 of the post-processing described in Section 2.5 of the main article. 
Let $\Delta_T = \min_{i=1, \ldots, N} \Big\{ \Big(\ubar{f}_T^i \Big)^{2/3} \cdot \delta_T^i \Big\}$ 
where $\ubar{f}_T^i = \min \Big( |f_{\eta_{i+1}}-2f_{\eta_{i}}+f_{\eta_{i-1}}|, |f_{\eta_{i+2}}-2f_{\eta_{i+1}}+f_{\eta_{i}}| \Big)$ and 
$\delta_T^i = \min \Big( |\eta_i-\eta_{i-1}|, |\eta_{i+1}-\eta_{i}| \Big)$.
Assume that $T^{1/3} R_T^{1/3} = o\Big(\Delta_T \Big)$ where $\big\| \dbtilde{f}^L - f \big\|_T^2 = O_p(R_T)$ is as in Theorem \ref{thm20}.
Then we have 
\begin{equation} 
\mathbb{P} \; \bigg( \hat{N}=N, \quad \max_{i=1, \ldots, N} \bigg\{ |\hat{\eta}_i-\eta_i| \cdot \Big(\ubar{f}_T^i \Big)^{2/3} \bigg\} \leq C T^{1/3} R_T^{1/3} \bigg) \; \rightarrow \; 1,
\end{equation}
as $T \rightarrow \infty$ where $C$ is a constant. 
\end{Thm}

\textbf{Proof.} 
The proof proceeds the same as the proof of Theorem 3 of the main article.



\section{Threshold selection and additional simulation results} \label{extrasim}

\subsection{Simulation results for non-Gaussian and/or dependent noise} \label{c1}

In addition to the simulations in Section 4 of the main article, here we present the results for the cases when $\varepsilon_t$ is possibly dependent and/or non-Gaussian. Including the standard Gaussian noise, we consider the following six scenarios for $\varepsilon_t$:
\begin{enumerate}[label=(\roman*)] \label{noisesettings}
    \item standard Gaussian, \label{noise_norm}
    \item iid $t_5$ distribution with unit-variance, \label{noise_t5}
    \item a stationary Gaussian AR(1) process of $\phi = 0.3$, with zero-mean and unit-variance, \label{noise_ar03}
    \item the same setting as in (iii) except $\phi = 0.6$, \label{noise_ar06}
    \item a stationary AR(1) process of $\phi = 0.3$ with the noise term following $t_5$, \label{noise_t5_ar03}
    \item the same setting as in (v) except $\phi = 0.6$. \label{noise_t5_ar06}
\end{enumerate}
In summary, (ii) is iid but heavy-tailed, (iii) and (iv) are Gaussian AR(1) error with relatively mild and strong dependence, respectively, and (v) and (vi) are both heavy-tailed but different strength of dependence, where the summary of the simulation results can be found in Tables \ref{tab:t1}-\ref{tab:t5_ar06_2}. 

Following the theoretical results presented in Sections \ref{lemma_nonG_error}-\ref{nonG_error}, we need to set the minimum segment length to be an order of $\log(T)$. As already used in the main paper, we set $\lfloor 0.9 \log(T) \rfloor$ as a default minimum segment length. We follow the Algorithm 1 introduced in the main paper and use $\lambda^\text{Robust}$ as a default threshold, as it is designed to work well in all circumstances. 

The simulation results under this robust threshold selection are presented in  Tables \ref{tab:t1}-\ref{tab:t5_ar06_2} and TrendSegment generally outperforms over all scenarios of noise and over almost all simulation models considered in this paper. Among other competitors, only ID provides the option for heavy-tailed noise in their R package IDetect and other methods are set to their default settings.

\begin{table}[!ht]
\centering
\caption {Distribution of $\hat{N}-N$ for models (M1)-(M4) and all methods with the noise term $\varepsilon_t \overset{\text{iid}}{\sim} t_5$ over 100 simulation runs. Also the average MSE (Mean Squared Error) of the estimated signal $\hat{f}_t$, the average Hausdorff distance $d_H$ and the average computational time in seconds using an Intel Core i5 2.9 GHz CPU with 8 GB of RAM, all over 100 simulations. Bold: methods within 10\% of the highest empirical frequency of $\hat{N}-N=0$ or within 10\% of the lowest empirical average $d_H(\times 10^2)$. Note that TrendSegment is shortened to TS.}
\renewcommand{\arraystretch}{1} 
\begin{tabular}{rrrrrrrrrrrr}
  \hline
    & & & & & $\hat{N}-N$ & & & \\
  \cline{3-9}
Model & Method & $\leq$-3 & -2 & -1 & 0 & 1 & 2 & $\geq$3 & MSE & $d_H(\times 10^2)$ & time \\ 
\hline
   \multirow{6}*{(M1)} 
  & TS & 0 & 0 & 1 & \textbf{88} & 9 & 2 & 0 & 0.24 & 3.10 & 0.09 \\ 
  & NOT & 0 & 0 & 0 & \textbf{92} & 6 & 2 & 0 & 0.20 &  2.51 & 0.22 \\ 
  & ID & 0 & 0 & 0 & \textbf{91} & 9 & 0 & 0 & 0.14 &  \textbf{1.69} & 0.01 \\ 
  & TF & 0 & 0 & 0 & 0 & 0 & 0 & 100 & 0.10 &  4.40 & 3.22 \\ 
  & CPOP & 0 & 0 & 0 & 78 & 12 & 9 & 1 & 0.13 &  \textbf{1.44} & 0.04 \\ 
  & BUP & 100 & 0 & 0 & 0 & 0 & 0 & 0 & 2.63 & 10.61 & 0.35 \\ 
   \hline
\multirow{6}*{(M2)} 
   & TS & 0 & 0 & 4 & \textbf{83} & 9 & 2 & 2 & 0.13 & 2.05 & 0.24 \\ 
   & NOT & 0 & 0 & 3 & \textbf{85} & 11 & 0 & 1 & 0.098 & \textbf{1.69} &  0.29 \\ 
   & ID & 0 & 0 & 0 & \textbf{77} & 21 & 2 & 0 & 0.102 & \textbf{1.36} & 0.38 \\ 
   & TF & 0 & 0 & 0 & 0 & 0 & 0 & 100 & 0.067 & 2.29 & 31.41 \\ 
   & CPOP & 0 & 0 & 0 & 14 & 23 & 25 & 38 & 0.119 & \textbf{1.54} &  1.66 \\ 
   & BUP & 100 & 0 & 0 & 0 & 0 & 0 & 0 & 0.752 & 4.69 &  2.18 \\ 
   \hline
\multirow{6}*{(M3)} 
   & TS & 0 & 0 & 8 & 81 & 8 & 2 & 1 & 0.04 & 4.43 & 0.29 \\ 
   & NOT & 0 & 0 & 1 & \textbf{97} & 2 & 0 & 0 & 0.021 & \textbf{2.71} & 0.31 \\ 
   & ID & 0 & 0 & 0 & 85 & 10 & 2 & 3 & 0.023 & \textbf{2.40} & 0.03 \\ 
   & TF & 0 & 0 & 0 & 0 & 0 & 0 & 100 & 0.010 & 5.20 & 28.83 \\  
   & CPOP & 0 & 0 & 0 & 32 & 25 & 24 & 19 & 0.039 & \textbf{2.51} & 13.06 \\ 
   & BUP & 0 & 0 & 0 & 2 & 26 & 46 & 26 & 0.032 & 5.39 & 2.18 \\ 
   \hline
\multirow{6}*{(M4)} 
  & TS & 0 & 0 & 0 & \textbf{91} & 7 & 0 & 2 & 0.11 & 3.39 & 0.09 \\ 
  & NOT & 0 & 0 & 0 & \textbf{98} & 2 & 0 & 0 & 0.08 &  \textbf{2.57} & 0.24 \\ 
  & ID & 0 & 0 & 0 & 87 & 12 & 1 & 0 & 0.08 &  \textbf{2.21} & 0.01 \\ 
  & TF & 0 & 0 & 0 & 0 & 0 & 0 & 100 & 0.05 &  5.54 & 8.73 \\ 
  & CPOP & 0 & 0 & 0 & 62 & 22 & 8 & 8 & 0.08 &  \textbf{2.24} & 0.38 \\ 
  & BUP & 2 & 73 & 24 & 1 & 0 & 0 & 0 & 0.52 & 10.80 & 0.57 \\ 
   \hline
\end{tabular} 
\label{tab:t1}
\end{table}

\begin{table}[!ht]
\centering
\caption {Distribution of $\hat{N}-N$ for models (M5)-(M8) and all methods with the noise term $\varepsilon_t \overset{\text{iid}}{\sim} t_5$ over 100 simulation runs. Also the average MSE (Mean Squared Error) of the estimated signal $\hat{f}_t$, the average Hausdorff distance $d_H$ and the average computational time in seconds using an Intel Core i5 2.9 GHz CPU with 8 GB of RAM, all over 100 simulations. Bold: methods within 10\% of the highest empirical frequency of $\hat{N}-N=0$ or within 10\% of the lowest empirical average $d_H(\times 10^2)$. Note that TrendSegment is shortened to TS.}
\renewcommand{\arraystretch}{1} 
\begin{tabular}{rrrrrrrrrrrr}
  \hline
    & & & & & $\hat{N}-N$ & & & \\
  \cline{3-9}
Model & Method & $\leq$-3 & -2 & -1 & 0 & 1 & 2 & $\geq$3 & MSE & $d_H(\times 10^2)$ & time \\ 
  \hline
 \multirow{6}*{(M5)} 
   & TS & 0 & 0 & 2 & \textbf{75} & 18 & 4 & 1 & 0.04 & 2.17 & 0.31 \\ 
   & NOT & 0 & 11 & 10 & 63 & 10 & 3 & 3 & 0.049 & \textbf{1.29} & 0.25 \\ 
   & ID & 0 & 0 & 0 & 0 & 0 & 7 & 93 & 0.332 & 9.58 & 0.03 \\ 
   & TF & 0 & 0 & 0 & 0 & 0 & 0 & 100 & 0.145 & 6.14 & 28.33 \\ 
   & CPOP & 0 & 0 & 0 & 4 & 6 & 20 & 70 & 0.064 & 2.61 & 3.07 \\ 
   & BUP & 0 & 0 & 0 & 32 & 44 & 20 & 4 & 0.097 & 4.62 & 2.22 \\ 
   \hline
\multirow{6}*{(M6)} 
   & TS & 0 & 4 & 1 & \textbf{88} & 3 & 1 & 3 & 0.02 & \textbf{1.23} & 0.35 \\ 
   & NOT & 6 & 10 & 26 & 44 & 6 & 2 & 6 & 0.071 & 3.53 & 0.24 \\ 
   & ID & 0 & 3 & 0 & 0 & 19 & 0 & 78 & 0.129 & 4.73 & 0.03 \\ 
   & TF & 0 & 0 & 0 & 0 & 0 & 0 & 100 & 0.136 & 9.88 & 30.26 \\ 
   & CPOP & 0 & 0 & 0 & 8 & 19 & 20 & 53 & 0.053 & 3.15 & 2.45 \\ 
   & BUP & 0 & 0 & 0 & 0 & 0 & 0 & 100 & 0.132 & 9.23 & 2.47 \\ 
   \hline
\multirow{6}*{(M7)} 
   & TS & 5 & 15 & 28 & \textbf{36} & 10 & 4 & 2 & 0.16 & 8.51 & 0.13 \\ 
   & NOT & 0 & 6 & 16 & \textbf{30} & 36 & 11 & 1 & 0.079 &  5.12 &  0.22 \\ 
   & ID & 6 & 3 & 9 & 18 & 19 & 15 & 30 & 0.385 & 12.40 & 0.01 \\ 
   & TF & 0 & 0 & 0 & 0 & 0 & 0 & 100 & 0.098 &  6.08 & 23.86 \\ 
   & CPOP & 0 & 0 & 0 & 0 & 4 & 5 & 91 & 0.102 & \textbf{3.01} &  0.81 \\ 
   & BUP & 69 & 28 & 3 & 0 & 0 & 0 & 0 & 0.266 & 12.12 &  1.47 \\ 
   \hline
\multirow{6}*{(M8)} 
   & TS & 0 & 0 & 0 & \textbf{99} & 0 & 0 & 1 & 0.00 & \textbf{0.50} & 0.19 \\ 
   & NOT & 0 & 0 & 0 & \textbf{100} & 0 & 0 & 0 & 0.001 & \textbf{0.00} &  0.17 \\ 
   & ID & 0 & 0 & 0 & \textbf{99} & 1 & 0 & 0 & 0.001 & \textbf{0.00} & 0.03 \\ 
   & TF & 0 & 0 & 0 & 65 & 12 & 9 & 14 & 0.003 & 14.63 & 36.03 \\ 
   & CPOP & 0 & 0 & 0 & 35 & 0 & 34 & 31 & 0.042 & 20.53 &  3.91 \\ 
   & BUP & 0 & 0 & 0 & 0 & 0 & 0 & 100 & 0.014 & 46.80 &  2.62 \\ 
   \hline
\end{tabular} 
\label{tab:t2}
\end{table}



\begin{table}[!ht]
\centering
\caption {Distribution of $\hat{N}-N$ for models (M1)-(M4) and all methods with the noise term $\epsilon_t$ being $AR(1)$ process of $\phi = 0.3$ over 100 simulation runs. Also the average MSE (Mean Squared Error) of the estimated signal $\hat{f}_t$, the average Hausdorff distance $d_H$ and the average computational time in seconds using an Intel Core i5 2.9 GHz CPU with 8 GB of RAM, all over 100 simulations. Bold: methods within 10\% of the highest empirical frequency of $\hat{N}-N=0$ or within 10\% of the lowest empirical average $d_H(\times 10^2)$. Note that TrendSegment is shortened to TS.}
\renewcommand{\arraystretch}{1} 
\begin{tabular}{rrrrrrrrrrrr}
  \hline
    & & & & & $\hat{N}-N$ & & & \\
  \cline{3-9}
Model & Method & $\leq$-3 & -2 & -1 & 0 & 1 & 2 & $\geq$3 & MSE & $d_H(\times 10^2)$ & time \\ 
  \hline
\multirow{6}*{(M1)} 
  & TS & 0 & 1 & 13 & \textbf{82} & 4 & 0 & 0 & 0.39 & 3.65 & 0.07 \\ 
  & NOT & 0 & 0 & 0 & \textbf{87} & 8 & 2 & 3 & 0.35 &  \textbf{3.10} & 0.23 \\ 
  & ID & 0 & 0 & 0 & 62 & 27 & 9 & 2 & 0.27 &  \textbf{2.70} & 0.02 \\ 
  & TF & 3 & 0 & 0 & 0 & 0 & 0 & 97 & 0.61 &  6.18 & 3.31 \\ 
  & CPOP & 0 & 0 & 0 & 53 & 35 & 10 & 2 & 0.23 &  \textbf{2.52} & 0.05 \\ 
  & BUP & 100 & 0 & 0 & 0 & 0 & 0 & 0 & 2.64 & 10.96 & 0.36 \\  
   \hline
\multirow{6}*{(M2)} 
   & TS & 4 & 9 & 30 & 57 & 0 & 0 & 0 & 0.20 & 2.56 & 0.24 \\  
   & NOT & 0 & 0 & 8 & \textbf{83} & 6 & 2 & 1 & 0.182 & 2.11 &  0.31 \\ 
   & ID & 0 & 0 & 0 & 69 & 24 & 5 & 2 & 0.155 & \textbf{1.75} & 0.40 \\ \ 
   & TF & 0 & 0 & 0 & 0 & 0 & 0 & 100 & 0.600 & 2.38 & 32.03 \\ 
   & CPOP & 0 & 0 & 0 & 1 & 6 & 8 & 85 & 0.163 & \textbf{1.98} &  1.50 \\ 
   & BUP & 100 & 0 & 0 & 0 & 0 & 0 & 0 & 0.717 & 4.63 &  2.39 \\ 
   \hline
\multirow{6}*{(M3)} 
   & TS & 0 & 0 & 17 & 79 & 4 & 0 & 0 & 0.05 & 4.79 & 0.30 \\ 
   & NOT & 0 & 0 & 1 & \textbf{89} & 7 & 2 & 1 & 0.045 & 3.81 & 0.32 \\ 
   & ID & 0 & 0 & 1 & \textbf{83}  & 14 & 1 & 1 & 0.037 & 2.98 & 0.03 \\ 
   & TF & 0 & 0 & 0 & 0 & 0 & 0 & 100 & 0.258 & 6.24 & 28.76 \\ 
   & CPOP & 0 & 0 & 0 & 76 & 10 & 9 & 5 & 0.022 & \textbf{2.14} & 15.35 \\ 
   & BUP & 0 & 0 & 0 & 0 & 6 & 23 & 71 & 0.040 & 5.59 & 2.31 \\ 
   \hline
\multirow{6}*{(M4)} 
  & TS & 0 & 0 & 2 & \textbf{95} & 3 & 0 & 0 & 0.15 & 3.61 & 0.09 \\ 
  & NOT & 0 & 0 & 0 & \textbf{86} & 9 & 3 & 2 & 0.16 &  \textbf{3.50} & 0.23 \\ 
  & ID & 0 & 0 & 0 & 84 & 14 & 0 & 2 & 0.13 &  \textbf{2.87} & 0.01 \\ 
  & TF & 1 & 0 & 1 & 0 & 1 & 0 & 97 & 0.64 &  6.76 & 8.67 \\ 
  & CPOP & 0 & 0 & 0 & 51 & 24 & 15 & 10 & 0.11 &  \textbf{3.17} & 0.39 \\ 
  & BUP & 1 & 61 & 38 & 0 & 0 & 0 & 0 & 0.50 & 10.43 & 0.58 \\ 
   \hline
\end{tabular} 
\label{tab:ar1}
\end{table}

\begin{table}[!ht]
\centering
\caption {Distribution of $\hat{N}-N$ for models (M5)-(M8) and all methods with the noise term $\epsilon_t$ being $AR(1)$ process of $\phi = 0.3$ over 100 simulation runs. Also the average MSE (Mean Squared Error) of the estimated signal $\hat{f}_t$, the average Hausdorff distance $d_H$ and the average computational time in seconds using an Intel Core i5 2.9 GHz CPU with 8 GB of RAM, all over 100 simulations. Bold: methods within 10\% of the highest empirical frequency of $\hat{N}-N=0$ or within 10\% of the lowest empirical average $d_H(\times 10^2)$. Note that TrendSegment is shortened to TS.}
\renewcommand{\arraystretch}{1} 
\begin{tabular}{rrrrrrrrrrrr}
  \hline
    & & & & & $\hat{N}-N$ & & & \\
  \cline{3-9}
Model & Method & $\leq$-3 & -2 & -1 & 0 & 1 & 2 & $\geq$3 & MSE & $d_H(\times 10^2)$ & time \\ 
\hline
 \multirow{6}*{(M5)} 
   & TS & 0 & 0 & 0 & \textbf{84} & 15 & 1 & 0 & 0.05 & \textbf{1.85} & 0.32 \\ 
   & NOT & 0 & 6 & 13 & \textbf{74} & 4 & 3 & 0 & 0.062 & \textbf{1.59} & 0.26 \\ 
   & ID & 0 & 0 & 0 & 1 & 4 & 17 & 78 & 0.347 & 8.87 & 0.03 \\ 
   & TF & 0 & 0 & 0 & 0 & 0 & 0 & 100 & 0.192 & 6.16 & 28.45 \\ 
   & CPOP & 0 & 0 & 0 & 2 & 14 & 20 & 64 & 0.059 & \textbf{2.02} & 3.64 \\ 
   & BUP & 0 & 0 & 0 & 11 & 32 & 30 & 27 & 0.131 & 5.19 & 2.32 \\ 
   \hline
\multirow{6}*{(M6)} 
   & TS & 0 & 6 & 0 & \textbf{93} & 1 & 0 & 0 & 0.02 & \textbf{1.34} & 0.35 \\ 
   & NOT & 6 & 18 & 28 & 31 & 4 & 2 & 11 & 0.094 & 5.30 & 0.25 \\ 
   & ID & 1 & 10 & 0 & 0 & 21 & 0 & 68 & 0.149 & 6.75 & 0.04 \\  
   & TF & 0 & 0 & 0 & 0 & 0 & 0 & 100 & 0.324 & 9.93 & 29.97 \\ 
   & CPOP & 0 & 0 & 0 & 7 & 32 & 28 & 23 & 0.043 & \textbf{1.04} & 3.58 \\ 
   & BUP & 0 & 0 & 0 & 0 & 0 & 0 & 100 & 0.159 & 9.09 & 2.82 \\ 
   \hline
\multirow{6}*{(M7)} 
   & TS & 20 & 47 & 24 & 7 & 2 & 0 & 0 & 0.23 & 11.74 & 0.12 \\ 
   & NOT & 5 & 12 & 19 & \textbf{24} & 22 & 7 & 11 & 0.158 &  7.69 &  0.24 \\ 
   & ID & 11 & 3 & 15 & \textbf{22} & 18 & 16 & 15 & 0.405 & 14.22 & 0.01 \\ 
   & TF & 3 & 0 & 0 & 0 & 0 & 0 & 97 & 0.623 &  7.01 & 23.25 \\ 
   & CPOP & 0 & 0 & 0 & 0 & 0 & 1 & 99 & 0.162 & \textbf{5.27} &  0.85 \\ 
   & BUP & 54 & 43 & 3 & 0 & 0 & 0 & 0 & 0.283 & 11.92 &  1.55 \\ 
   \hline
\multirow{6}*{(M8)} 
   & TS & 0 & 0 & 0 & \textbf{100} & 0 & 0 & 0 & 0.00 & \textbf{0.00} & 0.19 \\ 
   & NOT & 0 & 0 & 0 & \textbf{93} & 3 & 3 & 1 & 0.005 & \textbf{2.02} &  0.19 \\ 
   & ID & 0 & 0 & 0 & \textbf{100} & 0 & 0 & 0 & 0.003 & \textbf{0.00} & 0.51 \\ 
   & TF & 0 & 0 & 0 & 0 & 0 & 0 & 100 & 0.551 & 49.94 & 35.81 \\ 
   & CPOP & 0 & 0 & 0 & 30 & 10 & 3 & 57 & 0.035 & 19.71 &  7.55 \\ 
   & BUP & 0 & 0 & 0 & 0 & 0 & 0 & 100 & 0.025 & 46.73 &  2.72 \\ 
   \hline
\end{tabular} 
\label{tab:ar2}
\end{table}



\begin{table}[!ht]
\centering
\caption {Distribution of $\hat{N}-N$ for models (M1)-(M4) and all methods with the noise term $\epsilon_t$ being $AR(1)$ process of $\phi = 0.6$ over 100 simulation runs. Also the average MSE (Mean Squared Error) of the estimated signal $\hat{f}_t$, the average Hausdorff distance $d_H$ and the average computational time in seconds using 10 cores of Apple M1 Pro with 16 GB of RAM on mac OS, all over 100 simulations. Bold: methods within 10\% of the highest empirical frequency of $\hat{N}-N=0$ or within 10\% of the lowest empirical average $d_H(\times 10^2)$. Note that TrendSegment is shortened to TS.}
\renewcommand{\arraystretch}{1} 
\begin{tabular}{rrrrrrrrrrrr}
  \hline
    & & & & & $\hat{N}-N$ & & & \\
  \cline{3-9}
Model & Method & $\leq$-3 & -2 & -1 & 0 & 1 & 2 & $\geq$3 & MSE & $d_H(\times 10^2)$ & time \\ 
  \hline
\multirow{6}*{(M1)} 
 & TS & 2 & 2 & 22 & \textbf{67} & 7 & 0 & 0 & 0.82 & \textbf{4.82} & 0.07 \\ 
 & NOT & 0 & 1 & 4 & \textbf{50} & 15 & 9 & 21 & 0.86 & \textbf{4.19} & 0.09 \\ 
 & ID & 0 & 0 & 0 & 5 & 16 & 20 & 59 & 0.70 & \textbf{4.29} & 0.01 \\ 
 & TF & 28 & 0 & 0 & 1 & 0 & 1 & 70 & 1.44 & 14.23 & 1.84 \\ 
 & CPOP & 0 & 0 & 0 & 4 & 11 & 18 & 67 & 0.76 & \textbf{4.31} & 0.05 \\ 
 & BUP & 100 & 0 & 0 & 0 & 0 & 0 & 0 & 2.55 & 11.22 & 0.15 \\ 
    \hline
\multirow{6}*{(M2)} 
  & TS & 30 & 34 & 26 & 8 & 2 & 0 & 0 & 0.50 & 3.69 & 0.23 \\ 
 & NOT & 0 & 4 & 13 & 21 & 19 & 18 & 25 & 0.51 & 2.94 & 0.14 \\ 
 & ID & 2 & 5 & 4 & \textbf{33} & 23 & 17 & 16 & 0.41 & 3.20 & 0.02 \\ 
 & TF & 0 & 0 & 0 & 0 & 0 & 0 & 100 & 1.23 & \textbf{2.38} & 12.86 \\ 
 & CPOP & 0 & 0 & 0 & 0 & 0 & 0 & 100 & 0.54 & \textbf{2.48} & 0.87 \\ 
 & BUP & 100 & 0 & 0 & 0 & 0 & 0 & 0 & 0.70 & 4.43 & 0.70 \\ 
    \hline
\multirow{6}*{(M3)} 
  & TS & 0 & 4 & 23 & \textbf{45} & 16 & 8 & 4 & 0.14 & 6.77 & 0.31 \\ 
 & NOT & 0 & 0 & 4 & 24 & 7 & 6 & 59 & 0.21 & \textbf{5.95} & 0.20 \\ 
 & ID & 1 & 5 & 11 & 28 & 11 & 16 & 28 & 0.12 & 6.08 & 0.04 \\ 
 & TF & 0 & 0 & 0 & 0 & 0 & 0 & 100 & 0.58 & 6.23 & 19.54 \\ 
 & CPOP & 0 & 0 & 0 & 0 & 0 & 0 & 100 & 0.38 & 6.08 & 3.31 \\ 
 & BUP & 0 & 0 & 0 & 0 & 0 & 0 & 100 & 0.13 & \textbf{5.92} & 1.22 \\ 
    \hline
\multirow{6}*{(M4)} 
   & TS & 0 & 1 & 16 & \textbf{57} & 18 & 4 & 4 & 0.42 & \textbf{5.63} & 0.09 \\ 
  & NOT & 0 & 0 & 3 & 26 & 16 & 11 & 44 & 0.56 & \textbf{5.61} & 0.12 \\ 
  & ID & 0 & 0 & 8 & 41 & 24 & 17 & 10 & 0.35 & \textbf{4.78} & 0.01 \\ 
  & TF & 25 & 0 & 2 & 0 & 1 & 0 & 72 & 1.46 & 13.64 & 5.63 \\ 
  & CPOP & 0 & 0 & 0 & 0 & 2 & 0 & 98 & 0.59 & 5.72 & 0.26 \\ 
  & BUP & 1 & 37 & 58 & 4 & 0 & 0 & 0 & 0.57 & 9.45 & 0.31 \\ 
   \hline
\end{tabular}
\label{tab:ar1_06}
\end{table}


\begin{table}[!ht]
\centering
\caption {Distribution of $\hat{N}-N$ for models (M5)-(M8) and all methods with the noise term $\epsilon_t$ being $AR(1)$ process of $\phi = 0.6$ over 100 simulation runs. Also the average MSE (Mean Squared Error) of the estimated signal $\hat{f}_t$, the average Hausdorff distance $d_H$ and the average computational time in seconds using 10 cores of Apple M1 Pro with 16 GB of RAM on mac OS, all over 100 simulations. Bold: methods within 10\% of the highest empirical frequency of $\hat{N}-N=0$ or within 10\% of the lowest empirical average $d_H(\times 10^2)$. Note that TrendSegment is shortened to TS.}
\renewcommand{\arraystretch}{1} 
\begin{tabular}{rrrrrrrrrrrr}
  \hline
    & & & & & $\hat{N}-N$ & & & \\
  \cline{3-9}
Model & Method & $\leq$-3 & -2 & -1 & 0 & 1 & 2 & $\geq$3 & MSE & $d_H(\times 10^2)$ & time \\ 
  \hline
\multirow{6}*{(M5)} 
  & TS & 0 & 0 & 10 &\textbf{40} & 32 & 10 & 8 & 0.14 & \textbf{3.62} & 0.33 \\ 
 & NOT & 0 & 3 & 10 & 11 & 11 & 10 & 55 & 0.22 & 4.46 & 0.18 \\ 
 & ID & 2 & 3 & 1 & 4 & 6 & 5 & 79 & 0.42 & 7.28 & 0.04 \\ 
 & TF & 0 & 0 & 0 & 0 & 0 & 0 & 100 & 0.40 & 6.18 & 18.88 \\ 
 & CPOP & 0 & 0 & 0 & 0 & 0 & 0 & 100 & 0.47 & 5.96 & 2.18 \\ 
 & BUP & 0 & 0 & 0 & 0 & 0 & 0 & 100 & 0.24 & 5.90 & 1.24 \\ 
      \hline
\multirow{6}*{(M6)} 
  & TS & 0 & 3 & 0 & \textbf{65} & 12 & 11 & 9 & 0.07 & \textbf{3.09} & 0.36 \\ 
 & NOT & 13 & 8 & 16 & 22 & 6 & 3 & 32 & 0.19 & 8.10 & 0.15 \\ 
 & ID & 39 & 26 & 2 & 0 & 19 & 1 & 13 & 0.28 & 24.64 & 0.04 \\ 
 & TF & 0 & 0 & 0 & 0 & 0 & 0 & 100 & 0.69 & 9.91 & 197.40 \\ 
 & CPOP & 0 & 0 & 0 & 0 & 0 & 0 & 100 & 0.52 & 9.50 & 2.91 \\ 
 & BUP & 0 & 0 & 0 & 0 & 0 & 0 & 100 & 0.30 & 9.41 & 1.36 \\ 
      \hline
\multirow{6}*{(M7)} 
  & TS & 25 & 25 & 26 & 14 & 5 & 4 & 1 & 0.40 & 11.82 & 0.13 \\ 
 & NOT & 3 & 4 & 4 & 11 & 13 & 3 & 62 & 0.49 & 8.03 & 0.11 \\ 
 & ID & 10 & 3 & 9 & \textbf{20} & 19 & 15 & 24 & 0.47 & 13.15 & 0.02 \\ 
 & TF & 24 & 0 & 2 & 0 & 0 & 0 & 74 & 1.47 & 13.14 & 9.90 \\ 
 & CPOP & 0 & 0 & 0 & 0 & 0 & 0 & 100 & 0.59 & \textbf{7.02} & 0.44 \\ 
 & BUP & 3 & 30 & 42 & \textbf{21} & 4 & 0 & 0 & 0.35 & 8.97 & 0.46 \\ 
        \hline
\multirow{6}*{(M8)} 
  & TS & 0 & 0 & 0 & 63 & 7 & 26 & 4 & 0.03 & 10.42 & 0.19 \\ 
 & NOT & 0 & 0 & 0 & 7 & 8 & 4 & 81 & 0.19 & 37.77 & 0.10 \\ 
 & ID & 0 & 0 & 0 & \textbf{96} & 3 & 0 & 1 & 0.01 & \textbf{0.61} & 0.03 \\ 
 & TF & 0 & 0 & 0 & 0 & 0 & 0 & 100 & 1.10 & 49.94 & 15.54 \\ 
 & CPOP & 0 & 0 & 0 & 0 & 1 & 0 & 99 & 0.38 & 45.54 & 2.08 \\ 
 & BUP & 0 & 0 & 0 & 0 & 0 & 0 & 100 & 0.11 & 47.44 & 0.85 \\ 
   \hline
\end{tabular}
\label{tab:ar2_06}
\end{table}


\begin{table}[!ht]
\centering
\caption {Distribution of $\hat{N}-N$ for models (M1)-(M4) and all methods with the $\varepsilon_t$ being $AR(1)$ process of $\phi = 0.3$ with noise term following $t_5$ over 100 simulation runs. Also the average MSE (Mean Squared Error) of the estimated signal $\hat{f}_t$, the average Hausdorff distance $d_H$ and the average computational time in seconds using 10 cores of Apple M1 Pro with 16 GB of RAM on mac OS, all over 100 simulations. Bold: methods within 10\% of the highest empirical frequency of $\hat{N}-N=0$ or within 10\% of the lowest empirical average $d_H(\times 10^2)$. Note that TrendSegment is shortened to TS.}
\renewcommand{\arraystretch}{1} 
\begin{tabular}{rrrrrrrrrrrr}
  \hline
    & & & & & $\hat{N}-N$ & & & \\
  \cline{3-9}
Model & Method & $\leq$-3 & -2 & -1 & 0 & 1 & 2 & $\geq$3 & MSE & $d_H(\times 10^2)$ & time \\ 
\hline
   \multirow{6}*{(M1)}
  & TS & 0 & 0 & 0 & \textbf{100} & 0 & 0 & 0 & 0.07 & \textbf{1.72} & 0.10 \\ 
 & NOT & 0 & 0 & 0 & 90 & 7 & 2 & 1 & 0.06 & \textbf{1.61} & 0.09 \\ 
 & ID & 0 & 0 & 0 & 49 & 29 & 9 & 13 & 0.06 & 2.08 & 0.01 \\ 
 & TF & 17 & 0 & 0 & 0 & 1 & 0 & 82 & 0.13 & 9.94 & 1.64 \\ 
 & CPOP & 0 & 0 & 0 & \textbf{100} & 0 & 0 & 0 & 0.03 & \textbf{0.87} & 0.05 \\ 
 & BUP & 100 & 0 & 0 & 0 & 0 & 0 & 0 & 3.14 & 12.19 & 0.14 \\ 
     \hline
\multirow{6}*{(M2)} 
  & TS & 0 & 0 & 2 & \textbf{94} & 2 & 0 & 2 & 0.04 & 1.32 & 0.25 \\ 
 & NOT & 0 & 0 & 0 & \textbf{95} & 4 & 1 & 0 & 0.03 & \textbf{1.10} & 0.16 \\ 
 & ID & 0 & 0 & 0 & 47 & 25 & 14 & 14 & 0.08 & \textbf{1.18} & 0.02 \\ 
 & TF & 0 & 0 & 0 & 0 & 0 & 0 & 100 & 0.14 & 2.38 & 12.86 \\ 
 & CPOP & 0 & 0 & 0 & \textbf{100} & 0 & 0 & 0 & 0.04 & \textbf{0.82} & 1.38 \\ 
 & BUP & 100 & 0 & 0 & 0 & 0 & 0 & 0 & 1.22 & 5.08 & 0.69 \\ 
     \hline
\multirow{6}*{(M3)} 
  & TS & 0 & 0 & 0 & \textbf{99} & 0 & 0 & 1 & 0.01 & 2.52 & 0.31 \\ 
 & NOT & 0 & 0 & 0 & 88 & 9 & 2 & 1 & 0.01 & 2.04 & 0.24 \\ 
 & ID & 0 & 0 & 0 & 56 & 29 & 9 & 6 & 0.01 & 2.42 & 0.02 \\ 
 & TF & 0 & 0 & 0 & 0 & 0 & 0 & 100 & 0.04 & 6.23 & 20.08 \\ 
 & CPOP & 0 & 0 & 0 & \textbf{99} & 0 & 1 & 0 & 0.00 & \textbf{0.67} & 21.78 \\ 
 & BUP & 5 & 82 & 13 & 0 & 0 & 0 & 0 & 0.14 & 11.18 & 1.10 \\ 
     \hline
\multirow{6}*{(M4)} 
  & TS & 0 & 0 & 0 & \textbf{100} & 0 & 0 & 0 & 0.03 & \textbf{2.00} & 0.10 \\ 
 & NOT & 0 & 0 & 0 & 88 & 8 & 4 & 0 & 0.03 & \textbf{1.81} & 0.18 \\ 
 & ID & 0 & 0 & 0 & 54 & 27 & 15 & 4 & 0.05 & \textbf{2.02} & 0.01 \\ 
 & TF & 5 & 0 & 0 & 0 & 0 & 0 & 95 & 0.13 & 7.75 & 5.95 \\ 
 & CPOP & 0 & 0 & 0 & \textbf{100} & 0 & 0 & 0 & 0.03 & \textbf{1.62} & 0.34 \\ 
 & BUP & 85 & 15 & 0 & 0 & 0 & 0 & 0 & 0.85 & 12.55 & 0.30 \\ 
   \hline
\end{tabular}
\label{tab:t5_ar03_1}
\end{table}

\begin{table}[!ht]
\centering
\caption {Distribution of $\hat{N}-N$ for models (M5)-(M8) and all methods with the $\varepsilon_t$ being $AR(1)$ process of $\phi = 0.3$ with noise term following $t_5$ over 100 simulation runs. Also the average MSE (Mean Squared Error) of the estimated signal $\hat{f}_t$, the average Hausdorff distance $d_H$ and the average computational time in seconds using 10 cores of Apple M1 Pro with 16 GB of RAM on mac OS, all over 100 simulations. Bold: methods within 10\% of the highest empirical frequency of $\hat{N}-N=0$ or within 10\% of the lowest empirical average $d_H(\times 10^2)$. Note that TrendSegment is shortened to TS.}
\renewcommand{\arraystretch}{1} 
\begin{tabular}{rrrrrrrrrrrr}
  \hline
    & & & & & $\hat{N}-N$ & & & \\
  \cline{3-9}
Model & Method & $\leq$-3 & -2 & -1 & 0 & 1 & 2 & $\geq$3 & MSE & $d_H(\times 10^2)$ & time \\ 
  \hline
 \multirow{6}*{(M5)} 
  & TS & 0 & 0 & 0 & \textbf{98} & 0 & 0 & 2 & 0.01 & \textbf{0.66} & 0.34 \\ 
 & NOT & 0 & 12 & 7 & 66 & 6 & 4 & 5 & 0.03 & \textbf{0.74} & 0.17 \\ 
 & ID & 0 & 0 & 0 & 0 & 0 & 2 & 98 & 0.20 & 5.02 & 0.03 \\ 
 & TF & 0 & 0 & 0 & 0 & 0 & 0 & 100 & 0.27 & 5.92 & 18.55 \\ 
 & CPOP & 0 & 0 & 0 & 50 & 38 & 9 & 3 & 0.02 & 1.43 & 3.27 \\ 
 & BUP & 16 & 84 & 0 & 0 & 0 & 0 & 0 & 0.16 & \textbf{0.89} & 1.12 \\ 
     \hline
\multirow{6}*{(M6)} 
  & TS & 0 & 0 & 1 & \textbf{97} & 0 & 0 & 2 & 0.01 & \textbf{0.20} & 0.37 \\ 
 & NOT & 0 & 3 & 44 & 48 & 2 & 0 & 3 & 0.05 & \textbf{0.80} & 0.14 \\ 
 & ID & 0 & 0 & 0 & 0 & 0 & 0 & 100 & 0.09 & \textbf{0.39} & 0.03 \\ 
 & TF & 0 & 0 & 0 & 0 & 0 & 0 & 100 & 0.07 & 9.90 & 21.12 \\ 
 & CPOP & 0 & 0 & 0 & 72 & 24 & 4 & 0 & 0.01 & \textbf{0.09} & 2.01 \\ 
 & BUP & 23 & 36 & 30 & 11 & 0 & 0 & 0 & 0.18 & \textbf{0.27} & 1.42 \\ 
     \hline
\multirow{6}*{(M7)} 
  & TS & 32 & 35 & 26 & 6 & 0 & 0 & 1 & 0.13 & 11.69 & 0.13 \\ 
 & NOT & 0 & 0 & 0 & \textbf{74} & 20 & 4 & 2 & 0.01 & \textbf{0.81} & 0.11 \\ 
 & ID & 2 & 2 & 3 & 10 & 16 & 33 & 34 & 0.35 & 10.75 & 0.01 \\ 
 & TF & 0 & 0 & 0 & 0 & 0 & 0 & 100 & 0.14 & 6.24 & 11.24 \\ 
 & CPOP & 0 & 0 & 0 & 10 & 30 & 28 & 32 & 0.02 & \textbf{0.65} & 0.76 \\ 
 & BUP & 100 & 0 & 0 & 0 & 0 & 0 & 0 & 0.25 & 12.50 & 0.53 \\ 
     \hline
\multirow{6}*{(M8)} 
  & TS & 0 & 0 & 0 & \textbf{97} & 0 & 0 & 3 & 0.00 & \textbf{1.50} & 0.21 \\ 
 & NOT & 0 & 0 & 0 & 88 & 4 & 7 & 1 & 0.00 & \textbf{3.46} & 0.09 \\ 
 & ID & 0 & 0 & 0 & \textbf{100} & 0 & 0 & 0 & 0.00 & \textbf{0.00} & 0.03 \\ 
 & TF & 0 & 0 & 0 & 0 & 0 & 0 & 100 & 0.11 & 49.93 & 15.78 \\ 
 & CPOP & 0 & 0 & 0 & \textbf{99} & 0 & 1 & 0 & 0.00 & \textbf{0.05} & 5.50 \\ 
 & BUP & 0 & 0 & 0 & 0 & 100 & 0 & 0 & 0.00 & 39.45 & 0.84 \\ 
   \hline
\end{tabular}
\label{tab:t5_ar03_2}
\end{table}


\begin{table}[!ht]
\centering
\caption {Distribution of $\hat{N}-N$ for models (M1)-(M4) and all methods with the $\varepsilon_t$ being $AR(1)$ process of $\phi = 0.6$ with noise term following $t_5$ over 100 simulation runs. Also the average MSE (Mean Squared Error) of the estimated signal $\hat{f}_t$, the average Hausdorff distance $d_H$ and the average computational time in seconds using 10 cores of Apple M1 Pro with 16 GB of RAM on mac OS, all over 100 simulations. Bold: methods within 10\% of the highest empirical frequency of $\hat{N}-N=0$ or within 10\% of the lowest empirical average $d_H(\times 10^2)$. Note that TrendSegment is shortened to TS.}
\renewcommand{\arraystretch}{1} 
\begin{tabular}{rrrrrrrrrrrr}
  \hline
    & & & & & $\hat{N}-N$ & & & \\
  \cline{3-9}
Model & Method & $\leq$-3 & -2 & -1 & 0 & 1 & 2 & $\geq$3 & MSE & $d_H(\times 10^2)$ & time \\ 
\hline
   \multirow{6}*{(M1)} 
  & TS & 0 & 0 & 1 & \textbf{99} & 0 & 0 & 0 & 0.15 & \textbf{2.18} & 0.10 \\ 
 & NOT & 0 & 0 & 0 & 55 & 12 & 8 & 25 & 0.17 & \textbf{2.90} & 0.10 \\ 
 & ID & 0 & 0 & 0 & 7 & 9 & 11 & 73 & 0.15 & 3.98 & 0.01 \\ 
 & TF & 43 & 0 & 1 & 0 & 0 & 0 & 56 & 0.29 & 18.34 & 1.74 \\ 
 & CPOP & 0 & 0 & 0 & \textbf{100} & 0 & 0 & 0 & 0.10 & \textbf{1.21} & 0.06 \\ 
 & BUP & 100 & 0 & 0 & 0 & 0 & 0 & 0 & 3.03 & 11.77 & 0.15 \\ 
     \hline
\multirow{6}*{(M2)}  
  & TS & 1 & 7 & 19 & 72 & 0 & 0 & 1 & 0.11 & 2.11 & 0.24 \\ 
 & NOT & 0 & 0 & 0 & 59 & 8 & 7 & 26 & 0.09 & 1.68 & 0.17 \\ 
 & ID & 0 & 0 & 0 & 15 & 11 & 14 & 60 & 0.11 & 1.80 & 0.03 \\ 
 & TF & 0 & 0 & 0 & 0 & 0 & 0 & 100 & 0.25 & 2.38 & 12.96 \\ 
 & CPOP & 0 & 0 & 0 & \textbf{83} & 15 & 1 & 1 & 0.07 & \textbf{1.13} & 1.34 \\ 
 & BUP & 100 & 0 & 0 & 0 & 0 & 0 & 0 & 1.15 & 5.42 & 0.71 \\ 
     \hline
\multirow{6}*{(M3)} 
  & TS & 0 & 5 & 17 & 77 & 0 & 0 & 1 & 0.04 & 4.38 & 0.31 \\ 
 & NOT & 0 & 0 & 0 & 16 & 7 & 8 & 69 & 0.05 & 5.16 & 0.22 \\ 
 & ID & 0 & 0 & 0 & 24 & 18 & 14 & 44 & 0.03 & 4.20 & 0.03 \\ 
 & TF & 0 & 0 & 0 & 0 & 0 & 0 & 100 & 0.10 & 6.23 & 19.46 \\ 
 & CPOP & 0 & 0 & 0 & \textbf{96} & 2 & 2 & 0 & 0.01 & \textbf{1.39} & 19.42 \\ 
 & BUP & 0 & 43 & 49 & 8 & 0 & 0 & 0 & 0.10 & 9.61 & 1.10 \\ 
     \hline
\multirow{6}*{(M4)} 
  & TS & 0 & 0 & 3 & \textbf{97} & 0 & 0 & 0 & 0.08 & \textbf{2.82} & 0.09 \\ 
 & NOT & 0 & 0 & 0 & 22 & 14 & 10 & 54 & 0.12 & 4.68 & 0.18 \\ 
 & ID & 0 & 0 & 0 & 27 & 19 & 21 & 33 & 0.09 & 3.71 & 0.01 \\ 
 & TF & 38 & 0 & 0 & 1 & 1 & 0 & 60 & 0.29 & 17.07 & 6.15 \\ 
 & CPOP & 0 & 0 & 0 & \textbf{95} & 3 & 2 & 0 & 0.05 & \textbf{2.03} & 0.35 \\ 
 & BUP & 72 & 28 & 0 & 0 & 0 & 0 & 0 & 0.79 & 12.41 & 0.31 \\ 
   \hline
\end{tabular}
\label{tab:t5_ar06_1}
\end{table}


\begin{table}[!ht]
\centering
\caption {Distribution of $\hat{N}-N$ for models (M5)-(M8) and all methods with the $\varepsilon_t$ being $AR(1)$ process of $\phi = 0.6$ with noise term following $t_5$ over 100 simulation runs. Also the average MSE (Mean Squared Error) of the estimated signal $\hat{f}_t$, the average Hausdorff distance $d_H$ and the average computational time in seconds using 10 cores of Apple M1 Pro with 16 GB of RAM on mac OS, all over 100 simulations. Bold: methods within 10\% of the highest empirical frequency of $\hat{N}-N=0$ or within 10\% of the lowest empirical average $d_H(\times 10^2)$. Note that TrendSegment is shortened to TS.}
\renewcommand{\arraystretch}{1} 
\begin{tabular}{rrrrrrrrrrrr}
  \hline
    & & & & & $\hat{N}-N$ & & & \\
  \cline{3-9}
Model & Method & $\leq$-3 & -2 & -1 & 0 & 1 & 2 & $\geq$3 & MSE & $d_H(\times 10^2)$ & time \\ 
\hline
   \multirow{6}*{(M5)} 
  & TS & 0 & 2 & 15 & \textbf{81} & 1 & 0 & 1 & 0.03 & \textbf{1.11} & 0.33 \\ 
 & NOT & 0 & 5 & 10 & 37 & 7 & 4 & 37 & 0.05 & 2.80 & 0.18 \\ 
 & ID & 0 & 0 & 0 & 0 & 0 & 1 & 99 & 0.19 & 4.57 & 0.04 \\ 
 & TF & 0 & 0 & 0 & 0 & 0 & 0 & 100 & 0.26 & 6.02 & 18.55 \\ 
 & CPOP & 0 & 0 & 0 & 29 & 38 & 28 & 5 & 0.03 & \textbf{1.55} & 3.44 \\ 
 & BUP & 13 & 86 & 1 & 0 & 0 & 0 & 0 & 0.16 & \textbf{1.32} & 1.17 \\ 
     \hline
\multirow{6}*{(M6)} 
  & TS & 0 & 0 & 0 & \textbf{99} & 0 & 0 & 1 & 0.01 & \textbf{0.10} & 0.36 \\ 
 & NOT & 1 & 2 & 36 & 46 & 4 & 3 & 8 & 0.06 & 1.98 & 0.14 \\ 
 & ID & 0 & 2 & 0 & 0 & 8 & 1 & 89 & 0.12 & 3.02 & 0.04 \\ 
 & TF & 0 & 0 & 0 & 0 & 0 & 0 & 100 & 0.15 & 9.90 & 20.25 \\ 
 & CPOP & 0 & 0 & 0 & 81 & 17 & 1 & 1 & 0.01 & \textbf{0.15} & 2.92 \\ 
 & BUP & 0 & 2 & 24 & 74 & 0 & 0 & 0 & 0.08 & \textbf{0.24} & 1.22 \\ 
     \hline
\multirow{6}*{(M7)} 
  & TS & 73 & 25 & 1 & 0 & 0 & 0 & 1 & 0.20 & 12.44 & 0.12 \\ 
 & NOT & 0 & 0 & 0 & \textbf{7} & 5 & 16 & 72 & 0.08 & \textbf{4.50} & 0.11 \\ 
 & ID & 0 & 0 & 2 & 3 & 7 & 8 & 80 & 0.24 & 10.13 & 0.01 \\ 
 & TF & 0 & 0 & 0 & 0 & 0 & 0 & 100 & 0.29 & 6.26 & 11.19 \\ 
 & CPOP & 0 & 0 & 0 & 2 & 17 & 24 & 57 & 0.07 & \textbf{4.48} & 0.84 \\ 
 & BUP & 100 & 0 & 0 & 0 & 0 & 0 & 0 & 0.26 & 12.51 & 0.52 \\ 
     \hline
\multirow{6}*{(M8)} 
  & TS & 0 & 0 & 0 & \textbf{99} & 0 & 0 & 1 & 0.00 & \textbf{0.50} & 0.20 \\ 
 & NOT & 0 & 0 & 0 & 3 & 2 & 13 & 82 & 0.04 & 40.33 & 0.10 \\ 
 & ID & 0 & 0 & 0 & 86 & 4 & 1 & 9 & 0.00 & \textbf{2.89} & 0.02 \\ 
 & TF & 0 & 0 & 0 & 0 & 0 & 0 & 100 & 0.22 & 49.93 & 15.67 \\ 
 & CPOP & 0 & 0 & 0 & \textbf{97} & 0 & 3 & 0 & 0.00 & \textbf{0.72} & 8.40 \\ 
 & BUP & 0 & 0 & 0 & 0 & 80 & 20 & 0 & 0.00 & 40.07 & 0.83 \\ 
   \hline
\end{tabular}
\label{tab:t5_ar06_2}
\end{table}


\clearpage
\subsection{Choice of the threshold} \label{choicelambda}
In this section, we describe the details of how the thresholds, $\lambda^\text{Naïve}$ and $\lambda^\text{Robust}$, are built under different scenarios of noise introduced in Section \ref{c1}. 

\paragraph{The naive threshold selection}
We first explain how the best performing threshold constant $C$ in the naive threshold. To cover all the noise scenario settings including dependent and/or non-Gaussian noise, here we use a simpler version of the follwoing naïve threshold:
\begin{equation}
    \lambda^\text{Naïve} = C \sqrt{2 \log T},
\end{equation}
by considering that the $\sigma$ in $\lambda^\text{Naïve} = C \sigma \sqrt{2 \log T}$ can be absorbed into the constant $C$. To find the best performing constant $C$ over different noise scenarios introduced in Section \ref{c1}, we repeat the simulations with a range of $C$, $[0.5, 3.5]$. The performance can be evaluated by the accuracy of detecting number and location of change-points. For the number of change-point, we define
\begin{equation} \label{c_eta}
    C^\eta_\text{min} = \text{Median}(\tilde{C}^\eta_{\text{min}, 1}, \ldots, \tilde{C}^\eta_{\text{min}, 8}),  
\end{equation}
where $\tilde{C}^\eta_{\text{min}, j}$ is the minimum of those constants $C$ that give the maximum number of the case $\{\hat{N}-N=0\}$ for the $j^\text{th}$ model from 100 simulation runs. The minimum condition is actually used when there is more than one constant giving the same maximum number of $\{\hat{N}-N=0\}$. Similarly, we can define $C^\eta_\text{med}$ and $C^\eta_\text{max}$ by replacing the minimum condition with median and maximum respectively. For evaluating the performance of change-point location, we define
\begin{equation} \label{c_dh}
    C^{d_H}_\text{max} = \text{Median}(\tilde{C}^{d_H}_{\text{max}, 1}, \ldots, \tilde{C}^{d_H}_{\text{max}, 8}),  
\end{equation}
where $\tilde{C}^{d_H}_{\text{max}, j}$ is the maximum of those constants $C$ that give the minimum value of the average Hausdorff distance for the $j^\text{th}$ model computed from 100 simulation runs. Note that in contrast to that the maximum number of case $\{\hat{N}-N=0\}$ is considered in \eqref{c_eta}, the minimum value of Hausdorff distance is used in  \eqref{c_dh} as the smaller the Hausdorff distance, the better the estimation of the change-point locations. Similar to \eqref{c_eta}, the maximum condition actually works when there is more than one constant giving the same minimum average Hausdorff distance, and $C^{d_H}_\text{med}$ and $C^{d_H}_\text{min}$ can be defined by replacing the maximum condition with median and minimum respectively.



\begin{table}[ht!]
\centering
\caption {The default thresholding constant $C$ with its range examined for five scenarios.}
\begin{tabular}{ccccccc}
\hline
$\varepsilon_t$ & $C_\text{min}^\eta$ & $C_\text{med}^\eta$ & $C_\text{max}^\eta$ & $C_\text{min}^{d_H}$ & $C_\text{med}^{d_H}$ & $C_\text{max}^{d_H}$\\
\hline
(i) & 1.2 & 1.4 & 1.5 & 1.4 & 1.5 & 1.7\\
(ii) & 1.4 & 1.5 & 1.6 & 1.3 & 1.5 & 1.5\\
(iii) & 1.5 & 1.5 & 1.5 & 1.4 & 1.4 & 1.4\\
(iv) & 1.8 & 1.8 & 1.8 & 1.3 & 1.3 & 1.3\\
(v) & 1.0 & 1.6 & 1.9 & 1.0 & 1.4 & 1.9 \\
(vi) & 1.4 & 1.5 & 1.7 & 1.4 & 1.5 & 1.7 \\
\hline
\end{tabular}
\label{rd1}
\end{table}

The best performing constants over all noise scenarios are reported in Table \ref{rd1}. Compared to the iid Gaussian noise, it seems that a larger threshold constant tends to chosen when the noise is heavy-tailed and/or dependent. 
Also, compared to the other noise scenarios, when the noise is dependent but generated with Gaussian innovation ((iii) and (iv)), the best performing constant has a narrower range of $C_\text{max}^\cdot$ - $C_\text{min}^\cdot$. 

The naïve threshold, $\lambda^\text{Naïve}$, is an essential element in building the robust threshold, $\lambda^\text{Robust}$, as shown in Step 5 of Algorithm 1 in the main paper. For this, we use the default constants $C_\text{med}^\eta$ in Table \ref{rd1}, where there is not much difference in simulation performance presented in Tables \ref{tab:t1}-\ref{tab:t5_ar06_2} when $C_\text{max}^\eta$ is used instead. 




\paragraph{The robust threshold selection}
We now describe the details of how $\lambda^\text{Robust}$ is built. 
We first justify using the ratio, 
\begin{equation} \label{ratio}
    \frac{\lambda^\text{Naïve}}{1.3 \hat{\mathcal{I}} \sqrt{2 \log T}}
\end{equation}
in the process of building the kurtosis function $g$ in Step 5 of Algorithm 1 in the main paper. 

\begin{figure}[ht!] 
     \centering
    \begin{subfigure}[t]{0.9\textwidth} 
        \raisebox{-\height}{\includegraphics[width=\textwidth, height=0.33\textwidth]{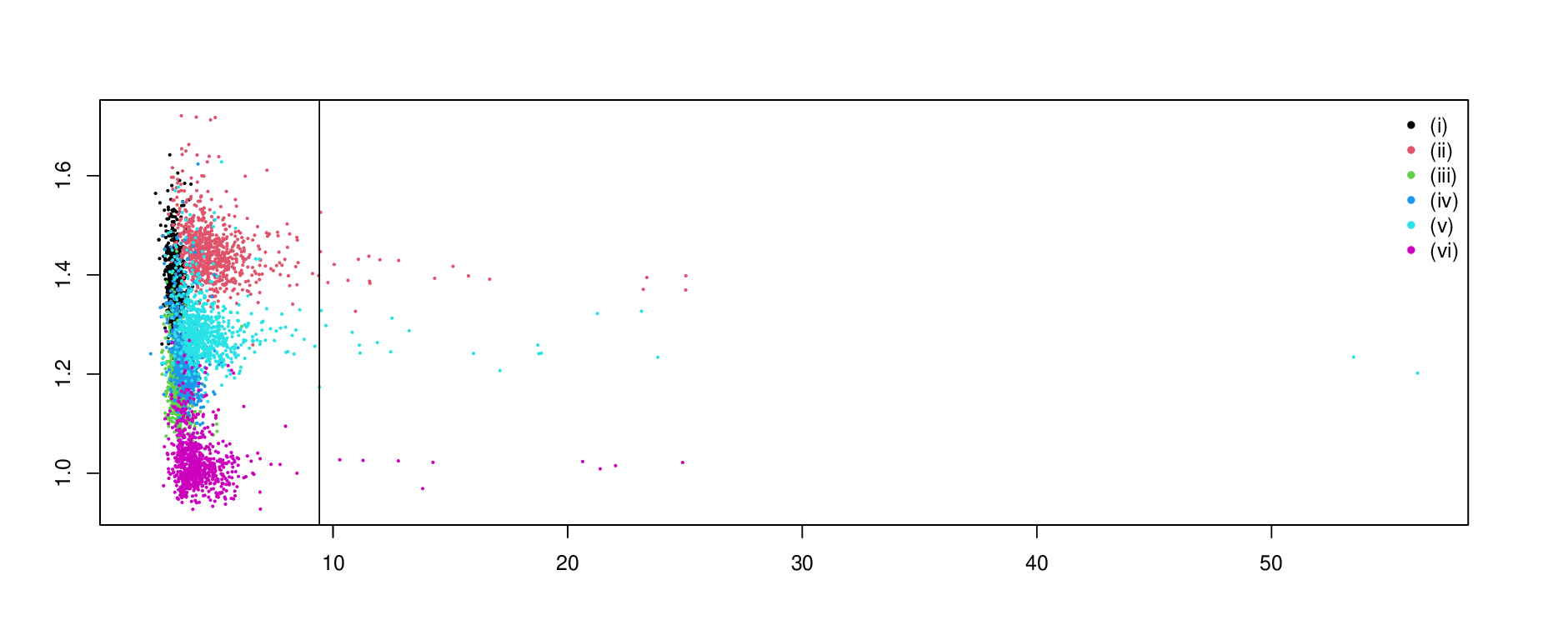}}
        \caption{The black solid line represents $99\%$ quantile of $\hat{\mathcal{\kappa}}$, $9.418$.}
        \label{fig:r1}
    \end{subfigure}
    \begin{subfigure}[t]{0.9\textwidth} 
        \raisebox{-\height}{\includegraphics[width=\textwidth, height=0.33\textwidth]{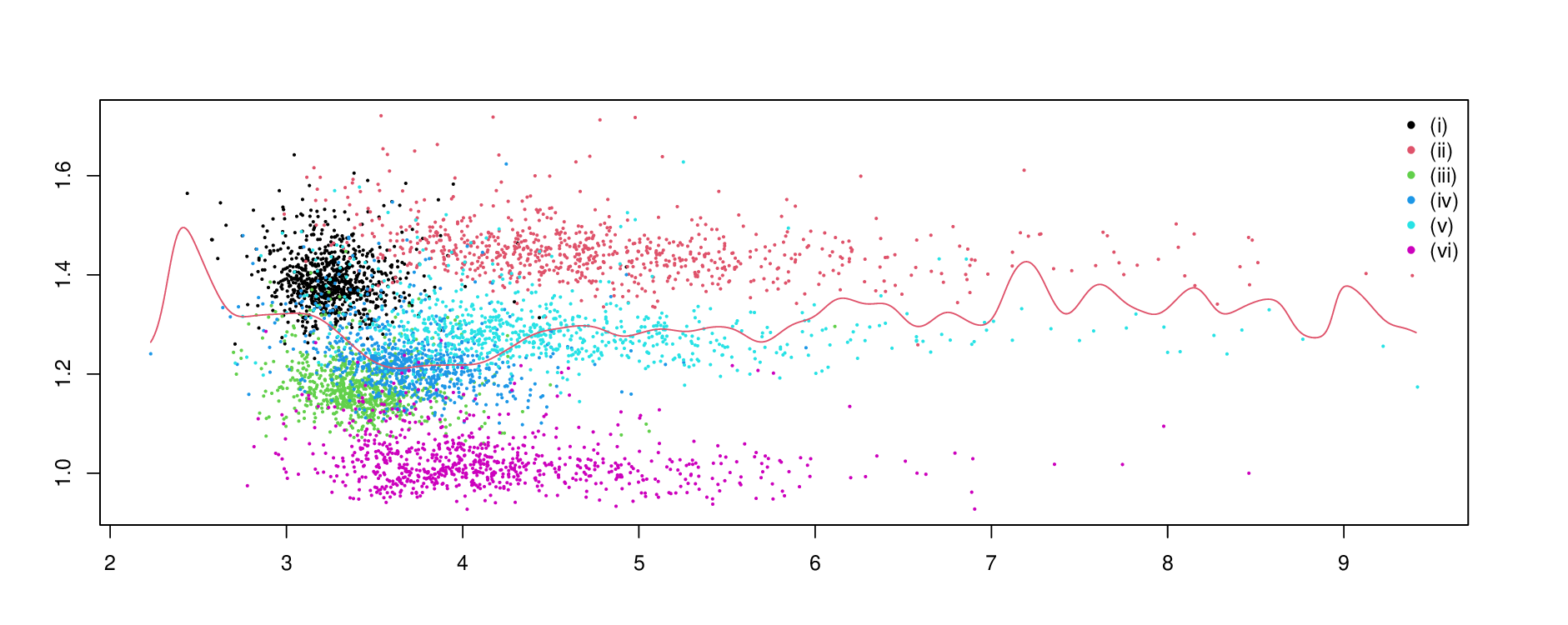}}
        \caption{The left-side of the black line in (a) with the estimated non-parametric fit $\hat{g}_1(\hat{\mathcal{\kappa}})$ (red solid).}
        \label{fig:r2}
    \end{subfigure}
    \begin{subfigure}[t]{0.9\textwidth} 
        \raisebox{-\height}{\includegraphics[width=\textwidth, height=0.33\textwidth]{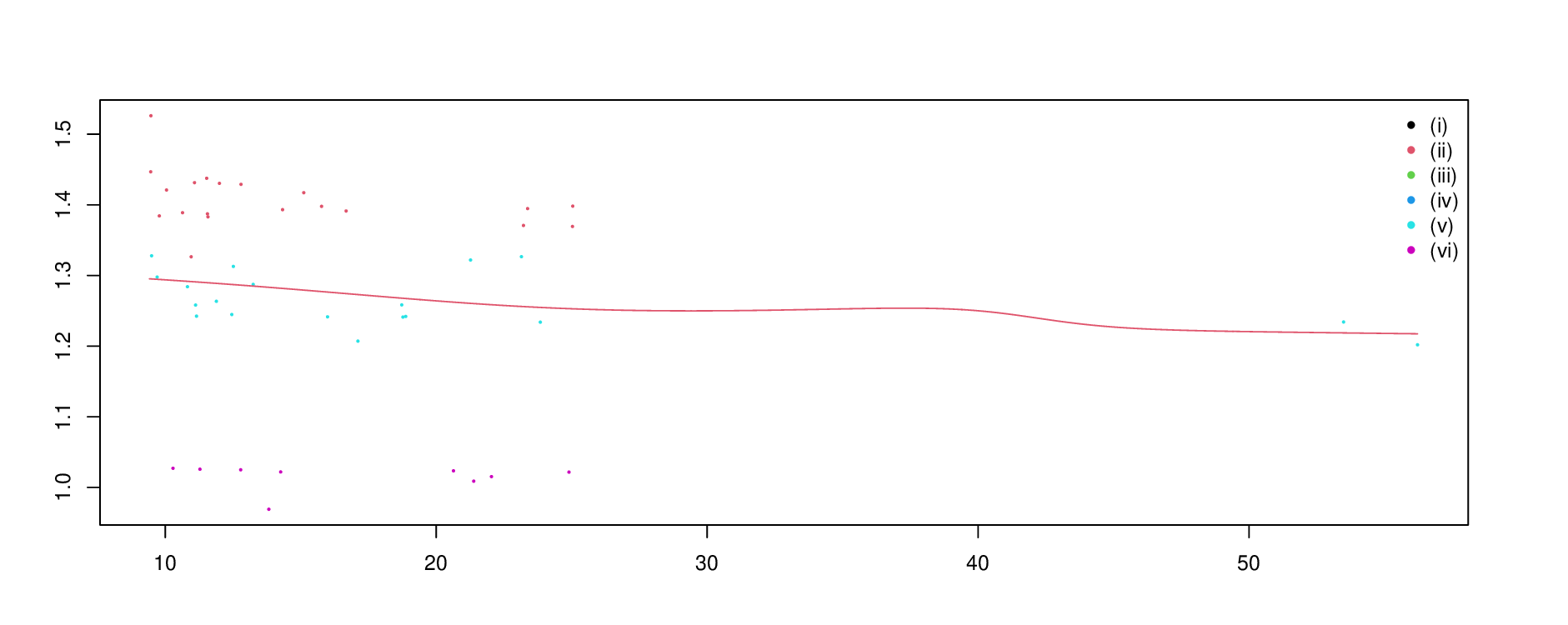}}
        \caption{The right-side of the black line in (a) with the estimated non-parametric fit $\hat{g}_2(\hat{\mathcal{\kappa}})$ (red solid).}
        \label{fig:r3}
    \end{subfigure}
     \caption{The estimated kurtosis, $\hat{\mathcal{\kappa}}$, of $\hat{\varepsilon}_t$ (x-axis) and the ratio in \eqref{ratio} (y-axis) over all models considered and over 6 different noise scenarios, (i)-(vi), presented in Section \ref{c1}, where each combination of model and noise scenario has $N=100$ repetitions (dots) in (a). $\hat{\varepsilon}_t$ is obtained from the pre-fit in Algorithm 1 of the main paper.}
\label{fig:fkurt_all}
\end{figure}

We first recall that the ratio in \eqref{ratio} corresponds to the kurtosis function $g(\mathcal{K})$ in the following robust threshold:
\begin{equation}\label{lambdarob}
    \lambda^\text{Robust} = C \mathcal{I} g(\mathcal{K}) \sqrt{2 \log T}.
\end{equation}
Figure \ref{fig:fkurt_all} shows that $g(\hat{\mathcal{K}})$ behaves like constant over a range of the $\hat{\mathcal{K}}$ under all models and noise scenarios we considered. This is due to the condition on the minimum segment length imposed for stable and good performance. With this condition, we found that the constant-like behaviour is also observed in case noise has relatively extreme heavy-tail e.g. $t_{2.1}$, however we do not include such an extreme case in estimating the non-parametric function $g$.

We are now ready to estimate the function $g(\cdot)$. To avoid the situation that the estimation of non-parametric fit is affected a lot by extremely large size of $\hat{\mathcal{K}}$, we split $\hat{\mathcal{K}}$ into two with the $99\%$ quantile of $\hat{\mathcal{\kappa}}$ as shown in Figure \ref{fig:fkurt_all}. Then we estimate the non-parametric regression fit for each interval, $g_1(\mathcal{K})$ and $g_2(\mathcal{K})$ respectively, and use these functions for computing the robust threshold. 

\clearpage

\section{Additional data application results} \label{extradat}

\subsection{Monthly average sea ice extent of Arctic and Antarctic data}\label{extra_seaice}
\begin{figure}[!ht] 
     \centering
    \begin{subfigure}[t]{0.9\textwidth} 
        \raisebox{-\height}{\includegraphics[width=\textwidth, height=0.48\textwidth]{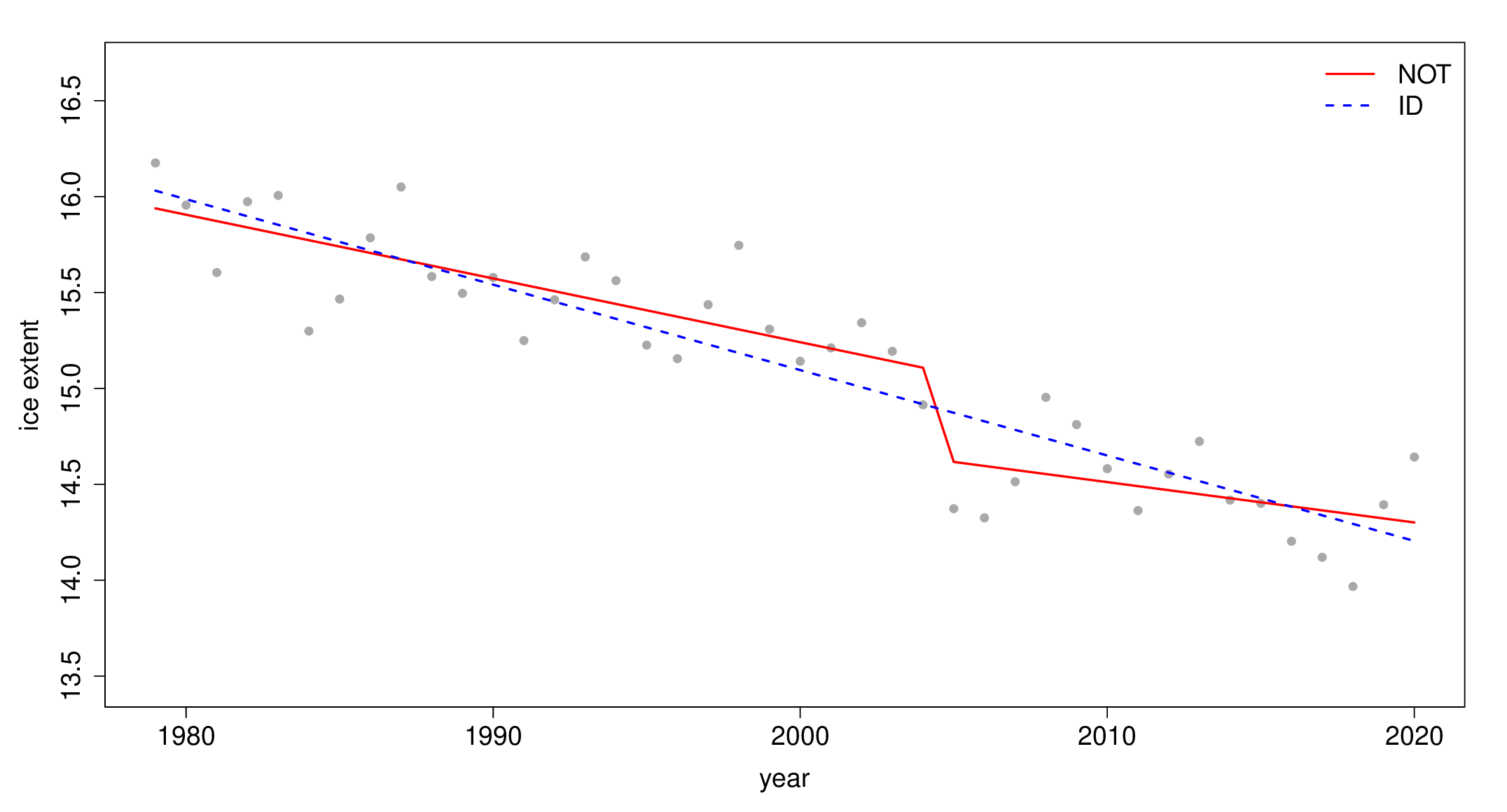}}
        \caption{NOT and ID}
        \label{fig:AF1}
    \end{subfigure}
    \begin{subfigure}[t]{0.9\textwidth} 
        \raisebox{-\height}{\includegraphics[width=\textwidth, height=0.48\textwidth]{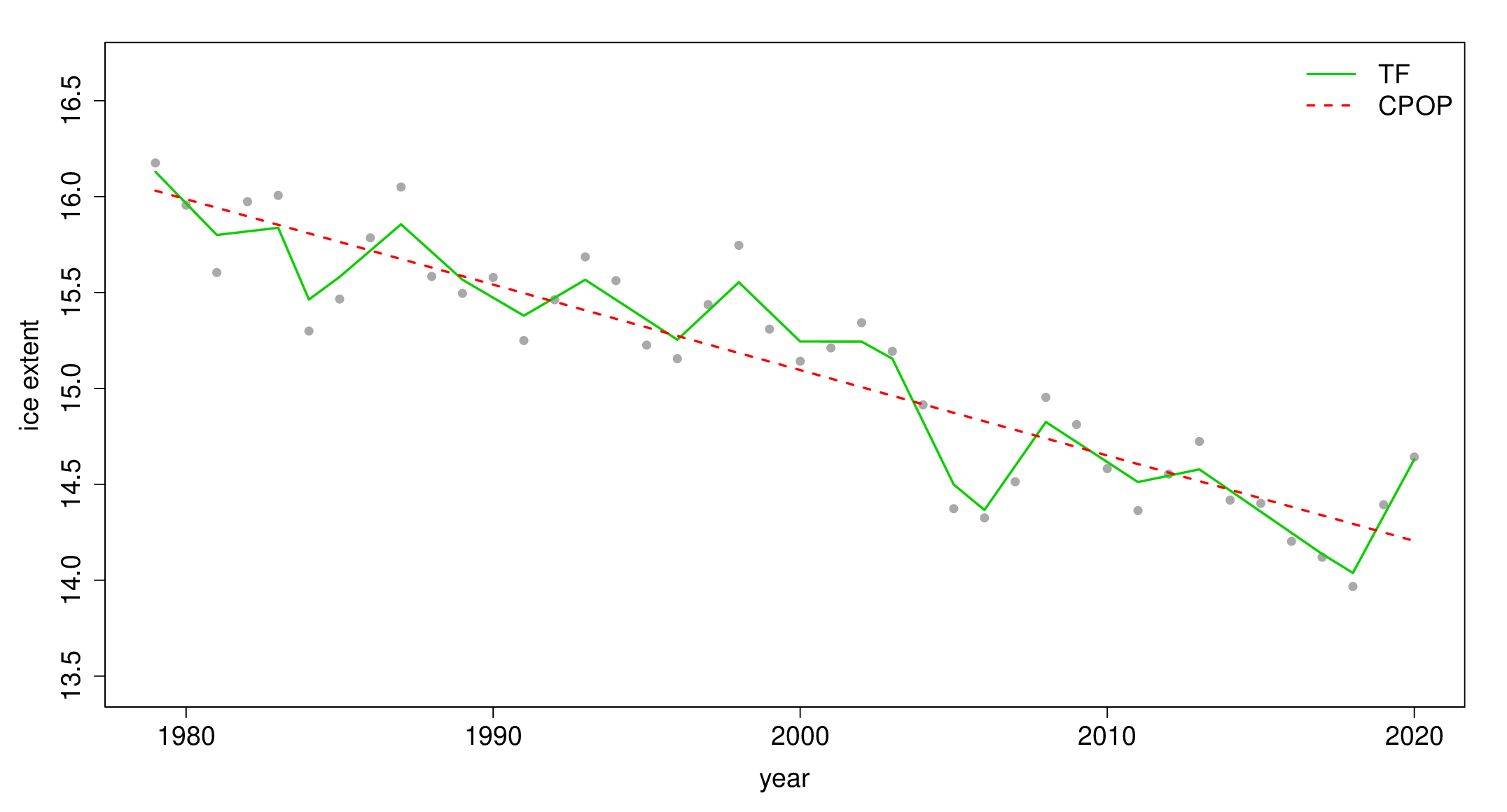}}
        \caption{TF and CPOP}
        \label{fig:AF2}
    \end{subfigure}
    \caption {Change-point analysis for the monthly average sea ice extent of the Arctic in February from 1979 to 2020 in Section 5.2. (a) the data series (grey dots) and estimated signal with change-points returned by NOT (\full) and ID (\dashed), (b) estimated signal with change-points returned by TF (\full) and CPOP (\chain).}
\end{figure}
\pagebreak
\begin{figure}[ht!] 
     \centering
    \begin{subfigure}[t]{0.9\textwidth} 
        \raisebox{-\height}{\includegraphics[width=\textwidth]{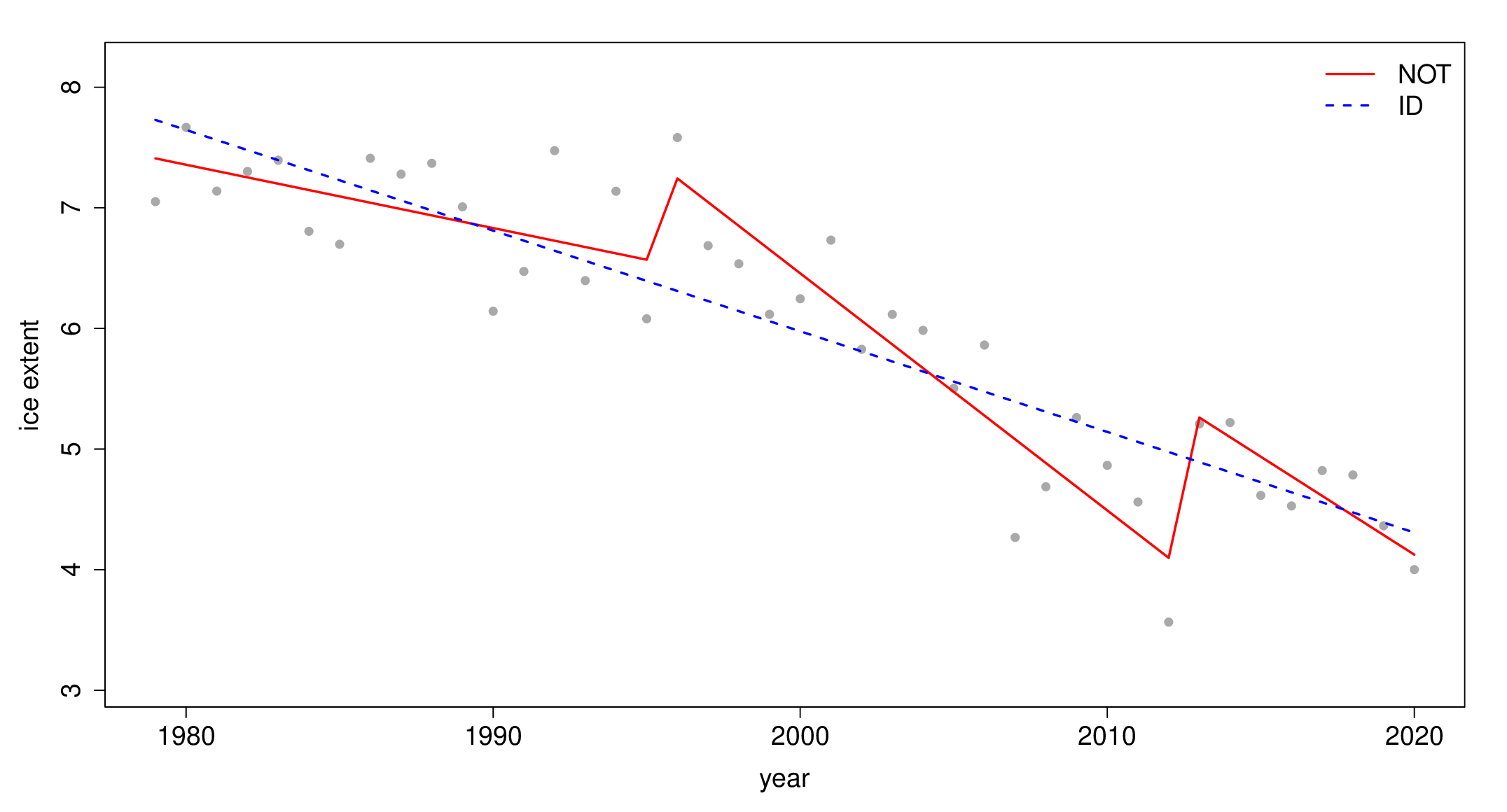}}
        \caption{NOT and ID}
        \label{fig:AS1}
    \end{subfigure}
    \begin{subfigure}[t]{0.9\textwidth} 
        \raisebox{-\height}{\includegraphics[width=\textwidth]{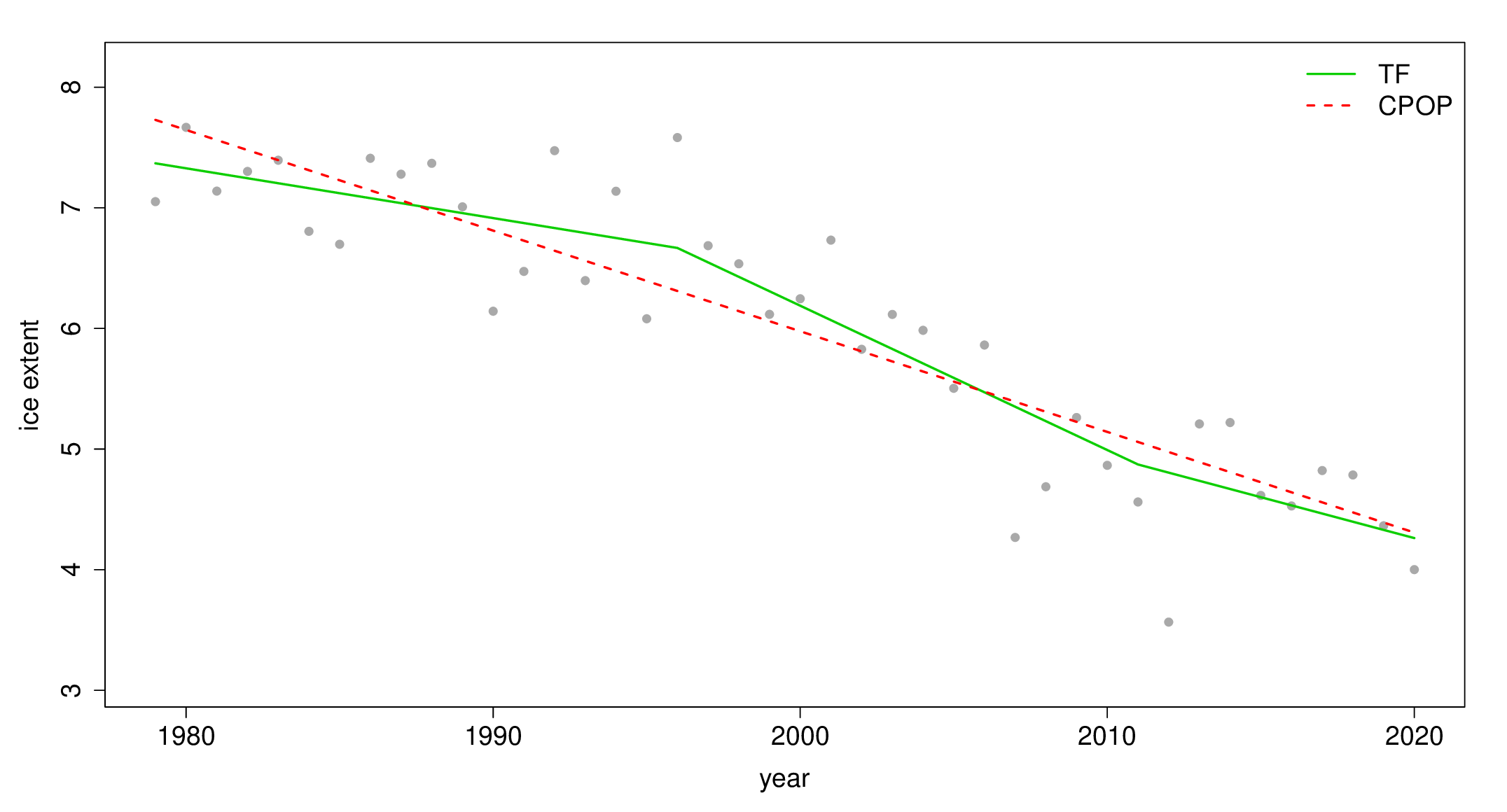}}
        \caption{TF and CPOP}
        \label{fig:AS2}
    \end{subfigure}
    \caption {Change-point analysis for the monthly average sea ice extent of the Arctic in September from 1979 to 2020 in Section 5.2. (a) the data series (grey dots) and estimated signal with change-points returned by NOT (\full) and ID (\dashed), (b) estimated signal with change-points returned by TF (\full) and CPOP (\chain).}
\end{figure}
\begin{figure}[ht!] 
     \centering
    \begin{subfigure}[t]{0.9\textwidth} 
        \raisebox{-\height}{\includegraphics[width=\textwidth]{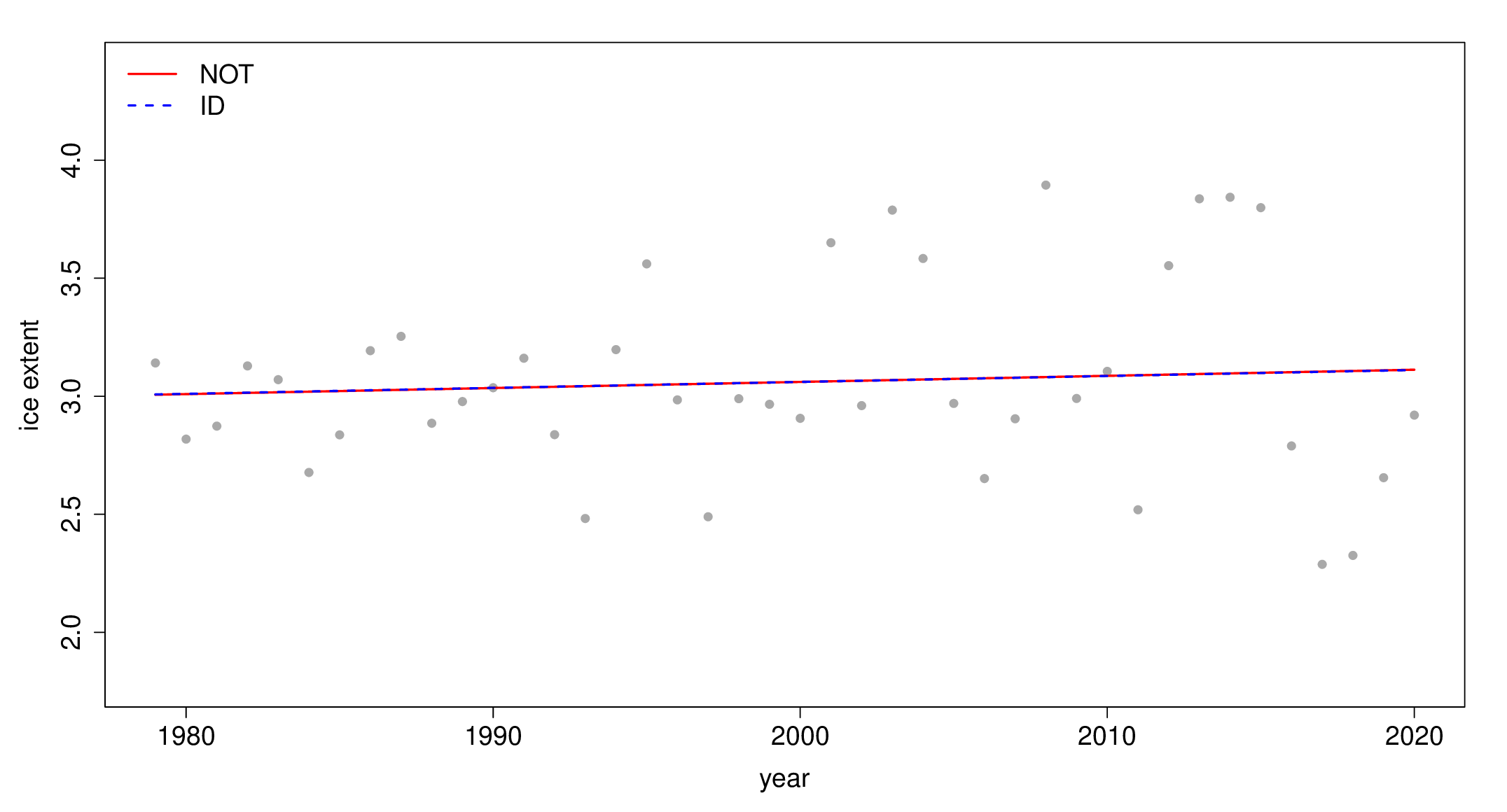}}
        \caption{NOT and ID}
        \label{fig:ATF1}
    \end{subfigure}
    \begin{subfigure}[t]{0.9\textwidth} 
        \raisebox{-\height}{\includegraphics[width=\textwidth]{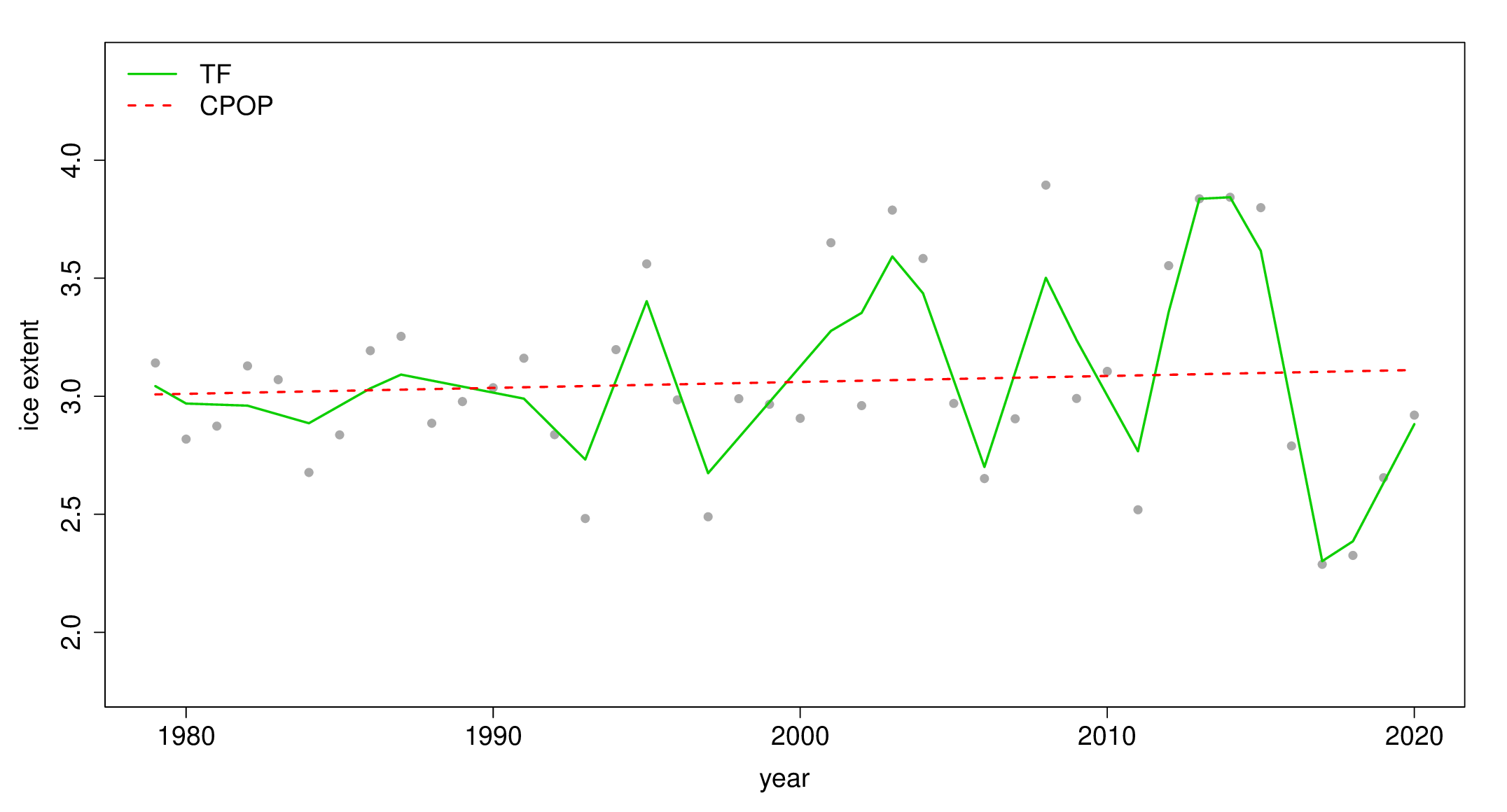}}
        \caption{TF and CPOP}
        \label{fig:ATF2}
    \end{subfigure}
    \caption {Change-point analysis for the monthly average sea ice extent of the Antarctic in February from 1979 to 2020 in Section 5.2. (a) the data series (grey dots) and estimated signal with change-points returned by NOT (\full) and ID (\dashed), (b) estimated signal with change-points returned by TF (\full) and CPOP (\chain).}
\end{figure}
\begin{figure}[ht!] 
     \centering
    \begin{subfigure}[t]{0.9\textwidth} 
        \raisebox{-\height}{\includegraphics[width=\textwidth]{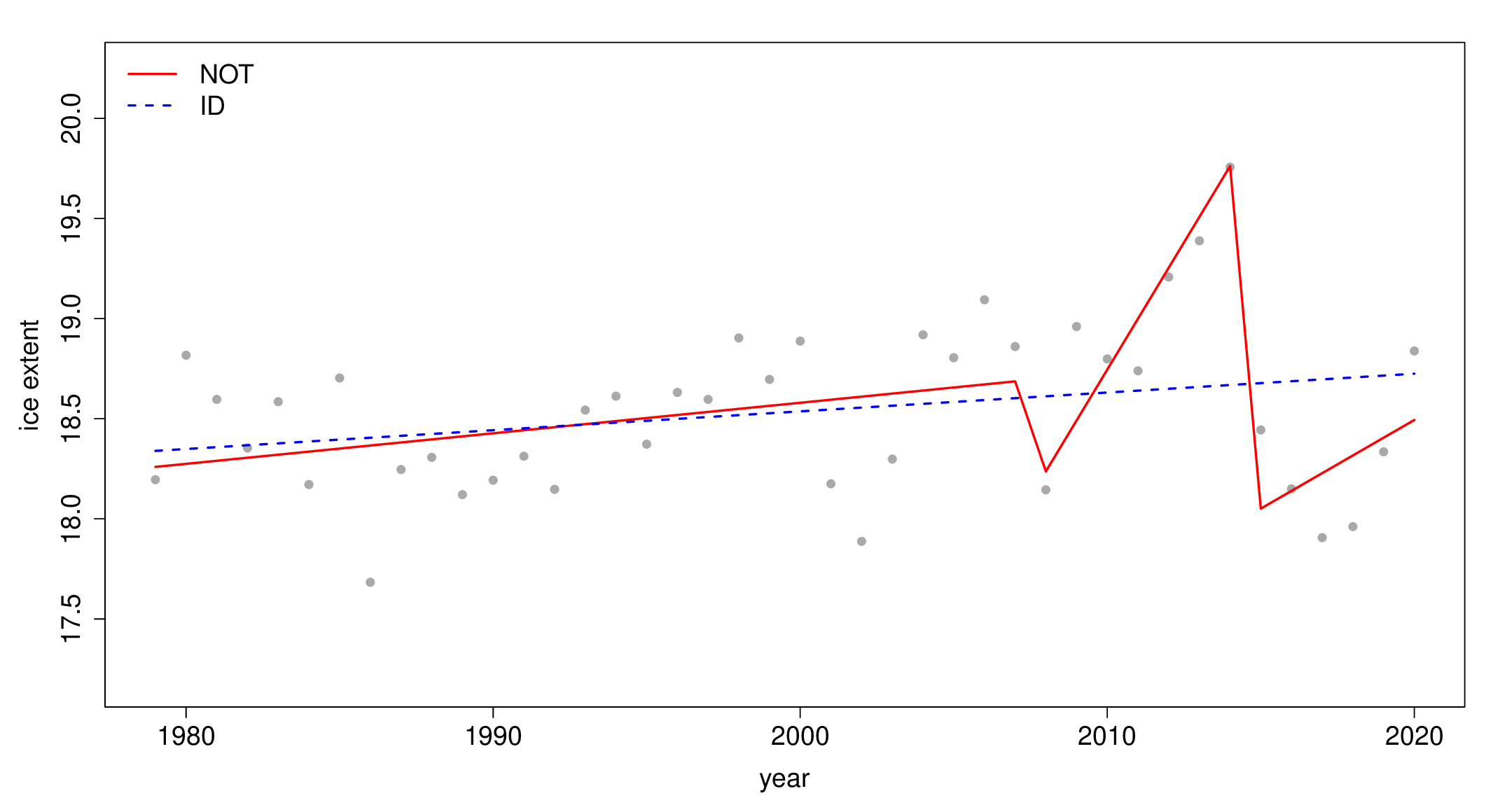}}
        \caption{NOT and ID}
        \label{fig:ATS1}
    \end{subfigure}
    \begin{subfigure}[t]{0.9\textwidth} 
        \raisebox{-\height}{\includegraphics[width=\textwidth]{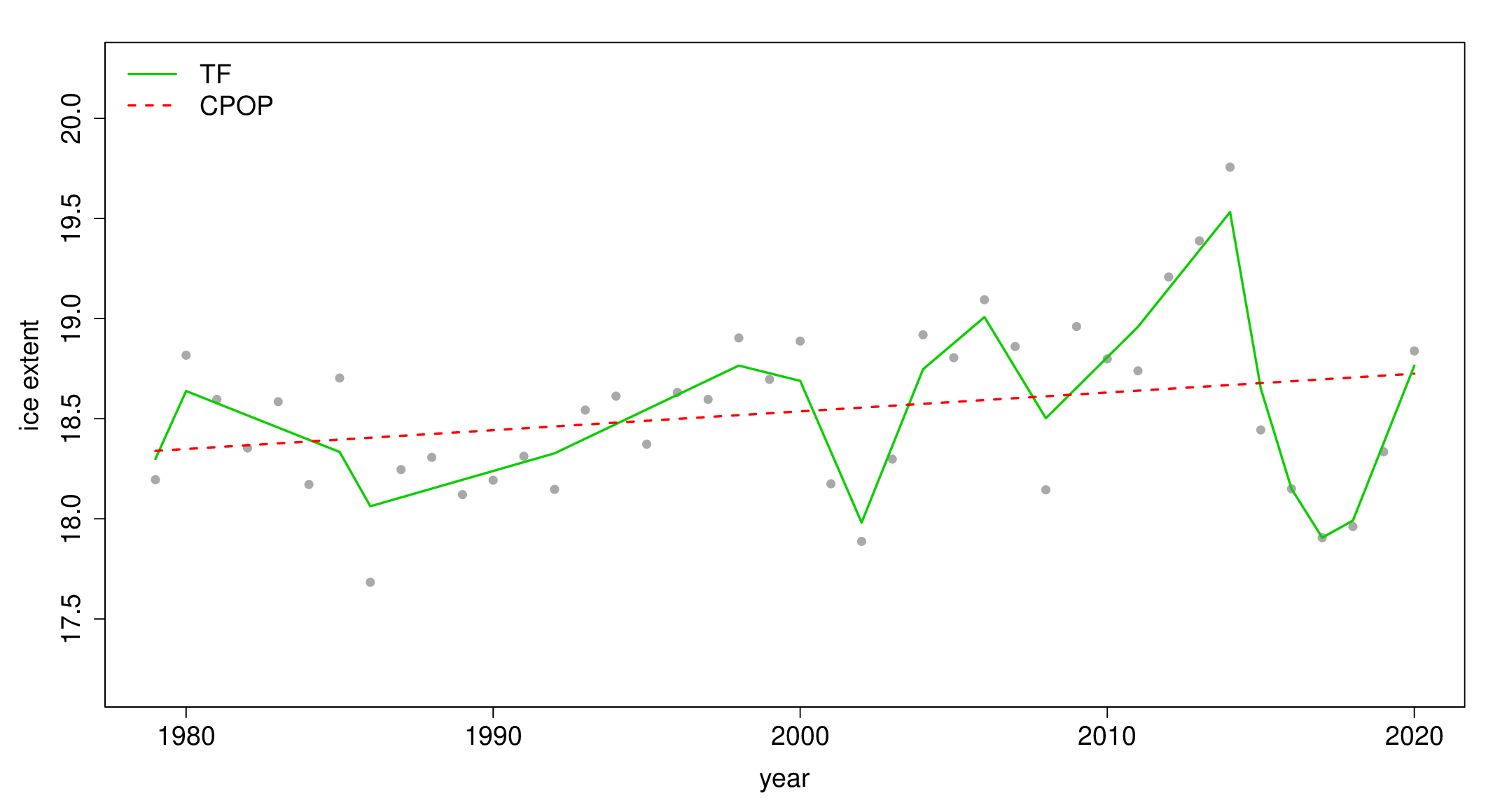}}
        \caption{TF and CPOP}
        \label{fig:ATS2}
    \end{subfigure}
    \caption {Change-point analysis for the monthly average sea ice extent of the Antarctic in September from 1979 to 2020 in Section 5.2. (a) the data series (grey dots) and estimated signal with change-points returned by NOT (\full) and ID (\dashed), (b) estimated signal with change-points returned by TF (\full) and CPOP (\chain).}
\end{figure}


\clearpage

\subsection{Nitrogen oxides concentrations}

\begin{figure}[ht!]
\begin{center}
\includegraphics[width=0.98\textwidth]{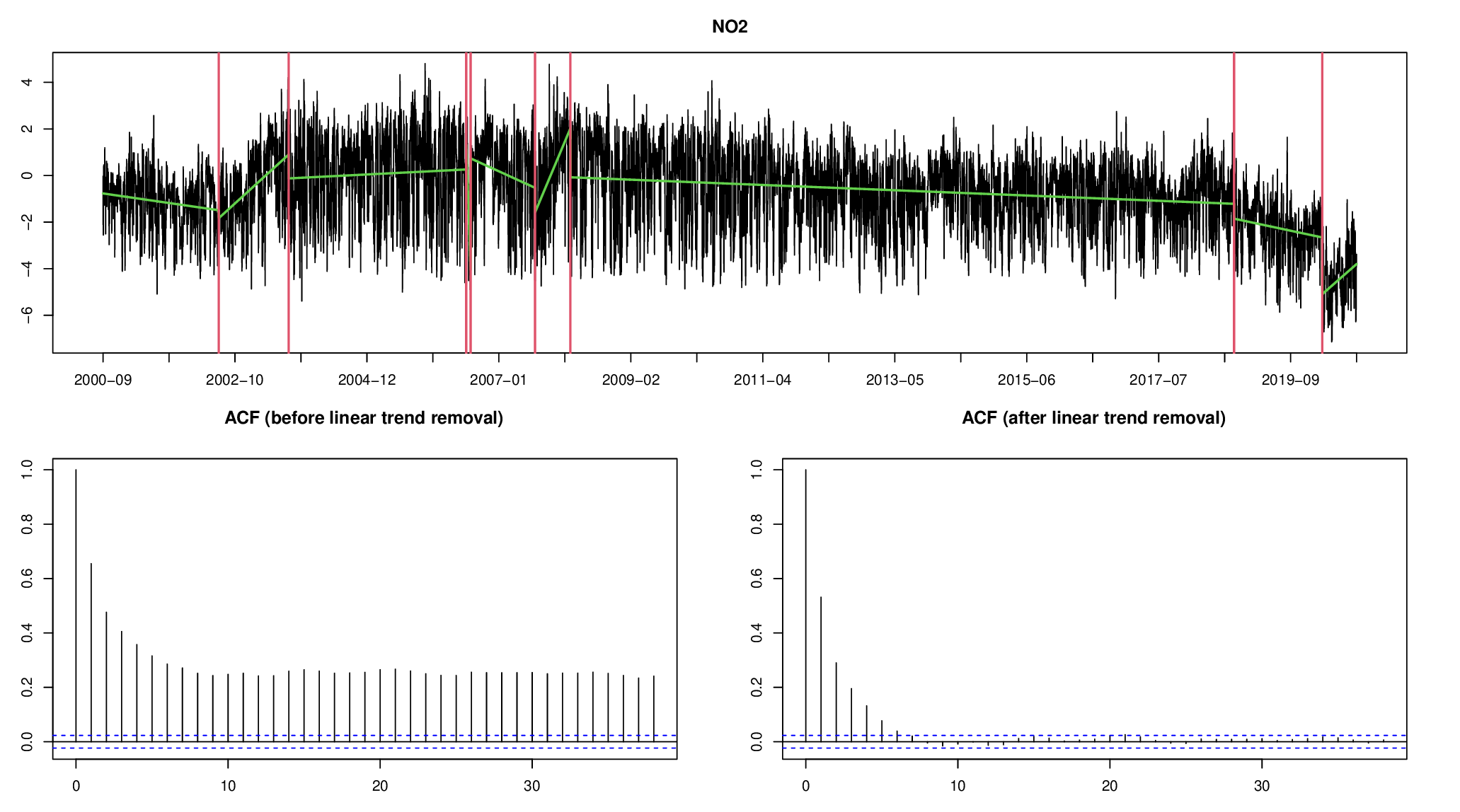}
\end{center}
\vspace*{-5mm}
 \caption{(Top) Daily average concentrations of $\text{NO}_2$ (black) with the detected change-points by TrendSegment (red) and the estimated piecewise-linear trend (green). (Bottom) Autocorrelation function of $\text{NO}_2$ before (left) and after (right) linear trend removal.}
  \label{fig:no2}
\end{figure}

In this section, we demonstrate that our TrendSegment algorithm shows a good performance on a real-world dataset that possibly has some nonnegligible autocorrelation. London air quality data is recently studied by \citet{cho2020multiple} in the context of proposing a methodology for detecting multiple changes in mean of a possibly autocorrelated time series. Using the same data but in a different context, we now detect changes in linear trend. We use the daily average concentrations of nitrogen dioxides ($\text{NO}_2$) measured from September 1, 2000 to September 30, 2020 at Marylebone Road in London, United Kingdom, which results in $T=7139$ time points. The data is downloaded from \url{https://github.com/haeran-cho/wem.gsc}, where the original data can be obtained from Defra (\url{https: //uk-air.defra.gov.uk/}). We follow the pre-processing steps used in \citet{cho2020multiple} by taking the square root transform and by removing weekly, seasonal and bank holiday effects. 

Considering that the data possibly has serial dependent and/or heavy-tailedness, we use the robust threshold ($\lambda^\text{Robust}$) introduced in Section 4.1.5 of the main paper. The top plot in Figure \ref{fig:no2} shows the detected change-points using the robust threshold selection. From the two bottom plots, we see that the persistent autocorrelations are not observable anymore after removing the linear trends, although a certain amount of autocorrelations still exists.

\section{Shape of the unbalanced wavelet basis} \label{geo}
We now explore the shape of the adaptively constructed unbalanced wavelet basis. 
First, we denote that $\psi^{(j, k)}$ is sometimes referred to as $\psi_{p,q,r}^{(j, k)}$.
One of the important properties of the unbalanced wavelet basis is that $\psi_{p, q, r}^{(j, k)}$ always has a shape of linear trend in regions that are previously merged and this linearity will also be preserved in future merges, as long as later transforms are performed under the ``two together'' rule. 
For example, two vectors $(\psi^{(0, 1)}, \psi^{(0, 2)})$ corresponding to the two smooth coefficients $s^1_{1, T}$ and $s^{[2]}_{1, T}$, have linear trends in the region $[1, T]$ as they form an orthonormal basis of the subspace $\{(x_1, x_2, \ldots, x_T)\;| \; x_1-x_2=x_2-x_3=\cdots=x_{T-1}-x_{T} \}$ of $\mathbb{R}^{T}$.
This is due to the fact that the local orthonormal transforms continue in a way of extending the geometric dimension of subspace in which an orthonormal basis lives. 

Through an illustrative example, we now show how a basis vector $\psi_{p, q, r}^{(j, k)}$ keeps its linearity in subregions that are already merged in previous scales, which includes a geometric interpretation of the TGUW transformation.
Suppose that the initial data sequence is $\bs^0=(X_1, \ldots, X_5)$ and the initial weight vectors of constancy and linearity are $\b{w}^{c}_0=(1,1,1,1,1)^\top $ and $\b{w}^{l}_0=(1,2,3,4,5)^\top $, respectively.
As we have the data sequence of length 5, the complete TGUW transform consists of 3 orthonormal transformations and the most important task for each transform is finding an appropriate orthonormal matrix.

\textbf{First merge}. Assume that $(X_3, X_4, X_5)$ is chosen as the first triplet to be merged. 
To find the values of the transform matrix $\Lambda$, 

\begin{equation} \label{e36}
\Lambda = \begin{pmatrix} 
&\ell_{1, 1} & \ell_{1, 2} & \ell_{1, 3} & \\ 
&\ell_{2, 1} & \ell_{2, 2} & \ell_{2, 3} & \\
& a & b & c &
\end{pmatrix} =  
\begin{pmatrix} 
 & \b{\ell}_1^\top  & \\ & \b{\ell}_2^\top  & \\ & \b{h}^\top 
\end{pmatrix},
\end{equation}
we first seek the detail filter, $\b{h}$, which satisfies the conditions (1) $\b{h}^\top \b{w}^{c}_{0, 3:5} = 0$, (2) $\b{h}^\top \b{w}^{l}_{0, 3:5} = 0$ and (3) $\b{h}^\top \b{h}=1$, where $\b{w}^{\cdot}_{0, p:r}$ is the subvector of length $r-p+1$. 
Thus, $\b{h}$ is obtained as a normal vector to the plane $\{(x, y, z)\;| \; x-2y+z=0 \}$. 
Then, two low filter vectors ($\b{\ell}_1$ and $\b{\ell}_2$) are obtained under the conditions, (1) $\b{\ell}_1^\top \b{h} = 0$, (2) $\b{\ell}_2^\top \b{h} = 0$, (3) $\b{\ell}_1^\top \b{\ell}_2 = 0$ and (4) $\b{\ell}_1^\top \b{\ell}_1 = \b{\ell}_2^\top \b{\ell}_2= 1$ which implies that $\b{\ell}_1$ and $\b{\ell}_2$ form an arbitrary orthonormal basis of the plane $\{(x, y, z)\;| \;x-2y+z=0 \}$ and this guarantees the linear trend of $\b{\ell}_1$ and $\b{\ell}_2$.
Now, the orthonormal transform updates the data sequence and weight vectors as follows,
\begin{align} \label{e37}
\begin{split}
\bs^0=(X_1, \ldots, X_5) &\quad \rightarrow \quad \bs=(X_1, X_2, s^{[1]}_{3,5}, s^{[2]}_{3,5}, d_{3,4,5}), \\
\b{w}^{c}_0=(1,1,1,1,1)^\top  &\quad \rightarrow \quad {\b{w}^{c}}=(1,1,e_{c_1},e_{c_2},0)^\top , \\
\b{w}^{l}_0=(1,2,3,4,5)^\top  &\quad \rightarrow \quad {\b{w}^{l}}=(1,2,e_{l_1},e_{l_2},0)^\top , \\
\end{split}
\end{align}
where the constants $(e_{c_1},e_{c_2})$ and $(e_{l_1},e_{l_2})$ are obtained by $\Lambda \b{w}^{c}_{0, 3:5} = (e_{c_1}, e_{c_2}, 0)^\top $ and $\Lambda \b{w}^{l}_{0, 3:5} = (e_{l_1}, e_{l_2}, 0)^\top $, respectively. As $\b{\ell}_1$ and $\b{\ell}_2$ form an orthonormal basis of the plane $\{(x, y, z)\;| \;x-2y+z=0 \}$, $e_{c_1},e_{c_2}$ and $e_{l_1},e_{l_2}$ are unique constants which represent $\b{w}^{c}_{0, 3:5}$ and $\b{w}^{l}_{0, 3:5}$ as a linear span of basis vectors $\b{\ell}_1$ and $\b{\ell}_2$ as follows:
\begin{align} \label{e38}
\begin{split}
\b{w}^{c}_{0, 3:5} = e_{c_1}\b{\ell}_1 + e_{c_2}\b{\ell}_2, \quad \b{w}^{l}_{0, 3:5} = e_{l_1}\b{\ell}_1 + e_{l_2}\b{\ell}_2. \\
\end{split}
\end{align}

Importantly, the orthonormal transform matrix $\Psi_{T \times T}$ introduced in (5) (i.e. an orthonormal basis in $\mathbb{R}^5$ in this example) is constructed by recursively updating its initial input $\Psi_0=\mathbf{I}_{5 \times 5}$ through local orthonormal transforms. 
For example, if $(p, q, r)^\text{th}$ elements in $\bs$ are selected to be merged, then we extract the corresponding $(p, q, r)^\text{th}$ columns of $\Psi^\top $ and update them through the matrix multiplication with $\Lambda$ used in that merge. 
Therefore, the first orthonormal transform performed in \eqref{e37} updates the initial matrix $\Psi_0^\top $ by multiplying $\Lambda$ to the corresponding $(3, 4, 5)^{th}$ columns of $\Psi_0^\top $ which returns the following,
\begin{equation} \label{e39}
\Psi^\top  =  
\begin{pmatrix} 
 & 1 & 0 & 0 & 0 & 0 & \\ 
 & 0 & 1 & 0 & 0 & 0 & \\
 & 0 & 0 & \ell_{1, 1} & \ell_{2, 1} & a & \\
 & 0 & 0 & \ell_{1, 2} & \ell_{2, 2} & b & \\
 & 0 & 0 & \ell_{1, 3} & \ell_{2, 3} & c & 
\end{pmatrix}.
\end{equation}
The $5^{\text{th}}$ column of $\Psi^\top $ is now fixed (not going to be updated again) as it corresponds to the detail coefficient but other four columns corresponding to the smooth coefficients in $\b{s}$ would be updated as the merging continues. 

\textbf{Second merge.} Suppose that $(X_2, s^{[1]}_{3,4,5}, s^{[2]}_{3,4,5})$ are selected to be merged next under the ``two together'' rule. 
Then we need to find the following orthonormal transform matrix, 
\begin{equation} \label{e391}
\Lambda^* = \begin{pmatrix} 
&\ell^*_{1, 1} & \ell^*_{1, 2} & \ell^*_{1, 3} &\\ 
&\ell^*_{2, 1} & \ell^*_{2, 2} & \ell^*_{2, 3} &\\
& a^* & b^* & c^* &
\end{pmatrix} =  
\begin{pmatrix} 
 & {\b{\ell}_1^*}^\top  & \\ & {\b{\ell}_2^*}^\top  & \\ & {\b{h}^*}^\top  &
\end{pmatrix},
\end{equation}
where its elements would be different from those in \eqref{e36}.
The detail filter ${\b{h}^*}^\top =(a^*, b^*, c^*)$ is constructed from the corresponding weight vectors, $\b{w}^c_{2:4} = (1,e_{c_1}, e_{c_2})^\top $ and $\b{w}^l_{2:4} = (2,e_{l_1},e_{l_2})^\top $, by satisfying the conditions (1) ${\b{h}^*}^\top \b{w}^c_{2:4} = 0$, (2) ${\b{h}^*}^\top \b{w}^l_{2:4} = 0$ and (3) ${\b{h}^*}^\top {\b{h}^*}=1$.
The detail filter is a weight vector designed for indicating the strength of linearity in $(X_2, X_3, X_4, X_5)$ as $(e_{c_1}, e_{c_2})$ and $(e_{l_1},e_{l_2})$ already contain the information of three raw observations $(X_3, X_4, X_5)$. 
Then, two low filters, ${\b{\ell}_1^*}$ and ${\b{\ell}_2^*}$, are obtained by satisfying the conditions, ${\b{\ell}_1^*}^\top {\b{h}^*} = 0$, ${\b{\ell}_2^*}^\top {\b{h}^*} = 0$, ${\b{\ell}_1^*}^\top {\b{\ell}_2^*} = 0$ and ${\Lambda^*}^\top \Lambda^*=\mathbf{I}$. 
Now the data sequence and the weight vectors are updated as follows,
\begin{align*} 
\bs=(X_1, X_2, s^{[1]}_{3, 5}, s^{[2]}_{3, 5}, d_{3,4,5}) &\quad \rightarrow \quad \bs=(X_1, s^{[1]}_{2, 5}, s^{[2]}_{2, 5}, d_{2,2,5}, d_{3,4,5}), \\
{\b{w}_c}=(1,1,e_{c_1},e_{c_2},0)^\top  &\quad \rightarrow \quad {\b{w}_c}=(1,e^*_{c_1},e^*_{c_2},0,0)^\top , \numberthis \label{e3915} \\
{\b{w}_l}=(1,2,e_{l_1},e_{l_2},0)^\top  &\quad \rightarrow \quad {\b{w}_l}=(1,e^*_{l_1},e^*_{l_2},0,0)^\top ,
\end{align*}
and $\Psi^\top $ is also updated into
\begin{equation} \label{e392}
\Psi^\top  =  
\begin{pmatrix} 
  1 & 0 & 0 & 0 & 0 \\ 
  0 & \ell^*_{1, 1} & \ell^*_{2, 1} & a^* & 0 \\
  \begin{matrix} & \;0 & \\ & \;0 & \\ & \;0 & \end{matrix} 
 & \begin{pmatrix} \\ \ell^*_{1, 2}\b{\ell}_1 + \ell^*_{1, 3}\b{\ell}_2 \\ \\ \end{pmatrix} 
 & \begin{pmatrix} \\ \ell^*_{2, 2}\b{\ell}_1 + \ell^*_{2, 3}\b{\ell}_2 \\ \\ \end{pmatrix}
 & \begin{pmatrix} \\ b^*\b{\ell}_1 + c^*\b{\ell}_2 \\ \\  \end{pmatrix} 
 & \begin{matrix} & \;a & \\ & \;b & \\ & \;c & \end{matrix} 
\end{pmatrix}.
\end{equation}
At this scale, the $4^{\text{th}}$ column of $\Psi^\top $ is fixed. 
This corresponds to the Type 2 basis vector in \eqref{e91} whose non-zero subregion is composed of a single point ($a^*$) and a linear trend ($b^*\b{\ell}_1 + c^*\b{\ell}_2$).

Importantly, the orthonormal transform at this scale is performed in a way of returning an orthonormal basis of the expanded subspace e.g. $2^{\text{nd}}$ and $3^{\text{rd}}$ columns of \eqref{e392} (which are referred to as ${\b{\ell}_1^{**}}$ and ${\b{\ell}_2^{**}}$ in \eqref{e393}) are obtained as an arbitrary orthonormal basis of the subspace $\{(w,x,y,z)\;|\;w-x=x-y=y-z \}$ of $\mathbb{R}^4$. This is due to the semi-orthogonality of the transformation matrix $\mathbf{\Pi}$ in \eqref{e393} 
which extends the dimension from  $\mathbb{R}^3$ to  $\mathbb{R}^4$ but preserves the fact that ($\b{\ell}_1^*, \b{\ell}_2^*$) and ($\b{\ell}_1^{**}, \b{\ell}_2^{**}$) form an arbitrary orthonormal basis of the corresponding subspaces. 
This guarantees the properties, ${\b{\ell}_1^{**}}^{\top}{\b{\ell}_2^{**}}=0$ and ${\b{\ell}_1^{**}}^{\top}{\b{\ell}_1^{**}}={\b{\ell}_2^{**}}^{\top}{\b{\ell}_2^{**}}=1$, where
\begin{align*} 
&{\b{\ell}_1^{**}} = \begin{pmatrix} 
 \ell^*_{1, 1} \\ \begin{pmatrix} \\ \ell^*_{1, 2}\b{\ell}_1 + \ell^*_{1, 3}\b{\ell}_2 \\ \\ \end{pmatrix} 
\end{pmatrix} = 
\begin{pmatrix} 
 1 & 0 & 0 \\
 \begin{matrix} \;0 \\ \;0 \\ \;0 \end{matrix} 
 & \begin{pmatrix} \\ \b{\ell}_1 \\ \\ \end{pmatrix} 
 & \begin{pmatrix} \\ \b{\ell}_2 \\ \\ \end{pmatrix} 
\end{pmatrix}
\begin{pmatrix} 
 \ell^*_{1, 1} \\ \ell^*_{1, 2} \\ \ell^*_{1, 3} 
\end{pmatrix} =
\mathbf{\Pi}
\begin{pmatrix} 
 \ell^*_{1, 1} \\ \ell^*_{1, 2} \\ \ell^*_{1, 3} 
\end{pmatrix}, \\
&{\b{\ell}_2^{**}}= \begin{pmatrix} 
 \ell^*_{2, 1} \\ \begin{pmatrix} \\ \ell^*_{2, 2}\b{\ell}_1 + \ell^*_{2, 3}\b{\ell}_2 \\ \\ \end{pmatrix}
\end{pmatrix} = 
\begin{pmatrix} 
 1 & 0 & 0 \\
 \begin{matrix} \;0 \\ \;0 \\ \;0 \end{matrix} 
 & \begin{pmatrix} \\ \b{\ell}_1 \\ \\ \end{pmatrix} 
 & \begin{pmatrix} \\ \b{\ell}_2 \\ \\ \end{pmatrix} 
\end{pmatrix}
\begin{pmatrix} 
\ell^*_{2, 1} \\ \ell^*_{2, 2} \\ \ell^*_{2, 3}
\end{pmatrix} =
\mathbf{\Pi}
\begin{pmatrix} 
\ell^*_{2, 1} \\ \ell^*_{2, 2} \\ \ell^*_{2, 3}
\end{pmatrix}, \numberthis \label{e393} 
\end{align*}
and $\mathbf{\Pi}$ is obtained from the $2^{\text{nd}}$ to $4^{\text{th}}$ columns of \eqref{e39} and the selected rows correspond to the indices of smooth coefficients associated in the orthonormal transformation in \eqref{e391}.  

As is in \eqref{e38},  
now the extended subregions of the original weight vectors, $\b{w}^{c}_{0, 2:5}$ and $\b{w}^{l}_{0, 2:5}$, can also be presented as a linear combination of $\b{\ell}_1^{**}$ and $\b{\ell}_2^{**}$ as follows:
\begin{equation} \label{e3935}
\b{w}^{c}_{0, 2:5} = e^*_{c_1}{\b{\ell}_1^{**}} + e^*_{c_2}{\b{\ell}_2^{**}}, \quad \b{w}^{l}_{0, 2:5} = e^*_{l_1}{\b{\ell}_1^{**}} + e^*_{l_2}{\b{\ell}_2^{**}},
\end{equation}
where $\b{\ell}_1^{**}$ and $\b{\ell}_2^{**}$ form an orthonormal basis of the subspace $\{(w,x,y,z)\;|\;w-x=x-y=y-z \}$ of $\mathbb{R}^4$.
This can be simply shown by 1) expressing the weight vectors as a linear combination of two low filters,
\begin{align} \label{e394}
\begin{split}
&\b{w}^c_{2:4} = (1,e_{c_1}, e_{c_2})^\top  = e^*_{c_1}{\b{\ell}_1^*} + e^*_{c_2}{\b{\ell}_2^*}, \\
&\b{w}^l_{2:4} = (2,e_{l_1},e_{l_2})^\top  = e^*_{l_1}{\b{\ell}_1^*} + e^*_{l_2}{\b{\ell}_2^*}, 
\end{split}
\end{align}
and 2) performing the matrix multiplication with $\mathbf{\Pi}$ in \eqref{e393} to both sides of \eqref{e394}, 
\begin{align} \label{e395}
\begin{split}
& \text{LHS}:
\mathbf{\Pi}
\b{w}^c_{2:4}  = (1, e_{c_1}\b{\ell}_1 + e_{c_2}\b{\ell}_2)^\top  = (1,1,1,1)^\top  = \b{w}^{c}_{0, 2:5} ,\quad  \text{RHS}: e^*_{c_1}{\b{\ell}_1^{**}} + e^*_{c_2}{\b{\ell}_2^{**}}, \\
& \text{LHS}:
\mathbf{\Pi} 
\b{w}^l_{2:4}  = (2, e_{l_1}\b{\ell}_1 + e_{l_2}\b{\ell}_2)^\top  = (2,3,4,5)^\top  = \b{w}^{l}_{0, 2:5} ,\quad \text{RHS}: e^*_{l_1}{\b{\ell}_1^{**}} + e^*_{l_2}{\b{\ell}_2^{**}}.
\end{split}
\end{align}

\textbf{Last merge.}
In the same manner, after the last orthonormal transform is applied to $(X_1, s^{[1]}_{2, 5}, s^{[2]}_{2, 5})$, we end up with the finalised $\Psi^\top $ in which an orthonormal basis of the subspace $\{(v,w,x,y,z)\;|\;v-w=w-x=x-y=y-z \}$ of $\mathbb{R}^5$ is shown in its first and second columns where these two columns correspond to two basis vectors, $\psi^{(0, 1)}$ and $\psi^{(0, 2)}$, in (5). Regardless of the length of data ($T$), the first two columns of the finalised $\Psi^\top $ build two smooth coefficients ($s^{[1]}_{1, T}, s^{[2]}_{1, T}$) and always keep a linear trend with length $T$, while the shape of other columns of $\Psi^\top $ corresponding to the detail coefficients depends on the type of merge and follows one of the forms in \eqref{e91}.

As shown above, the non-uniqueness of the low filters has no effect on preserving the linearity of the subregions that are already merged. In simulation studies, we empirically found that the choice of low filters has no qualitative effect on the results as long as they are chosen by satisfying the orthonormality condition of the transform, thus we used a fixed type of function for choosing a set of low filters rather than choosing an arbitrary set of low filters that satisfies the orthonormal condition every run which also saves the computational costs.


\section{A practical way to implement the TGUW transformation} \label{tguw_prac}
In this section, we explore a way of implementing the TGUW transform. 
As briefly mentioned in Section 2.2.3, it is implemented by consecutively updating so-called weight vectors of constancy and linearity. 
These two weight vectors are initially used in the first stage of the TGUW transform for obtaining the detail filter $\b{h}$ and updated through the orthonormal transform.
In detail, Steps 1 and 5 of the TGUW algorithm presented in Section 2.2.3 can be reformulated by weight vectors as follows.

\textbf{Step 1}. At each scale $j$, find the set of triplets that are candidates for merging under the ``two together'' rule and compute the corresponding detail coefficients. 
Regardless of the type of merge, a detail coefficient $d_{p,q,r}^{\cdot}$ is, in general, obtained as
\begin{equation} \label{e220}
d_{p,q,r}^{\cdot} = a\bs_{p:r}^1 + b\bs_{p:r}^2 + c\bs_{p:r}^3,
\end{equation}
where $p \leq q < r$, $\bs_{p:r}^k$ is the $k^{\text{th}}$ smooth coefficient of the subvector $\bs_{p:r}$ with a length of $r-p+1$ and the constants $a, b, c$ are the elements of the detail filter $\b{h}=(a, b, c)^{\top}$. 
Specifically, the detail filter $\b{h}$ is established by solving the following equations, 
\begin{align} \label{e23}
\begin{split}
& a\b{w}^{c, 1}_{p:r} + b\b{w}^{c, 2}_{p:r} + c\b{w}^{c, 3}_{p:r} = 0, \\
& a\b{w}^{l, 1}_{p:r} + b\b{w}^{l, 2}_{p:r} + c\b{w}^{l, 3}_{p:r} = 0, \\
& a^2 + b^2 + c^2 = 1,
\end{split}
\end{align}
where $\b{w}^{\cdot, k}_{p:r}$ is $k^{\text{th}}$ non-zero element of the subvector $\b{w}^{\cdot}_{p:r}$ with a length of $r-p+1$, and $\b{w}^c$ and $\b{w}^l$ are weight vectors of constancy and linearity, respectively, in which the initial inputs have a form of ${\b{w}^{c}_0}=(1,1, \ldots, 1)^\top , {\b{w}^{l}_0}=(1, 2, \ldots, T)^\top $. 
The last condition in \eqref{e23} is to preserve the orthonormality of the transform and the detail filter $\b{h}$ becomes a unit normal vector of the plane $\{(x, y, z)\;| \; x-2y+z=0 \}$. 
The solution to \eqref{e23} is unique up to multiplication by $-1$ and this can be simply shown by solving the equations e.g. $a+b+c=0$, $a+2b+3c=0$ and $a^2 + b^2 + c^2 = 1$. 

More specifically, the detail coefficient in \eqref{e220} is formulated for each type of merging introduced in Section 2.2.1 as follows. \\
Type 1: merging three initial smooth coefficients $(s^0_{p, p}, s^0_{p+1, p+1}, s^0_{p+2, p+2})$,
\begin{equation} \label{e24}
d_{p, p+1, p+2} = a_{p, p+1, p+2}s^0_{p, p} + b_{p, p+1, p+2}s^0_{p+1, p+1} + c_{p, p+1, p+2}s^0_{p+2, p+2}.
\end{equation} 
Type 2: merging one initial and a paired smooth coefficient $(s^0_{p, p}, s^{[1]}_{p+1, r}, s^{[2]}_{p+1, r})$, 
\begin{equation} \label{e25}
d_{p, p, r} = a_{p, p, r}s^0_{p, p} + b_{p, p, r}s^{[1]}_{p+1, r} + c_{p, p, r}s^{[2]}_{p+1, r}, \quad \text{where} \quad p+2<r,
\end{equation}
similarly, when merging a paired smooth coefficient and one initial, $(s^{[1]}_{p, r-1}, s^{[2]}_{p, r-1}, s^0_{r, r})$, 
\begin{equation} \label{e26}
d_{p, r-1, r} = a_{p, r-1, r}s^{[1]}_{p, r-1} + b_{p, r-1, r}s^{[2]}_{p, r-1} + c_{p, r-1, r}s^0_{r, r}, \quad \text{where} \quad p+2<r.
\end{equation}
Type 3: merging two sets of (paired) smooth coefficients, $(s^{[1]}_{p, q}, s^{[2]}_{p, q})$ and $(s^{[1]}_{q+1, r}, s^{[2]}_{q+1, r})$,
\begin{equation} \label{e27}
\begin{aligned}[c]
d^{[1]}_{p, q, r} = a^1_{p, q, r}s^{[1]}_{p, q} + b^1_{p, q, r}s^{[2]}_{p, q} + c^1_{p, q, r}s^{[1]}_{q+1, r}\\
d^{[2]}_{p, q, r} = a^2_{p, q, r}s^{01}_{p, r} + b^2_{p, q, r}s^{02}_{p, r} + c^2_{p, q, r}s^{[2]}_{q+1, r}
\end{aligned}
\qquad\Longrightarrow\qquad
\begin{aligned}[c]
d_{p, q, r} = \max(|d^{[1]}_{p, q, r}|, |d^{[2]}_{p, q, r}|),\\
\end{aligned}
\end{equation}
where $q > p+1$ and $r > q+2$. 
Importantly, the two consecutive merges in \eqref{e27} are achieved by visiting the same two adjacent data regions twice. 
In this case, after the first detail coefficient, $d^{[1]}_{p, q, r}$, has been obtained,  we instantly update the corresponding triplets $\bs$, $\b{w}^{c}$ and $\b{w}^{l}$ via an orthonormal transform as defined in \eqref{e28} and \eqref{e29}. 
Therefore, the second detail filter, $(a^2_{p, q, r}, b^2_{p, q, r}, c^2_{p, q, r})$, is constructed with the updated $\b{w}^c$ and $\b{w}^l$ in a way that satisfies the conditions \eqref{e23}. \\

\textbf{Step 5}. For each $|d_{p,q,r}^{\cdot} |$ extracted in step 4, merge the corresponding smooth coefficients by updating the corresponding triplet in $\bs$, $\b{w}^{c}$ and $\b{w}^{l}$ through the orthonormal transform as follows,
\begin{align} \label{e280}
\begin{pmatrix} 
s^{[1]}_{p, r} \\ 
s^{[2]}_{p, r} \\ 
d^\cdot_{p, q, r} 
\end{pmatrix} = 
\begin{pmatrix} 
& & \b{\ell}_1^\top  & & \\ 
& & \b{\ell}_2^\top  & &  \\
& & \b{h}^\top  & & \\
\end{pmatrix}
\begin{pmatrix} 
\bs_{p:r}^1 \\
\bs_{p:r}^2\\
\bs_{p:r}^3
\end{pmatrix} =
\Lambda \begin{pmatrix} 
\bs_{p:r}^1 \\
\bs_{p:r}^2\\
\bs_{p:r}^3
\end{pmatrix} , \\ \label{e29}
\begin{pmatrix} 
{w}^{c, 1}_{p, r}\\ 
{w}^{c, 2}_{p, r}\\ 
0 
\end{pmatrix} =
\Lambda \begin{pmatrix} 
\b{w}^{c, 1}_{p:r} \\
\b{w}^{c, 2}_{p:r} \\
\b{w}^{c, 3}_{p:r}
\end{pmatrix} , \quad
\begin{pmatrix} 
{w}^{l, 1}_{p, r}\\ 
{w}^{l, 2}_{p, r}\\ 
0 
\end{pmatrix} =
\Lambda \begin{pmatrix} 
\b{w}^{l, 1}_{p:r} \\
\b{w}^{l, 2}_{p:r} \\
\b{w}^{l, 3}_{p:r}
\end{pmatrix}.
\end{align}
The key step is finding the  $3 \times 3$ orthonormal matrix, $\Lambda$, which is composed of one detail and two low-pass filter vectors in its rows. 
Firstly the detail filter $\b{h}^\top $ is determined to satisfy the conditions in \eqref{e23}, and then the two low-pass filters ($\b\ell_{1}^\top, \b\ell_{2}^\top $) are obtained by satisfying the orthonormality of $\Lambda$.
There is no uniqueness in the choice of ($\b\ell_{1}^\top, \b\ell_{2}^\top $), but as described in Section \ref{geo}, this has no effect on the orthonormal transformation itself.


\section{Extension to piecewise-quadratic signal} 
\label{tguw_quad}

\begin{figure}[ht!] 
\centering
\includegraphics[width=14cm, height=12cm]{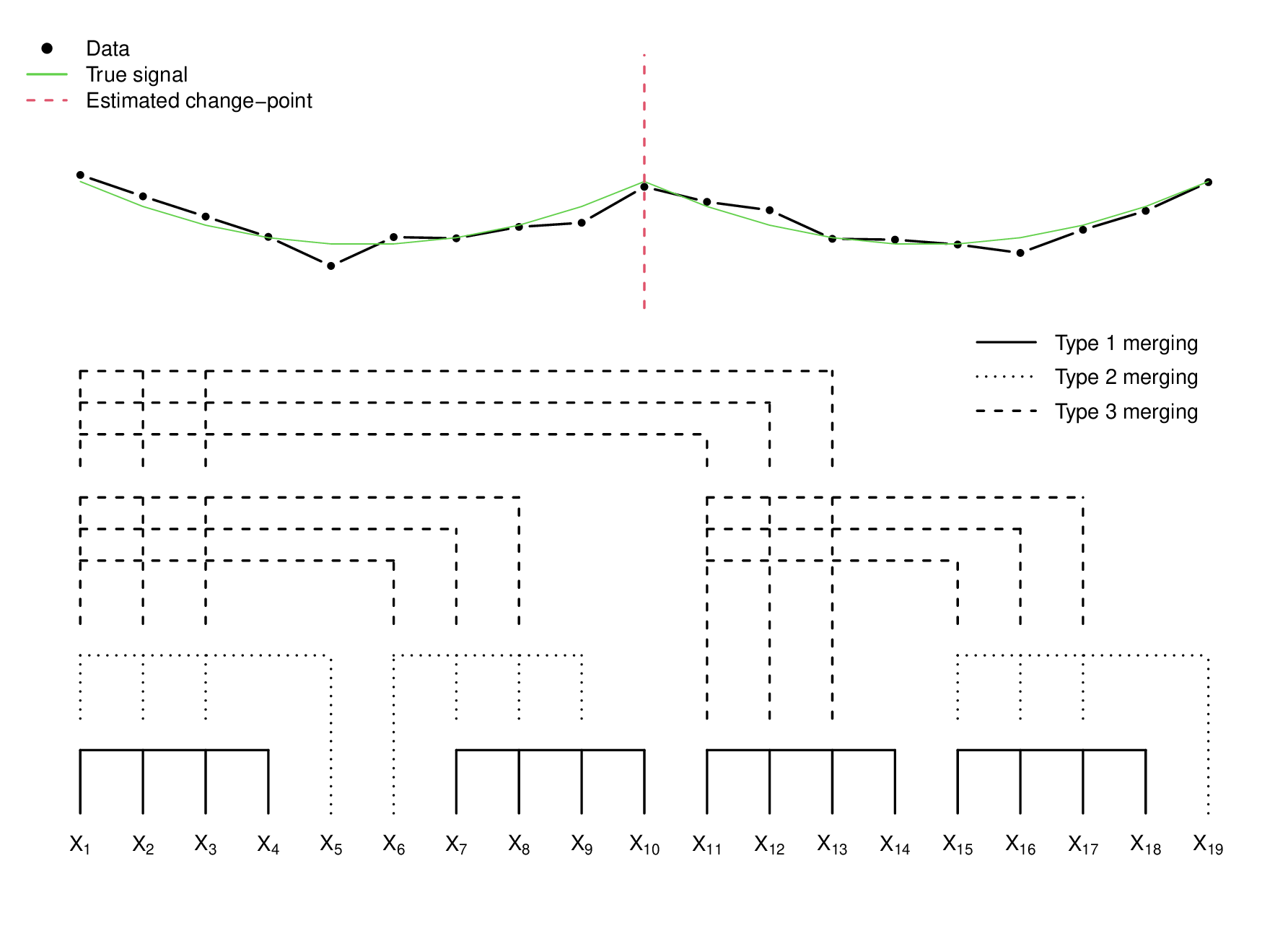}
 \caption {Example of data with one true change-point at $t=10$ in its underlying piecewise-quadratic signal (top) along with the tree structure constructed in TGUW transform by merging (bottom).}
\label{fig:quad}
\end{figure}

In this section, we explore how the TGUW transform can be extended to handle piecewise-quadratic signals. Considering the fact that we perform an orthonormal transformation to the chosen pair (triplet) to deal with piecewise-constant (piecewise-linear) signals, it is natural to perform a transform to the chosen quadraplet of the smooth conefficients in the process of establishing a data-adaptive unbalanced wavelet basis. In each merge, four adjacent smooth coefficients are selected and the orthonormal transformation converts them into one detail and three (updated) smooth coefficients. Those three updated smooth coefficients are tripled in the sense that they contain information about one local quadratic regression fit. Therefore, any such triplet of smooth coefficients cannot be separated when choosing quadruplet in any subsequent merges which can be called as ``three together'' rule (instead of ``two together'' rule invented for piecewise-linear model). We now give a simple example to illustrate how the TGUW transform for piecewise-quadratic siganal works. Figure \ref{fig:quad} shows the merging history of the modified TGUW transform which follows the ``three together'' rule. Three different types of merges are similary defined as for piecewise-linear signal except the fact that the merges are performed on quadraplet instead of triplet. The tree structure show that the modified TGUW transform performs well in detecting a single change-point in piecewise-quadratic scenario as the last type 3 merge is corresponding to the true change-point.

\end{appendix}

\bibliographystyle{apalike}
\bibliography{TSref}

\end{document}